\documentclass[a4paper,11pt]{article}
\usepackage[english]{babel}
\pdfoutput=1 

\usepackage{jheppub} 

\usepackage[T1]{fontenc} 

\usepackage[T1,T2A]{fontenc}
\usepackage[utf8x]{inputenc}

\usepackage{ytableau}
\ytableausetup{smalltableaux}
\RequirePackage[vcentermath]{youngtab}

\newcommand{\bi}{\begin{itemize}}
    \newcommand{\ei}{\end{itemize}}
\newcommand{\bea}{\begin{eqnarray}}
    \newcommand{\eea}{\end{eqnarray}}
\newcommand{\bt}{\begin{tabular}}
    \newcommand{\et}{\end{tabular}}
\newcommand{\bc}{\begin{center}}
    \newcommand{\ec}{\end{center}}

\newcommand{\be}{\begin{equation}}
    \newcommand{\ee}{\end{equation}}
\newcommand{\ba}{\begin{array}}
    \newcommand{\ea}{\end{array}}

\newcommand{\lb}[1]{\label{#1}}

\def\bbox{{\,\lower0.9pt\vbox{\hrule \hbox{\vrule height 0.2 cm
                \hskip 0.2 cm \vrule height 0.2 cm}\hrule}\,}}
\newcommand{\dsl}{\pa \kern-0.5em /}

\newcommand{\nn}{\nonumber \\}

\usepackage{etoolbox}

\makeatletter
\newenvironment{sqcases}{%
    \matrix@check\sqcases\env@sqcases
}{%
    \endarray\right.%
}
\def\env@sqcases{%
    \let\@ifnextchar\new@ifnextchar
    \left\lbrack
    \def\arraystretch{1.2}%
    \array{@{}l@{\quad}l@{}}%
}
\makeatother

\usepackage{tikz}
\newcommand*\circled[1]{\tikz[baseline=(char.base)]{
        \node[shape=circle,draw,inner sep=2pt] (char) {#1};}}

\makeatletter \@addtoreset{equation}{section} \makeatother

\def\slashchar#1{\setbox0=\hbox{$#1$}           
    \dimen0=\wd0                                 
    \setbox1=\hbox{/} \dimen1=\wd1               
    \ifdim\dimen0>\dimen1                        
    \rlap{\hbox to \dimen0{\hfil/\hfil}}      
    #1                                        
    \else                                        
    \rlap{\hbox to \dimen1{\hfil$#1$\hfil}}   
    /                                         
    \fi}

\pdfoutput=1

\title{\boldmath $\mathcal{N}=2$ superconformal higher-spin multiplets and their \textbf{hypermultiplet
couplings}}

\author[a,b,c]{Ioseph Buchbinder}
\author[a,d]{Evgeny~Ivanov,}
\author[a,d]{Nikita~Zaigraev}

\affiliation[a]{Bogoliubov Laboratory of Theoretical Physics,
JINR,\\141980 Dubna, Moscow region, Russia}

\affiliation[b]{Center
of Theoretical Physics, Tomsk State Pedagogical University,\\
634061, Tomsk,  Russia}

\affiliation[c]{Tomsk University of Control Systems and
Radioelectronics (TUSUR), \\634034, Tomsk, Russia}

\affiliation[d]{Moscow Institute of Physics and Technology,\\ 141700
Dolgoprudny, Moscow region, Russia}

\emailAdd{buchbinder@theor.jinr.ru} \emailAdd{eivanov@theor.jinr.ru}
\emailAdd{nikita.zaigraev@phystech.edu}

\abstract{We construct an off-shell $\mathcal{N}=2$
superconformal cubic vertex for the hypermultiplet coupled to an arbitrary
integer higher spin ${\bf s}$  gauge $\mathcal{N}=2$ supermultiplet
in a general $\mathcal{N}=2$ conformal supergravity background.
We heavily use $\mathcal{N}=2, 4D$ harmonic superspace that provides
an unconstrained superfield Lagrangian description. We start with
$\mathcal{N}=2$ global superconformal symmetry transformations of
the free hypermultiplet model and require invariance of the cubic vertices
of general form under these transformations and their gauged version. As a
result, we deduce $\mathcal{N}=2, 4D$ unconstrained analytic
 superconformal gauge potentials for an arbitrary integer ${\bf s}$. These are the basic
ingredients of the approach under consideration. We describe the
properties of the gauge potentials, derive the corresponding
superconformal and gauge transformation laws, and inspect the off-shell
contents of the thus obtained $\mathcal{N}=2$ superconformal higher-spin ${\bf s}$ multiplets in the Wess-Zumino gauges. The
spin ${\bf s}$ multiplet involves $8(2{\bf s} -1)_B + 8(2{\bf s}-1)_F$ essential off-shell degrees of freedom. The
cubic vertex has the generic  structure {\it higher spin gauge superfields
$\times$ hypermultiplet supercurrents}. We present the explicit form of the relevant supercurrents.}

\makeatletter
\gdef\@fpheader{}
\makeatother

\begin{document}
\maketitle
\flushbottom

\section{Introduction}

Superconformal field theories constitute an important subclass of field theories, with numerous applications in classical and quantum field theory, gravity and string theory (see, e.g., \cite{FP}, \cite{FMS}, \cite{G}). For
example, such theories can be treated as fixed points of the proper renormalization group flows and any quantum field theory can be recovered as a deformation of some conformal field theory (see,
e.g., \cite{Rychkov:2016iqz}).
One more
well known application of conformal theories, especially in
supergravities, is the method of conformal
compensators in diverse dimensions. It allows one to derive standard
Einstein gravity and the relevant non-conformal supergravities,
starting from the conformal (super)gravities coupled to the
appropriate matter compensating (super)fields.  These compensators ensure the
spontaneous breaking of conformal (super)groups to some subgroups
thereof (see, e.g., \cite{BK}, \cite{Freedman:2012zz}). The compensator
approach is a powerful way of constructing diverse
supergravity actions.

Higher spin theories are a natural generalization of the standard
(super)gauge theories and (super)gravities, and they attract vast
attention due to their intimate relationships with  (super)string
theory \cite{Vasiliev:2003cph,  Bekaert:2004qos, Bekaert:2022poo,
Gaberdiel:2015wpo, Ponomarev:2022vjb, Tseytlin:2002gz}. There
arises the natural task of constructing (super)conformal theories of
higher spins as the basis of the whole plethora of the
higher-spin theories. To know such superconformal extensions
is also of high importance for constructing higher spin theories in
AdS$_4$ and other conformally flat backgrounds. Indeed, these theories can be obtained by
gauging the proper subgroups of the
(super)conformal groups, like the standard $4D$ Poincar\'e
(super)symmetry in the case of flat (super)Minkowski background.

Free higher-spin theories in $4D$ Minkowski space were pioneered by Fronsdal and Fang and Fronsdal in refs. \cite{Fronsdal:1978rb,Fang:1978wz}.
 Their conformal generalizations were constructed by Fradkin and Tseytlin
\cite{Fradkin:1985am}. They introduced conformal higher spin fields and defined the corresponding gauge transformations. The actions constructed provided a higher-spin cousins of the Weyl tensor - squared actions. Since
then, various generalizations of these theories, including generalizations to curved gravity backgrounds, were intensively studied (see, e.g., \cite{ Metsaev:2007rw, Grigoriev:2016bzl, Joung:2012qy, Kuzenko:2019ill,
Kuzenko:2020jie, Vasiliev:2009ck, Kuzenko:2019eni, Basile:2022nou, Basile:2018eac, Manvelyan:2018qxd, Beccaria:2017nco, Nutma:2014pua}). Conformal higher-spin cubic vertices were for the first time constructed in refs.
\cite{Fradkin:1989md, Fradkin:1990ps}. After that,  cubic conformal higher-spin vertices were explicitly calculated and discussed by Segal in the context
of induced actions \cite{Segal:2002gd}. The latter paper inspired a wide
discussion of interacting conformal higher spin theories in the context of induced quantum actions \cite{Bekaert:2010ky, Bonezzi:2017mwr, Kuzenko:2022hdv}. Using the higher-spin conformal vertices, one can in
principle construct consistent interacting higher-spin actions as the induced actions and calculate them in the leading order.
Such interacting theories are higher-spin generalizations of Weyl gravity and involve higher
derivatives, so they are non-unitary on their own. These theories are  invariant under conformal gravity group in a general background.
Conformal invariance can be made manifest by choosing one or another specific background.
For instance, the invariance of the models considered in \cite{Kuzenko:2022hdv} is manifest in any conformally-flat background.

$\mathcal{N}=1, 4D$ supersymmetric generalization of conformal higher spins was considered in \cite{Fradkin:1990ps}
using the component approach. The free  off-shell $\mathcal{N}=1$ superconformal theories and their
couplings to chiral and some other $\mathcal{N}=1$ matter multiplets were constructed in refs. \cite{Kuzenko:2017ujh} and \cite{Buchbinder:2017nuc, Koutrolikos:2017qkx, Buchbinder:2018gle}
in $\mathcal{N}=1, 4D$ superspace. These authors, based on the earlier articles \cite{Kuzenko:1993jp, Kuzenko:1993jq, Kuzenko:1994dm}\footnote{The unconstrained prepotentials of $\mathcal{N}=1,4D$
non-conformal higher spin theories introduced in these papers were later discussed in  \cite{Gates:2013rka} from the standpoint of the super-Poincar\'e group representations.
These prepotentials were also shown to be correct variables for massive $\mathcal{N}=1, 4D$ higher spins \cite{Koutrolikos:2020tel}. The further analysis of these
prepotentails and their various actions was undertaken  in \cite{Buchbinder:2020yip, Koutrolikos:2022chj}.},
constructed unconstrained $\mathcal{N}=1$ higher-spin prepotentials (see also \cite{Howe:1981qj}),
found  their gauge and superconformal transformations,  investigated their component structure and derived invariant actions
in the flat $\mathcal{N}=1$ superspace. An extension of the actions constructed to $\mathcal{N}=1$
conformal supergravity (for multiplets with integer higher spin) was also proposed \cite{Kuzenko:2019ill} and it was shown that these actions
are gauge invariant only on conformally flat superspaces with vanishing super-Weyl tensor (including AdS superspace $AdS^{4|4}$, see the earlier ref. \cite{Kuzenko:2017ujh}). Some generalizations of these theories were considered in \cite{Kuzenko:2019eni,
Kuzenko:2020opc}. In ref. \cite{Hutomo:2017nce} an off-shell formulation of $\mathcal{N}=1$ higher-spin theories with the half-integer highest spin was given,
using the appropriate compensator supermultiplets.

$\mathcal{N}=2, 4D$ superconformal higher-spin theories (equally as their $\mathcal{N}$ extended versions) in an arbitrary conformally flat background
were elaborated in \cite{Kuzenko:2021pqm}\footnote{The $\mathcal{N}=2, 4D$ superconformal gravitino multiplet was described in \cite{Hutchings:2023iza}.}, based on the notion of
$\mathcal{N}=2$ conformal superspace \cite{Butter:2011sr}. $\mathcal{N}$ extended superconformal  actions were constructed there from superfield strengths
 (for applications of $\mathcal{N}=1$ higher spin superfield strengths
see \cite{Kuzenko:2017ujh} and later references \cite{Buchbinder:2018wzq, Gates:2019cnl}). The
appropriate Noether couplings to an on-shell hypermultiplet were
constructed in \cite{Kuzenko:2021pqm}. It is worth noting, however, that the component contents of $\mathcal{N}=2$ higher-spin superconformal multiplets and
the relevant off-shell cubic vertices were not explicitly  addressed in \cite{Kuzenko:2021pqm}\footnote{The authors thank the referee for pointing out that
the component contents of such multiplets can be figured out through reduction to $\mathcal{N}=1$ superfields (the component structure of which is well known)
as was presented in section 3.4 of \cite{Kuzenko:2021pqm}.}.

One of the ways to define gauge fields and their gauge transformations is to gauge the rigid symmetries of some free theory and to construct the corresponding cubic vertex.
The simplest cubic vertex, the\,$(s,0,0)$ vertex, is the product of the higher-spin $s$ gauge field and the Noether current bilinear in massless complex scalar fields.
The vertices of this type were widely studied. The most natural questions regarding them are: is it possible to make the $(s,0,0)$ vertices  gauge-invariant to all orders,
and is it possible to set up  such vertices in an arbitrary gravitational background?
\begin{enumerate}

\item Conformal cubic $(s,0,0)$ vertex (plus free scalar action) is gauge invariant only to the leading order, however cubic coupling of complex scalar field to an infinite set of conformal higher spins can be made gauge invariant to all orders by deforming the gauge transformation laws of higher spin fields \cite{Bekaert:2010ky, Segal:2002gd}.
The resulting gauge transformations are generically nonabelian and nonlinear. They mix  different higher-spin fields among themselves, while the scalar fields are transformed linearly and homogeneously.
In ref. \cite{Kuzenko:2022hdv} this construction was extended to an arbitrary conformally flat background in the  manifestly covariant  way, as well as to $\mathcal{N}=1$ superconformal case.

\item  One can easily construct the conserved spin 1 and spin 2 currents for the conformally-coupled complex scalar in an arbitrary curved background.
Respectively, one can build $(1,0,0)$ and $(2,0,0)$ vertices in
curved backgrounds. However, starting from $s\geq3$, the naive
attempts to construct conserved currents in a curved background gave
rise to the conclusion that the conservation can be achieved only
for the conformally flat case \cite{Beccaria:2017nco,
Kuzenko:2022hdv}. So the non-vanishing Weyl tensor provides an
obstruction to the existence of the $(s,0,0)$ vertices for $s\geq3$.
It was shown in ref. \cite{Beccaria:2017nco} (based on the suggestion of \cite{Grigoriev:2016bzl}) that,
by adding the vertex $(1,0,0)$ and modifying accordingly the gauge
transformation of the spin 1 field, one can achieve gauge invariance
in an arbitrary curved background too. Similar conclusions  were
drawn in ref. \cite{Kuzenko:2019eni, Kuzenko:2020jie}.
\end{enumerate}

In the present paper we analyze the off-shell $\mathcal{N}=2$ superconformal $(s,0,0)$ interactions in the flat and an arbitrary conformal supergravity backgrounds.
As we will see, already the $\mathcal{N}=2$ spin $\mathbf{3}$ superconformal multiplet simultaneously contains both spin 3 and spin 1, and this property
greatly simplifies the construction of the corresponding vertices in a curved background.

Our work extends to the superconformal case some of our previous results on the  off-shell
$\mathcal{N}=2, 4D$ higher spins and their hypermultiplet cubic
coupling \cite{Buchbinder:2021ite, Buchbinder:2022kzl,
Buchbinder:2022vra} (see also reviews \cite{Buchbinder:2022svx,
Ivanov23}). Since the hypermultiplet has
a  natural off-shell formulation in harmonic superspace (HSS)
\cite{HSS, HSS1, 18} in terms of ${\cal N}=2$ analytic harmonic
superfields, here we make use of just this formulation. We demonstrate
that the harmonic analyticity imposes severe constraints on the
admissible structure of the cubic interaction vertices of the hypermultiplet
and higher-spin conformal ${\cal N}=2$ gauge
superfields. We focus just on the construction of
$\mathcal{N}=2$ superconformal cubic couplings with the matter hypermultiplets \footnote{$\mathcal{N}=2$ generalizations of Fradkin-Tseytlin action in HSS
will be studied in a separate paper.}. To set up such cubic
couplings, we introduce the corresponding off-shell superconformal
spin $\mathbf{s}$ gauge multiplets \footnote{We use bold
$\mathbf{s}$ to denote $\mathcal{N}=2$ multiplet with the highest
spin s. For example, hypermultiplet corresponds to $\mathbf{s}=
\frac{1}{2}$, $\mathcal{N}=2$ Maxwell  multiplet to $\mathbf{s}= {1}$,
$\mathcal{N}=2$ Weyl multiplet to  $\mathbf{s}= {2}$. }, define the
corresponding minimal sets of analytic gauge potentials, derive
their rigid superconformal transformation laws and, at the linearized level, their gauge transformation laws.
We present the relevant Wess-Zumino gauges for the component fields.
We also expound  how to promote these cubic vertices to an arbitrary $\mathcal{N}=2$ conformal supergravity background:
one should consider an infinite tower of $\mathcal{N}=2$ superconformal higher-spin fields interacting with a hypermultiplet.
This allows one to define a nonabelian deformation of the gauge transformation algebra and demonstrate that the relevant interacting
theory is gauge-invariant to all orders.

The basic novel features of superconformal couplings of
$\mathcal{N}=2$ higher-spin gauge superfields to the hypermultiplet
in HSS compared to the non-conformal case \cite{Buchbinder:2022vra} can be schematically outlined as follows.
The difference arises already in the case
 of the spin ${\bf s}=2$ multiplet
(conformal ${\cal N}=2$ supergravity), where the
analyticity-preserving harmonic derivative ${\cal D}^{++}$, when
acting on the hypermultiplet superfields, is covariantized as
\be
{\cal D}^{++} \;\Rightarrow \;  {\cal D}^{++} + \kappa_2\hat{\cal
H}_{(s=2)}^{++}\,, \quad \hat{\cal H}^{++}_{(s=2)}= h^{++M}\partial_M\,,
\;\; M = (\alpha\dot\alpha\,, \alpha+\,, \dot\alpha+\,,++)\,.\lb{A}
\ee
Here there appears a new analytic gauge superfield
$h^{(+4)}$ \cite{18}. Its necessity can be substantiated
from requiring rigid conformal ${\cal N}=2$ invariance with respect
to which only the whole set of the potentials in the operator
$\hat{\cal H}_{(s=2)}$ turns out to be closed. In the spin
${\bf s}=3$ case, the covariantization is accomplished by the differential
operator of the second order,
\be
{\cal D}^{++} \;\Rightarrow \;
{\cal D}^{++} + \kappa_3 \hat{\cal H}_{(s=3)}^{++} J \,, \quad \hat{\cal
H}^{++}_{(s=3)}= h^{++MN}\partial_N\partial_M + h^{++}\,, \lb{B}
\ee
where the analytic gauge potentials $h^{++MN}$ satisfy some grading and
irreducibility conditions (see below), while $J$ is some matrix $U(1)$
generator. Once again, only the whole set of gauge potentials in
\eqref{B} is closed under a linear realization of rigid ${\cal N}=2$
superconformal group. So the latter plays the same restrictive role
for cubic superconformal vertices, as the rigid ${\cal N}=2$
supersymmetry for non-conformal vertices \cite{Buchbinder:2021ite,
Buchbinder:2022kzl, Buchbinder:2022vra}. The radical extension of
the number of gauge potentials for ${\bf s} \geq 3$
also gives rise to an essential extension of the gauge freedom
compared to the ${\bf s}=2$ case. This can be used to fully gauge
away many gauge potentials,
 \be
 \hat{\cal H}^{++}_{(s=3)} \;\Rightarrow \\
h^{++\alpha\dot\alpha M}\partial_M\partial_{\alpha\dot\alpha}\,. \lb{C}
\ee
The rigid ${\cal N}=2$ superconformal symmetry acts on this minimal set of potentials by transformations which are in general {\it nonlinear} in the
potentials. All these notable features directly generalize to ${\cal N}=2$ spins ${\bf s}>3$.

The paper is organized as follows. In section \ref{sec: harmonic superspace} we recall
the basic elements of harmonic superspace and describe free off-shell hypermultiplet. Section \ref{sec: sc hyper} contains discussion
of the $\mathcal{N}=2$ superconformal symmetry realization in harmonic superspace and expounds our strategy of construction
of the off-shell superconformal $\mathcal{N}=2$ multiplets in (curved) harmonic superspace.
In sections \ref{sec: maxwell} and \ref{sec: weyl} we present superconformal transformations for the spin $\mathbf{1}$ and spin $\mathbf{2}$
multiplets and the corresponding off-shell  $(\mathbf{1}, \mathbf{\tfrac{1}{2}}, \mathbf{\tfrac{1}{2}})$,
$(\mathbf{2}, \mathbf{\tfrac{1}{2}}, \mathbf{\tfrac{1}{2}})$ superconformal couplings. In section \ref{sec: weyl} we discuss the hypermultiplet in the background of $\mathcal{N}=2$ conformal supergravity.
Section \ref{sec: spin 3}
is devoted to the crucial new spin $\mathbf{3}$ case: we introduce a minimal set of analytic prepotentials, study their component structure and construct
off-shell $(\mathbf{3}, \mathbf{\frac{1}{2}}, \mathbf{\frac{1}{2}})$ vertices in an arbitrary conformal supergravity background.
In section \ref{sec: spin s} we generalize the spin $\mathbf{3}$ results to the general ${\cal N}=2$ spin $\mathbf{s}$.
In section \ref{sec: 8 fully consistent} we sketch some results on nonabelian (and nonlinear) deformation of higher-spin gauge algebra in the case of infinite tower of $\mathcal{N}=2$ conformal higher-spin fields
minimally interacting with the hypermultiplet. Such a theory possesses the exact invariance with respect to these nonabelian gauge transformations.
The concluding comments and the basic problems  for the future study  are contents of the last section \ref{sec: conclusion}. Appendix \ref{app: WZ gauge}
contains technical details of fixing Wess-Zumino gauge in the spin $\mathbf{3}$ case. In Appendix \ref{app: residual} we discuss some interesting reparametrization
freedom of free hypermultiplet. The superconformal transformation properties of the derivatives in the analytic superspace coordinates and those of some gauge potentials
(for ${\bf s} = 2, 3$) are collected in Appendix \ref{app: N2SCDeriv}.

\section{Harmonic superspace}\label{sec: harmonic superspace}

We deal with $\mathcal{N}=2$ harmonic superspace (HSS) \cite{HSS, HSS1, 18} parametrized by the coordinates in the analytic basis:
\begin{equation}
    Z : = (x^{\alpha\dot{\alpha}}, \theta^{+\hat{\alpha}}, \theta^{-\hat{\alpha}}, u^{\pm i}), \quad \hat{\alpha} = (\alpha, \dot\alpha)\,.
\end{equation}
In addition to the standard $4D$ superspace coordinates $(x, \theta^\pm)$, HSS involves additional $SU(2)/U(1)$ harmonic variables $u^{\pm }_i$, $i=1,2,$
satisfying the unitarity constraint $u^{+i}u^-_i = 1 $.

The crucial feature of HSS is the presence of the invariant subspace with half the number of Grassmann variables.  This analytic superspace is parametrized by the coordinates:
\begin{equation}
    \zeta: = (x^{\alpha\dot{\alpha}}, \theta^{+\hat{\alpha}}, u^{\pm i}).
\end{equation}
For the description of massive hypermultiplet and its higher spin couplings it is also necessary to introduce an auxiliary $x^5$ coordinate, see, e.g., \cite{Buchbinder:2022vra}.
The analytic superspace is closed under the tilde-conjugation defined as:
\begin{equation}\label{eq: tilde coord}
    \widetilde{x^{\alpha\dot{\alpha}}} = x^{\alpha\dot{\alpha}},
    \qquad
    \widetilde{\theta^{\pm}_\alpha} = \bar{\theta}^\pm_{\dot{\alpha}},
    \qquad
    \widetilde{\bar{\theta}^{\pm}_{\dot{\alpha}}}
    =
    -
    \theta^{\pm}_\alpha,
    \qquad
    \widetilde{u^{\pm i}} = - u^\pm_i,
    \qquad
        \widetilde{u^{\pm}_i} =  u^{\pm i}.
\end{equation}

The covariant harmonic derivatives in the analytic basis are defined by,
\begin{subequations}
\begin{equation}
    \mathcal{D}^{++} := \partial^{++} - 4i \theta^{+\rho} \bar{\theta}^{+\dot{\rho}} \partial_{\rho\dot{\rho}} + \theta^{+\hat{\rho}} \partial^+_{\hat{\rho}}
    + i \theta^{+\hat{\rho}} \theta^+_{\hat{\rho}} \partial_5,
\end{equation}
\begin{equation}
    \mathcal{D}^{--} := \partial^{--} - 4i \theta^{-\rho} \bar{\theta}^{-\dot{\rho}} \partial_{\rho\dot{\rho}} + \theta^{-\hat{\rho}} \partial^-_{\hat{\rho}}
    + i \theta^{-\hat{\rho}} \theta^-_{\hat{\rho}} \partial_5,
\end{equation}
\begin{equation}\label{eq:D0}
    \mathcal{D}^0 = \partial^0 + \theta^{+\hat{\rho}} \partial^-_{\hat{\rho}}
        -
        \theta^{-\hat{\rho}} \partial^+_{\hat{\rho}},
\end{equation}
\end{subequations}
and satisfy $su(2)$ algebra relations:
\begin{equation}
    [\mathcal{D}^{++} , \mathcal{D}^{--} ] = \mathcal{D}^0,
    \qquad
    [\mathcal{D}^0, \mathcal{D}^{\pm\pm}] = \pm 2 \mathcal{D}^{\pm\pm} .
\end{equation}
Here we used the following  notations for the partial derivatives in harmonic variables:
\begin{equation}
    \partial^{++} = u^{+i} \frac{\partial}{\partial u^{-i}},
    \qquad
    \partial^{--} = u^{-i} \frac{\partial}{\partial u^{+i}},
    \qquad
    \partial^0 =  u^{+i} \frac{\partial}{\partial u^{+i}}
    -
     u^{-i} \frac{\partial}{\partial u^{-i}}\,,
\end{equation}
\begin{equation}
    [   \partial^{++},  \partial^{--} ] =   \partial^{0}.
\end{equation}
Other partial derivatives are defined in the standard way, e.g., $\partial_{\alpha\dot\alpha} = \frac{\partial}{\partial x^{\alpha\dot\alpha}}$, etc
\footnote{The relation with the vector notation is the same as in \cite{18}, $x^{\dot\alpha\beta} = x^m(\tilde{\sigma}_m)^{\dot\alpha\beta}, \quad
\partial_m = (\tilde{\sigma}_m)^{\dot\alpha\beta}\partial_{\dot\alpha\beta}$,
\quad $\partial_{\alpha\dot\beta} = \frac{1}{2}\sigma^m_{\alpha\dot\beta}\partial_m$.}.

The action of tilde-conjugation on various derivatives follows directly from the definitions \eqref{eq: tilde coord}:
\begin{equation}
    \widetilde{\partial_{\alpha\dot{\alpha}}} =
    \partial_{\alpha\dot{\alpha}},
    \quad
    \widetilde{\partial^{-}_\alpha} = - \partial^-_{\dot{\alpha}},
    \qquad
    \widetilde{\partial^-_{\dot{\alpha}}} = \partial^-_\alpha,
    \qquad
    \widetilde{\partial^{\pm\pm}} = \partial^{\pm\pm}.
\end{equation}

Harmonic superspace provides efficient tools to deal with $\mathcal{N}=2$ supersymmetric theories, both on the classical and quantum levels.
The hypermultiplet and the most general hypermultiplet self-couplings \cite{Galperin:1985de}, $\mathcal{N}=2$ Yang-Mills theory, different $\mathcal{N}=2$  supergravities
(see a recent review \cite{Ivanov:2022vwc}), as well as $\mathcal{N}=2$ generalizations of Fronsdal theory \cite{Buchbinder:2021ite}, are
adequately  described in  $\mathcal{N}=2$ HSS approach. The pivotal feature of the HSS approach is that all the basic $\mathcal{N}=2$ superfields are analytic,
thus manifesting the crucial role of the \textit{harmonic Grassmann analyticity principle} in ${\cal N}=2$ supersymmetric theories.

\subsection{Free hypermultiplet}

Since our main subject will be $\mathcal{N}=2$ superconformal interactions of higher-spin superfields with hypermultiplet, we start by giving
all the necessary details of the HSS formulation of hypermultiplet.  It is described by an analytic unconstrained superfield $q^{+}(\zeta)$ with an infinite
number of auxiliary fields off shell. The free hypermultiplet action reads \cite{18}:
\begin{equation}\label{eq:hyper action}
    S_{free}= - \frac{1}{2} \int d\zeta^{(-4)}\, q^{+a} \mathcal{D}^{++} q^+_a
    =
    - \int d \zeta^{(-4)}\; \tilde{q}^+ \mathcal{D}^{++} q^+.
\end{equation}
Here we used the notation:
\begin{equation}
    q^{+a} = (\tilde{q}^+, q^+) ,
    \qquad
    q^+_a = \epsilon_{ab} q^{+b} =\begin{pmatrix}
        q^+
        \\
        -\tilde{q}^+
    \end{pmatrix}.
\end{equation}
The superfield $q^{+a}$ forms a doublet of the Pauli-G\"ursey group ${\rm SU}(2)_{PG}$. The ${\rm SU}(2)_{PG}$ - covariant notation
is  useful when constructing higher-spin vertices.

The free hypermultiplet equation of motions is:
\begin{equation}\label{eq: hyper eom}
    \mathcal{D}^{++} q^{+a} = 0.
\end{equation}
The discussion of the on-shell field content of hypermultiplet can be found, e.g., in section 5.1 of ref. \cite{Buchbinder:2022vra}.
In what follows we will merely use  the superfield aspects of the HSS description of the hypermultiplet.
Here we only remark that the hypermultiplet contains a doublet of complex scalars, so it can interact with both even and odd spins.

\section{$\mathcal{N}=2$ superconformal symmetry of hypermultiplet} \label{sec: sc hyper}

The realization of $\mathcal{N}=2$ superconformal symmetry on the HSS coordinates is given in \cite{18, Galperin:1985zv}.
We will be interested in constructing $\mathcal{N}=2$ superconformal cubic $(\mathbf{s}, \mathbf{\tfrac{1}{2}}, \mathbf{\tfrac{1}{2}})$ vertices
of the higher spin gauge superfields  with the hypermultiplet.
To this end, we need to introduce the appropriate set of analytic higher-spin superconformal gauge potentials and define their superconformal transformation laws.
The hypermultiplet superconformal transformation law is well known, so from requiring the invariance of the interaction one can determine transformation properties of
the higher spin gauge potentials.
Based on the experience of dealing with the non-conformal case \cite{Buchbinder:2022vra, Buchbinder:2022kzl}, we will use, as a departure point,
the most general type of interaction with higher derivatives
and determine the minimal set of the analytic higher-spin gauge potentials closed under $\mathcal{N}=2$ superconformal symmetry.
In this section we first discuss $\mathcal{N}=2$ superconformal symmetry of the free massless hypermultiplet and then explain
our general strategy of constructing $\mathcal{N}=2$ higher-spin superconformal couplings.

 We start with the general one-derivative hypermultiplet transformations\footnote{Such transformations can be realized
 on the HSS coordinates, see \cite{18}. We basically consider superfield transformations in their active form, since we are interested in their generalization
 to the case of higher-spin symmetries which cannot be realized on the coordinates.}:
\begin{equation}\label{eq:superconformal transfromations}
    \delta q^{+a}
    = - \hat{\Lambda} q^{+a}
    -
    \frac{1}{2}\Omega q^{+a},
\end{equation}
where\footnote{$M = (\alpha\dot{\alpha}, \alpha+, \dot{\alpha}+, ++, 5)$;\quad $P(\alpha\dot{\alpha}) = P(++) = P(5) = 0$, $P(\hat{\alpha}+) = 1 $.}:
\begin{equation}
    \hat{\Lambda} :=
    \lambda^M \partial_M
    =
     \lambda^{\alpha\dot{\alpha}} \partial_{\alpha\dot{\alpha}}
    +
    \lambda^{+\hat{\alpha}} \partial^-_{\hat{\alpha}}
    +
    \lambda^{++}\partial^{--}
    +
    \lambda^5 \partial_5,
\end{equation}
\begin{equation}\label{eq: Omega}
    \Omega
    :=
    (-1)^{P(M)} \partial_M \lambda^M
    =
    \partial_{\alpha\dot{\alpha}}\lambda^{\alpha\dot{\alpha}}
    -
    \partial^-_{\hat{\alpha}}
    \lambda^{+\hat{\alpha}}
    +
    \partial^{--} \lambda^{++}.
\end{equation}
Here $\hat{\Lambda}$ is the first-order  differential operator, $\Omega$ is the weight factor constructed out of the parameters
$\lambda^M$. Since the superfield $q^{+a}$ is analytic, we impose the condition that the transformations \eqref{eq:superconformal transfromations} preserve the analyticity, which implies
the parameters $\lambda^M$ to be  unconstrained analytic:
\begin{equation}
    \partial^+_{\hat{\rho}} \lambda^{\alpha\dot{\alpha}} =0,
    \qquad
    \partial^+_{\hat{\rho}} \lambda^{+\hat{\alpha}}  =0,
    \qquad
    \partial^+_{\hat{\rho}} \lambda^{++} =0,
    \qquad
        \partial^+_{\hat{\rho}} \lambda^{5} =0.
\end{equation}
Also we assume  $x^5$-independence of the transformation parameters, since $x^5$ is an auxiliary coordinate needed merely for the description of massive hypermultiplet.
Unlike the rigid symmetries considered in our previous papers \cite{Buchbinder:2022kzl, Buchbinder:2022vra}, here we allow for a nontrivial coordinate dependence
of $\lambda^M$ in the rigid case. This will lead to an extended algebra of rigid hypermultiplet symmetries
with  a larger number of independent transformation parameters.

Varying the free action \eqref{eq:hyper action} with respect to the transformations \eqref{eq:superconformal transfromations} with generic analytic
parameters $\lambda^M$ yields:
\begin{equation}\label{s=2 action trnsform}
    \begin{split}
    \delta S_{free} =
     \frac{1}{2} \int d\zeta^{(-4)}\; q^{+a} [\mathcal{D}^{++},\, \hat{\Lambda}] q^+_a.
     \end{split}
\end{equation}
The precise form of the commutator in \eqref{s=2 action trnsform} is as follows:
\begin{equation}\label{eq: s=2 commutator}
    \begin{split}
        [\mathcal{D}^{++}, \,\hat{\Lambda}]
        =
        & \left(\mathcal{D}^{++}\lambda^{\alpha\dot{\alpha}}
        + 4i \lambda^{+\alpha} \bar{\theta}^{+\dot{\alpha}}
        + 4i \theta^{+\alpha} \bar{\lambda}^{+\dot{\alpha}}
        \right) \partial_{\alpha\dot{\alpha}}
        \\&+
        \left(\mathcal{D}^{++}\lambda^{+\hat{\alpha}} - \lambda^{++} \theta^{+\hat{\alpha}} \right)\partial^-_{\hat{\alpha}}
        +
        \mathcal{D}^{++} \lambda^{++}\partial^{--}
        +
        \lambda^{++} \mathcal{D}^0
        \\& +
        \left(\mathcal{D}^{++}\lambda^5 - 2 \lambda^{+\hat{\rho}} \theta^+_{\hat{\rho}} \right)\partial_5
        +
        \lambda^{++} \theta^{-\hat{\alpha}} \partial^+_{\hat{\alpha}}.
    \end{split}
\end{equation}

Taking into account the relations $\mathcal{D}^0 q^{+a} = q^{+a}$ and $q^{+a}q^+_a =0$, we derive the condition of invariance of the action \eqref{eq:hyper action} as
\begin{equation}\label{eq: comm s=2}
        [\mathcal{D}^{++},\, \hat{\Lambda}] = \lambda^{++} \mathcal{D}^0,
\end{equation}
or, in terms of the parameters $\lambda^M$,
\begin{equation}\label{eq: spin 2 system}
    \begin{cases}
        \mathcal{D}^{++}\lambda^{\alpha\dot{\alpha}}
        + 4i \lambda^{+\alpha} \bar{\theta}^{+\dot{\alpha}}
        + 4i \theta^{+\alpha} \bar{\lambda}^{+\dot{\alpha}} = 0,
        \\
        \mathcal{D}^{++}\lambda^{+\alpha} - \lambda^{++} \theta^{+\alpha} = 0,
        \\
            \mathcal{D}^{++}\lambda^{+\dot{\alpha}} - \lambda^{++} \bar{\theta}^{+\dot{\alpha}} = 0,
        \\
            \mathcal{D}^{++} \lambda^{++} = 0,
        \\
        \mathcal{D}^{++}\lambda^5 - 2 \lambda^{+\hat{\rho}} \theta^+_{\hat{\rho}}=0.
    \end{cases}
\end{equation}
The general solution of the system \eqref{eq: spin 2 system} is just the sought {\bf $\mathcal{N}=2$ superconformal  transformations}:
\begin{equation}\label{eq:superconformal symmetry}
    \begin{cases}
        \lambda_{sc}^{\alpha\dot{\alpha}}
        =&
            a^{\alpha\dot{\alpha}}
        -
        4i \left( \epsilon^{\alpha i} \bar{\theta}^{+\dot{\alpha}} + \theta^{+\alpha} \bar{\epsilon}^{\dot{\alpha}i} \right) u^-_i

        +
        x^{\dot{\alpha}\rho} k_{\rho\dot{\rho}} x^{\dot{\rho}\alpha}
        + a x^{\alpha\dot{\alpha}}
        \\&-
        4i \theta^{+\alpha} \bar{\theta}^{+\dot{\alpha}} \lambda^{(ij)}u^-_i u^-_j
        -
        4i \left(x^{\alpha\dot{\rho}}\eta_{\dot{\rho}}^i \bar{\theta}^{+\dot{\alpha}}
        +
        \theta^{+\alpha} \eta^i_{\rho} x^{\rho\dot{\alpha}}
        \right) u^-_i,
        \\
        \lambda^{+\alpha}_{sc}
        =&
        \epsilon^{\alpha i} u^+_i
        +
        \frac{1}{2}
        \theta^{+\alpha} (a + i b)
        +
        x^{\alpha\dot{\beta}} k_{\beta\dot{\beta}} \theta^{+\beta}
        +
        x^{\alpha\dot{\alpha}}  \eta^i_{\dot{\alpha}} u^+_i
        \\&
        +
        \theta^{+\alpha}
        \left( \lambda^{(ij)}u^+_i u^-_j
        +
        4i \theta^{+\rho} \eta^i_{\rho} u^-_i
        \right),
        \\
        \bar{\lambda}^{+\dot{\alpha}}_{sc}
        =&
        \epsilon^{\dot{\alpha} i} u^+_i
        +
        \frac{1}{2}
        \bar{\theta}^{+\dot{\alpha}}  (a - i b)
        +
        x^{\dot{\alpha}\beta} k_{\beta\dot{\beta}} \bar{\theta}^{+\dot{\beta}}
        +
        x^{\alpha\dot{\alpha}}  \eta^i_{\alpha} u^+_i
        \\&
        +
        \bar{\theta}^{+\dot{\alpha}}
        \left( \lambda^{(ij)}u^+_i u^-_j
        -
        4i \bar{\theta}^{+\dot{\rho}} \eta^i_{\dot{\rho}} u^-_i
        \right),
        \\
            \lambda^{++}_{sc} =&
             \lambda^{ij} u^+_i u^+_j
        +
        4i \theta^{+\alpha} \bar{\theta}^{+\dot{\alpha}} k_{\alpha\dot{\alpha}}
        +
        4i \left( \theta^{+\alpha} \eta^i_{\alpha} + \eta^i_{\dot{\alpha}} \bar{\theta}^{+\dot{\alpha}}  \right) u^+_i.
    \end{cases}
\end{equation}
Respectively, the weight factor \eqref{eq: Omega} is expressed as:
\begin{equation}\label{eq:conformal weight}
    \Omega_{sc} = 2a + 2 k_{\beta\dot{\beta}} x^{\beta\dot{\beta}} - 2 \lambda^{(ij)}u^+_i u^-_j - 8i \left( \theta^{+\alpha} \eta^i_{\alpha} +\eta^i_{\dot{\alpha}} \bar{\theta}^{+\dot{\alpha}} \right) u^-_i.
\end{equation}
It satisfies the useful relation:
\begin{equation}
        \mathcal{D}^{++} \Omega_{sc} =
        -
        2\lambda^{++}_{sc}.
\end{equation}

As follows from \eqref{eq: s=2 commutator}, the last condition in the system \eqref{eq: spin 2 system} appears only if $\partial_5 q^{+a} \neq 0$. So
 we are led to impose the constraint $\partial_5 q^{+a} =0$, i.e. limit our consideration to the massless hypermultiplet.
This is consistent with the well known fact that all theories with exact (super)conformal symmetry are massless (see, e.g., \cite{BK}).

\medskip

Symmetry \eqref{eq:superconformal symmetry} extends rigid $\mathcal{N}=2$ supersymmetry of the free massless hypermultiplet, which was generalized to
the higher-spin symmetries in \cite{Buchbinder:2022vra, Buchbinder:2022kzl}:
\begin{equation}
    \underbrace{\{a^{\alpha\dot{\alpha}}, \epsilon^{\hat{\alpha}i}\}}_{{\cal N}=2\; \text{supersymmetry}}
    \to
    \underbrace{\{a^{\alpha\dot{\alpha}}, \epsilon^{\hat{\alpha}i}, a, b, k_{\alpha\dot{\alpha}}, \eta^{\hat{\alpha}i}, \lambda^{(ij)} \}}_{{\cal N}=2\; \text{superconformal symmetry}}.
\end{equation}
The analytic parameters \eqref{eq:superconformal symmetry} are those of $\mathcal{N}=2$ superconformal symmetry in the realization on the coordinates of analytic superspace \cite{18, Galperin:1985zv}
(here we omit Lorentz transformations). The transformation parameters can be attributed as:

\begin{itemize}
    \item $a^{\alpha\dot{\alpha}}$ - global translations;
    \item $\epsilon^{\hat{\alpha} i}$ - rigid $\mathcal{N}=2$ supersymmetry;
    \item $a$ - dilatations;
    \item $b$ - $U(1)$ R-symmetry;
    \item $k_{\alpha\dot{\alpha}}$ - special conformal transformations;
    \item $\eta^{\hat{\alpha}i}$ - rigid  $\mathcal{N}=2$ conformal supersymmetry;
    \item $\lambda^{(ij)}$ - $SU(2)_R$ symmetry.
\end{itemize}

One can directly check that these transformations satisfy the relations of $\mathfrak{su(2,2|2)}$ superalgebra, that is $\mathcal{N}=2$ superconformal algebra.
We will require the cubic couplings to be invariant under these transformations.

\medskip

For completeness, we also quote how conformal transformations are implemented on non-analytic coordinates $\theta^-$.
Using the relation
\begin{equation}
    \theta^{+\hat{\alpha}} = \mathcal{D}^{++}   \theta^{-\hat{\alpha}}
\end{equation}
and the transformation laws $\delta^*\theta^{\pm \hat{\alpha}} = \lambda_{sc}^{\pm\hat{\alpha}}$, one obtains:
\begin{equation}
    \lambda_{sc}^{+\hat{\alpha}} = \mathcal{D}^{++} \lambda_{sc}^{-\hat{\alpha}} + [\hat{\Lambda}_{sc}, \mathcal{D}^{++}] \theta^{-\hat{\alpha}}
    =
    \mathcal{D}^{++} \lambda^{-\hat{\alpha}}_{sc}
    +
    \lambda_{sc}^{++}\theta^{-\hat{\alpha}},
\end{equation}
whence
\begin{subequations}
\begin{equation}
    \begin{split}
    \lambda^{-\alpha}_{sc} =\;& \epsilon^{\alpha i} u^-_i
    +
    \frac{1}{2} \theta^{-\alpha} (a+ib)
    +
    x^{\alpha\dot{\beta}}
 k_{\beta\dot{\beta}}   \theta^{-\beta}
 -
 2i (\theta^-)^2 \bar{\theta}^+_{\dot{\beta}} k^{\dot{\beta}\alpha}
 \\&+ \left( x^{\alpha\dot{\alpha}} + 4i \theta^{-\alpha}\bar{\theta}^{+\dot{\alpha}} \right) \eta^i_{\dot{\alpha}} u^-_i
 +
 4i \eta^i_{\beta} \theta^{-\beta} \left( \theta^{-\alpha} u^+_i - \theta^{+\alpha} u^-_i \right)
 \\&
 +
 \lambda^{ij} u^-_i \left( u^-_j \theta^{+\alpha} - u^+_j \theta^{-\alpha} \right),
 \end{split}
\end{equation}
\begin{equation}
    \begin{split}
    \lambda^{-\dot{\alpha}}_{sc}
    =\;&
    \bar{\epsilon}^{\dot{\alpha}i} u^-_i +
    \frac{1}{2} \bar{\theta}^{-\dot{\alpha}}(a-ib)
    +
    x^{\dot{\alpha}\beta} k_{\beta\dot{\beta}} \bar{\theta}^{-\dot{\beta}}
    -
    2i (\bar{\theta}^-)^2 \theta^+_{\beta} k^{\beta\dot{\alpha}}
    \\&
    +
    \left( x^{\alpha\dot{\alpha}} + 4i \theta^{+\alpha} \bar{\theta}^{-\dot{\alpha}} \right) \eta^i_{\alpha} u^-_i
    -
    4i \eta^i_{\dot{\beta}} \bar{\theta}^{-\dot{\beta}} \left( \bar{\theta}^{-\dot{\alpha}} u^+_i - \bar{\theta}^{+\dot{\alpha}} u^-_i \right)
     \\&
    +
    \lambda^{ij} u^-_i \left( u^-_j \bar{\theta}^{+\dot{\alpha}} - u^+_j \bar{\theta}^{-\dot{\alpha}} \right).
    \end{split}
\end{equation}
\end{subequations}

\bigskip

In the next sections, we shall consider transformations of the three types:
\begin{itemize}
    \item $\delta_{sc}$ - rigid $\mathcal{N}=2$ superconformal transformations;

        \item $\delta_{diff}$  - localized $\mathcal{N}=2$ superconformal transformations,
        i.e. local superdiffeomorphisms (gauge group of $\mathcal{N}=2$ Weyl supergravity). The $\delta_{sc}$ transformations
        form a subgroup of the  $\delta_{diff}$ ones, with the parameters constrained by eqs. \eqref{eq: spin 2 system}.
         Using such an identification, we can study invariance with respect to the more general transformations $\delta_{diff}$,
         by imposing additional constraints on the parameters in order  to reduce $\delta_{diff}$ to $\delta_{sc}$, if necessary;

    \item $\delta_{\lambda}$ - linearized  {\it gauge} transformations.
\end{itemize}

\subsection{The general strategy of construction of superconformal couplings \break
and multiplets}\label{sec: The general strategy of construction of superconformal couplings and multiplets}

While constructing superconformal cubic vertices, we will start with \textbf{\circled{1}} singling out the minimal set of gauge superfields $h^{++M_1\dots M_{s-1}}(\zeta)$
needed for ensuring the invariance of the most general coupling \footnote{Here we use the projection operator $P(s) := \frac{1+(-1)^s}{2}$. },
    \begin{equation}\label{eq: general interaction}
        \begin{split}
        S^{(s)}_{int} =& - \frac{\kappa_s}{2} \int d\zeta^{(-4)}\, q^{+a} h^{++M_1\dots M_{s-1}} \partial_{M_{s-1}}  \dots  \partial_{M_1} \left(J\right)^{P(s)} q^+_a
    \\& +
        \text{lower derivative terms}\,,
        \end{split}
    \end{equation}
     under the hypermultiplet rigid superconformal transformations\footnote{Though the analytic parameters
     of $\mathcal{N}=2$ superconformal transformations are given in \eqref{eq:superconformal symmetry}, in what follows we shall not stick  to their specific form
     and deal with arbitrary analytic parameters $\lambda^M(\zeta)$  associated with the transformations $\delta_{diff}$.}:
    \begin{equation}\label{eq: superconformal hyper 0}
        \delta_{diff} q^{+a} = - \hat{\Lambda} q^{+a} - \frac{1}{2} \Omega q^{+a}
    \end{equation}
and some appropriate transformations of the gauge superfields
\begin{equation}
    \delta_{diff} h^{++M_1\dots M_{s-1} } = \dots\, .
\end{equation}
  The generator $J$ is defined as:
 \begin{equation}\label{eq: generator J}
    J q^{+a} := i (\tau^3)^a_{\;b}q^{+b},
    \qquad
    (\tau^3)^a_{\;b}
    =
    {\scriptsize\begin{pmatrix}
            1 & 0 \\
            0 & -1
    \end{pmatrix}}.
 \end{equation}
The choice of interaction as in \eqref{eq: general interaction} is
largely motivated by the consideration of the non-conformal case
\cite{Buchbinder:2022kzl}\footnote{As was noticed in
\cite{Buchbinder:2022kzl}, the matrix generator $J$ perfectly
well works for all odd spins $s \geq 3$. The fact that the cubic
interaction of scalars with the gauge fields of higher odd spins has
certain peculiarities is well known (see, e.g.,
\cite{Fotopoulos:2007yq}).} and is strictly constrained by the analyticity of $q^{+a}$.

As the next steps, we \textbf{\circled{2}} analyze  the gauge freedom of the superconformal action constructed
\begin{equation}
   S^{(s)}_q = S_{free} + S_{int}^{(s)}
\end{equation}
 and \textbf{\circled{3}} determine the set of unremovable Wess-Zumino gauge fields in the potentials $h^{++M_1\dots M_{s-1}}(\zeta)$, and hence
 reveal the irreducible field contents of the full $\mathcal{N}=2$ off-shell superconformal multiplet.

\medskip

In the next sections we start with the well known spin $\mathbf{1}$ and spin $\mathbf{2}$ cubic hypermultiplet couplings
in order to illustrate how the above procedure works.
Then we apply the same procedure to the novel case of the superspin $\mathbf{3}$ superconformal gauge multiplet
and finally generalize the results to an arbitrary integer spin $\mathbf{s}$.

\section{$\mathcal{N}=2$ Maxwell supermultiplet and superconformal $(\mathbf{1}, \mathbf{\tfrac{1}{2}}, \mathbf{\tfrac{1}{2}})$ coupling}
\label{sec: maxwell}

The simplest example of $\mathcal{N}=2$ superconformal cubic interaction of the hypermultiplet is supplied by its coupling to the superspin ${\bf 1}$ gauge multiplet  \cite{18}.
The  spin $\mathbf{1}$  hypermultiplet vertex $(\mathbf{1}, \mathbf{\tfrac{1}{2}},\mathbf{\tfrac{1}{2}})$ has the  form:
\begin{equation}\label{eq: 1 12 12 vertex}
    S_{int}^{(s=1)} = - \frac{\kappa_1}{2}
    \int d\zeta^{(-4)}\, q^{+a} V^{++} J q^+_a
    =
    \kappa_1 \int d\zeta^{(-4)}\, i\,V^{++} \tilde{q}^+ q^+.
\end{equation}
Here $V^{++}(\zeta)$ is an arbitrary unconstrained analytic gauge superfield with the gauge transformation $\delta_\lambda V^{++}(\zeta) = {\cal D}^{++}\lambda(\zeta)$,
where $\lambda(\zeta)$ is an arbitrary analytic superfield parameter. The gauge potential  $V^{++}$ satisfies the reality condition $\widetilde{V^{++}} = V^{++}$.
According to our general strategy, this is the most general spin ${\bf 1}$ -- hypermultiplet coupling containing no derivatives.

\subsection{$\mathcal{N}=2$ superconformal symmetry}

Next we analyze the superconformal invariance  of the vertex \eqref{eq: 1 12 12 vertex}.
Under both the local superconformal hypermultiplet transformations \eqref{eq: superconformal hyper 0} and still unspecified  local superconformal transformation of $V^{++}$
the vertex \eqref{eq: 1 12 12 vertex} transforms as:
\begin{equation}\label{spin1Inv}
    \begin{split}
        \delta_{diff} S_{int}^{(s=1)}
        =&
        \frac{\kappa_1}{2}
        \int d\zeta^{(-4)}\, \left[(\hat{\Lambda}q^{+a}) V^{++} J q^+_a
        +
        q^{+a} V^{++} J (\hat{\Lambda}q^+_a)
        \right]
        \\&+
        \frac{\kappa_1}{2} \int d\zeta^{(-4)}\, \Omega q^{+a} V^{++} J q^+_a
        -
        \frac{\kappa_1}{2} \int d\zeta^{(-4)}\, q^{+a} \delta_{diff} V^{++} J q^+_a\,,
    \end{split}
\end{equation}
where $\hat{\Lambda}$ and $\Omega$ were defined in \eqref{eq:superconformal transfromations} and \eqref{eq: Omega}.
We will require vanishing of such a variation by choosing the appropriate spin $\mathbf{1}$ superconformal transformation law $\delta_{diff} V^{++} $.

The first line of \eqref{spin1Inv}, modulo a total derivative, can be rewritten as:
\begin{equation}
    \begin{split}
        (\hat{\Lambda}q^{+a}) V^{++} J q^+_a
        +
        q^{+a} V^{++} J (\hat{\Lambda}q^+_a)
        &=
        \hat{\Lambda} \left(q^{+a} V^{++} J q^+_a \right)
        -
        q^{+a} (\hat{\Lambda}V^{++}) J q^+_a
        \\&=
        - \Omega \left(q^{+a} V^{++} J q^+_a \right)
        -
        q^{+a} (\hat{\Lambda}V^{++}) J q^+_a.
    \end{split}
\end{equation}
The first term is canceled by the first term in second line of \eqref{spin1Inv}, and so the requirement of the invariance of the coupling \eqref{eq: 1 12 12 vertex}
implies
\begin{equation}\label{eq: superconformal vector}
    \delta_{diff} V^{++} = - \hat{\Lambda}V^{++}.
\end{equation}
We observe that in the spin $\mathbf{1}$ case                           the $\mathcal{N}=2$ diffeomorphism transformation of potential $V^{++}$ amounts to the transport term.
Thus the vertex \eqref{eq: 1 12 12 vertex} is invariant with respect to the total localized $\mathcal{N}=2$ superconformal transformations, not only to the rigid form of the latter.
Similar results will be found in the higher-spin case. To prevent a misunderstanding, recall that the free $q^{+a}$
action \eqref{eq:hyper action} {\it is not invariant} under {\it general} analytic diffeomorphisms, but only with respect to the superconformal subclass
of them. The same is true of course for the sum  $S_{free} + S^{(s=1)}_{int}$.

\subsection{Gauge freedom}

At the next step we analyze the gauge freedom. The sum
\begin{equation}
    S_{free} + S^{(s=1)}_{int}
    =
    - \frac{1}{2} \int d\zeta^{(-4)}\, q^{+a} \mathcal{D}^{++} q^+_a
     - \frac{\kappa_1}{2}
     \int d\zeta^{(-4)}\, q^{+a} V^{++} J q^+_a, \lb{Sums=1}
\end{equation}
is invariant under the gauge $s=1$ transformations:
\begin{equation}\label{eq: spin 1 gauge}
    \begin{cases}
        \delta_\lambda V^{++} = \mathcal{D}^{++} \lambda,
        \\
        \delta_\lambda q^{+a} = - \kappa_1 \lambda
        J  q^{+a}
    \end{cases}
\end{equation}
for an arbitrary analytic parameter $\lambda(\zeta)$. Thus the full symmetry of the action  \eqref{Sums=1}
is ${\cal N}=2$ superconformal symmetry and $U(1)$ gauge symmetry.

The conserved current superfield associated with \eqref{eq: spin 1 gauge} can be directly obtained by varying cubic vertex with respect to $V^{++}$ \cite{Buchbinder:2022vra}:
\begin{equation}
    \mathcal{J}^{++} = - \frac{1}{2} q^{+a} J q^+_a,
    \qquad
    \mathcal{D}^{++}    \mathcal{J}^{++} = 0 \; ({\rm on \,\,shell}).
\end{equation}
$\mathcal{N}=2$ superconformal transformations of $\mathcal{J}^{++}$ read:
\begin{equation}
    \delta_{sc} \mathcal{J}^{++}
    =
    - \hat{\Lambda} \mathcal{J}^{++}
    -
    \Omega \mathcal{J}^{++}.
\end{equation}

\subsection{Wess-Zumino gauge: $\mathcal{N}=2$ Maxwell multiplet}

Using the gauge freedom \eqref{eq: spin 1 gauge} one can choose Wess-Zumino gauge for the analytic spin $\mathbf{1}$ potential:
\begin{equation}\label{Maxwell WZ}
    \begin{split}
        V^{++}_{WZ} =& -4i \theta^{+\alpha} \bar{\theta}^{+\dot{\alpha}} A_{\alpha\dot{\alpha}} - i (\theta^+)^2 \bar{\phi} + i (\bar{\theta}^+)^2 \phi \\&+ 4 (\bar{\theta}^+)^2 \theta^{+\alpha} \psi^i_\alpha u^-_i - 4 (\theta^+)^2 \bar{\theta}^+_{\dot{\alpha}} \bar{\psi}^{\dot{\alpha}i} u^-_i + (\theta^+)^2 (\bar{\theta}^+)^2 D^{ij} u^-_i u^-_j\;,
    \end{split}
\end{equation}
which yields just the off-shell field content of massless $\mathcal{N}=2$ spin $\mathbf{1}$ multiplet, {\it viz.} a complex scalar, a doublet of gaugini,
Maxwell gauge field and a real triplet of auxiliary fields:
\begin{equation}
    \phi, \quad \psi^i_\alpha, \quad A_{\alpha\dot{\alpha}}\,,\quad D^{(ij)}\,.
\end{equation}

The residual gauge freedom is given by $\lambda(\zeta) = a(x)$ and it is realized as the gauge transformation of Maxwell field:
\begin{equation}
    \delta_\lambda A_{\alpha\dot{\alpha}} = \partial_{\alpha\dot{\alpha}} a.
\end{equation}
So the spin $\mathbf{1}$ multiplet has $\mathbf{8}_B+ \mathbf{8}_F$ off-shell degrees of freedom.

For the coordinate-independent parameter $a$ the transformations \eqref{eq: spin 1 gauge} reduce to rigid $U(1)$ symmetry of the free hypermultiplet action.
This manifests the Noether nature of such an interaction. One can obtain this vertex by gauging rigid $U(1)$ symmetry.

\medskip

Thus we conclude that the $(\mathbf{1}, \mathbf{\tfrac{1}{2}},\mathbf{\tfrac{1}{2}})$ vertex \eqref{eq: 1 12 12 vertex}
is invariant under general analytic $\mathcal{N}=2$ superdiffeomorphisms realized as in  \eqref{eq: superconformal hyper 0} and \eqref{eq: superconformal vector}.
The special choice of parameters \eqref{eq:superconformal symmetry} yields rigid $\mathcal{N}=2$ superconformal transformations which leave invariant
the total hypermultiplet action \eqref{Sums=1} as well.

\section{$\mathcal{N}=2$ Weyl supermultiplet and superconformal $(\mathbf{2}, \mathbf{\tfrac{1}{2}}, \mathbf{\tfrac{1}{2}})$ coupling}\label{sec: weyl}
The superconformal vertex for the ${\cal N}=2$ spin $\mathbf{2}$ gauge multiplet interacting  with hypermultiplet is also known \cite{18, Galperin:1987ek, Ivanov:2022vwc, Galperin:1987em}.
Here we reproduce it, following our general strategy. This will give insights in how to construct higher-spin interactions
in non-trivial $\mathcal{N}=2$ conformal supergravity backgrounds.  Though the further generalization to higher spins will require  introducing additional derivatives,
the spin ${\bf 2}$  example is still instructive for exhibiting the common features of our approach.

In the spin $\mathbf{2}$ case the most general first-derivative analytic cubic interaction with the hypermultiplet has the form:
\begin{equation}\label{eq: s=2 general coupling}
    S^{(s=2)}_{int} = - \frac{\kappa_2}{2} \int d\zeta^{(-4)} \, q^{+a} h^{++M} \partial_M q^+_a
    =
    - \frac{\kappa_2}{2} \int d\zeta^{(-4)} \, q^{+a} \hat{\mathcal{H}}^{++}_{(s=2)} q^+_a.
\end{equation}
Here we have introduced the set of unconstrained analytic gauge superfields,
\begin{equation}\label{eq: s=2 prepotentials}
    h^{++\alpha\dot{\alpha}}(\zeta),
    \qquad
    h^{++\alpha+}(\zeta),
    \qquad
        h^{++\dot{\alpha}+}(\zeta),
    \qquad
    h^{(+4)}(\zeta)\,,
\end{equation}
and composed  the first-order analytic differential operator out of them:
\begin{equation}\label{eq: spin 2 operator}
    \hat{\mathcal{H}}^{++}_{(s=2)} : = h^{++M} \partial_M
    =
        h^{++\alpha\dot{\alpha}}\partial_{\alpha\dot{\alpha}}
        +
    h^{++\alpha+}\partial^-_\alpha
    +
    h^{++\dot{\alpha}+} \partial^-_{\dot{\alpha}}
    +
    h^{(+4)}\partial^{--}.
\end{equation}
As compared to the analogous operator in the non-conformal case \cite{Buchbinder:2022kzl}, here we have added the new analytic potential $h^{(+4)}$
entering with the partial harmonic derivative $\partial^{--}$.
The necessity of such a modification will become clear later.
Due to the reality of the action \eqref{eq: s=2 general coupling}, the operator \eqref{eq: spin 2 operator} should also satisfy  the reality condition:
\begin{equation}
     \widetilde{\hat{\mathcal{H}}^{++}_{(s=2)}}  =  \hat{\mathcal{H}}^{++}_{(s=2)}
     \quad
     \Rightarrow
     \quad
      \widetilde{h^{++\alpha\dot{\alpha}}} = h^{++\alpha\dot{\alpha}},
     \qquad
     \widetilde{h^{++\alpha+}}
    =
     h^{++\dot{\alpha}+},
     \qquad
     \widetilde{h^{(+4)}}
     =
      h^{(+4)}
     .
\end{equation}

\subsection{$\mathcal{N}=2$ superconformal symmetry }
\label{sec: s=2 sc}


To start with, we require invariance  of the cubic vertex \eqref{eq: s=2 general coupling} under rigid $\mathcal{N}=2$ superconformal transformations.
The variation of the $(\mathbf{2}, \mathbf{\tfrac{1}{2}}, \mathbf{\tfrac{1}{2}})$ vertex with respect to local $\mathcal{N}=2$ superconformal transformations
with\textit{ arbitrary analytic superfield parameters} reads:
\begin{equation}\label{s=2 coupling transform}
    \begin{split}
        \delta_{diff} S^{(s=2)}_{int} =&
        \frac{1}{2} \int d\zeta^{(-4)}\;    q^{+a}   [\hat{\mathcal{H}}^{++}_{(s=2)} ,   \hat{\Lambda} ] q^+_a
        -
        \frac{1}{2} \int d\zeta^{(-4)}\; q^{+a}  \delta_{diff} \hat{\mathcal{H}}^{++}_{(s=2)} q^+_a.
    \end{split}
\end{equation}

The condition of invariance under local $\mathcal{N}=2$ superconformal transformations gives rise to the following transformation law for
$\hat{\mathcal{H}}^{++}_{(s=2)}$:
\begin{equation}
    \delta_{diff} \hat{\mathcal{H}}^{++}_{(s=2)}
    =
    [\hat{\mathcal{H}}^{++}_{(s=2)} , \,  \hat{\Lambda} ],
\end{equation}
or, in terms of the analytic potentials,
\begin{equation}\label{eq: s=2 gauge transformations}
    \delta_{diff} h^{++M} = - \hat{\Lambda} h^{++M} + h^{++N} \partial_N \lambda^M.
\end{equation}
The resulting $\mathcal{N}=2$ superconformal transformations for the spin $\mathbf{2}$ analytic potentials (corresponding to the choice \eqref{eq:superconformal symmetry})
mix various gauge potentials  among each other. This is an essential difference from the approach of ref. \cite{Kuzenko:2022hdv} where every potential is primary,
i.e. homogeneously transforms through itself \footnote{One can represent analytic operator  $\hat{\mathcal{H}}^{++}_{(s=2)}$
as $\hat{\mathcal{H}}^{++}_{(s=2)}  = (\mathcal{D}^+)^4 \left( \Upsilon \mathcal{D}^{--} \right)$ (eq. \eqref{HUpsilon}), where $\Upsilon$ is an unconstrained Mezincescu-type prepotential, see \cite{Kuzenko:1999pi, Butter:2010sc}.
The conformal transformations of $\Upsilon$ are homogeneous and so it can be treated as a superconformal primary superfield \cite{Kuzenko:2021pqm}
in some harmonic-independent gauge. However such a description shadows the geometric origin of prepotentials and introduces an additional gauge freedom of ``non-geometric'' type.
See also discussion in section \ref{s=2 superconformal current superfields}.  }.

For example, the transformations of $h^{++\alpha+}$ under rigid special conformal transformations (parameter $k_{\alpha\dot{\alpha}}$ in  \eqref{eq:superconformal symmetry})
and rigid conformal supersymmetry (parameter $\eta^i_{\dot{\alpha}}$) are:
\begin{subequations}
\begin{equation}
    \delta_{k_{\alpha\dot{\alpha}}} h^{++\alpha+} = -\hat{\Lambda} h^{++\alpha+}
    +
     h^{++\alpha\dot{\rho}} k_{\rho\dot{\rho}} \theta^{+\rho}
    +
    h^{++\rho+} x^{\alpha\dot{\rho}} k_{\rho\dot{\rho}},
\end{equation}
\begin{equation}\label{eq: 57b}
    \delta_{\eta^i_{\dot{\alpha}}} h^{++\alpha+}
    =
    -
    \hat{\Lambda}  h^{++\alpha+}
    + h^{++\alpha\dot{\alpha}} \eta^i_{\dot{\alpha}} u^+_i - 4i h^{++\beta+}\theta^+_\beta \eta^{\alpha i}u^-_i
    +
    h^{(+4)} x^{\alpha\dot{\alpha}} \eta^i_{\dot{\alpha}} u^-_i.
\end{equation}
\end{subequations}
From the transformation
\begin{equation}
    \delta_{k_{\alpha\dot{\alpha}}} h^{(+4)} = - \hat{\Lambda} h^{(+4)}
    +
    4i h^{++\alpha+} \bar{\theta}^{+\dot{\alpha}} k_{\alpha\dot{\alpha}}
    +
    4i \theta^{+\alpha} h^{++\dot{\alpha}+}
    k_{\alpha\dot{\alpha}}
\end{equation}
it is obvious that it is impossible to avoid introducing the extra potential $h^{(+4)}$
in addition to the potentials of $\mathcal{N}=2$ Einstein's supergravity. Indeed, equating it to zero would inevitably
break rigid ${\cal N}=2$ superconformal symmetry (only ${\cal N}=2$ rigid Poincar\`e supersymmetry would survive).

\subsection{Gauge freedom}

Now we shall analyze the gauge freedom of the action:
\begin{equation}
    S_{free} + S^{(s=2)}_{int}
    =
    -
    \frac{1}{2} \int d \zeta^{(-4)}\, q^{+a} \mathcal{D}^{++} q^+_a
    -
    \frac{\kappa_2}{2} \int d\zeta^{(-4)}\, q^{+a} h^{++M} \partial_M q^+_a. \lb{Spin2Total}
\end{equation}

It is well known that ${\cal N}=2$ Weyl multiplet is produced as a result of gauging $\mathcal{N}=2$ superconformal transformations,
so in this case one should identify $\delta^{(s=2)}_\lambda = \kappa_2 \delta_{diff}$. So we start with the hypermultiplet transformation
of the form:
\begin{equation}\label{eq: s=2 gauge}
    \delta_\lambda^{(s=2)} q^{+a}
    =
    -
    \kappa_2\,
    \hat{\mathcal{U}}_{(s=2)}
    q^{+a}
    = - \kappa_2 \hat{\Lambda}\, q^{+a}
    -
    \frac{\kappa_2}{2}\Omega q^{+a}.
\end{equation}
Here we treat all gauge parameters
\begin{equation}\label{eq: s=2 parameters}
    \lambda^{\alpha\dot{\alpha}}(\zeta),
    \qquad
    \lambda^{+\alpha}(\zeta),
    \qquad
    \lambda^{+\dot{\alpha}}(\zeta),
    \qquad
    \lambda^{++}(\zeta)
\end{equation}
as arbitrary unconstrained analytic functions.

In \eqref{s=2 action trnsform} we have derived the variation of the free hypermultiplet action under such transformations:
\begin{equation}\label{eq: s=2 variation}
    \delta_\lambda^{(s=2)} S_{free} =  \frac{\kappa_2}{2} \int d\zeta^{(-4)}\, q^{+a}[\mathcal{D}^{++}, \hat{\Lambda}] q^+_a.
\end{equation}
There we required it to vanish in order to derive the constraints on the parameters \eqref{eq: s=2 parameters} yielding the rigid superconformal symmetry of the free hypermultiplet.
Now, instead of nullifying this term,  we cancel it in the sum \eqref{Spin2Total} by picking up the special gauge transformation
of the $s=2$ operator\footnote{The second term $\lambda^{++} \mathcal{D}^0$ gives the identically vanishing contribution $q^{+a} \lambda^{++} \mathcal{D}^0 q^+_a = 0$.
However, this term  is necessary for ensuring the compatibility of the left- and right-hand parts of the transformation law \eqref{5.14}.
In particular, it cancels the contribution $\lambda^{++} [\partial^{++}, \partial^{--}] = \lambda^{++} \partial^0$ coming from the first term.}:
\begin{equation} \label{5.14}
    \delta_\lambda \hat{\mathcal{H}}^{++}_{(s=2)}
    =
    [\mathcal{D}^{++}, \hat{\Lambda}]
    -
    \lambda^{++}\mathcal{D}^0.
\end{equation}
This transformation law amounts to the linearized gauge transformations of the potentials $h^{++M}(\zeta)$:
\begin{equation}\label{eq: spin 2 gauge transformations}
    \begin{cases}
        \delta_\lambda h^{++\alpha\dot{\alpha}} &=  \mathcal{D}^{++}\lambda^{\alpha\dot{\alpha}}
        + 4i \lambda^{+\alpha} \bar{\theta}^{+\dot{\alpha}}
        + 4i \theta^{+\alpha} \bar{\lambda}^{+\dot{\alpha}},
        \\
        \delta_\lambda h^{++\alpha+} &= \mathcal{D}^{++} \lambda^{+\alpha} - \lambda^{++}\theta^{+\alpha},
        \\
        \delta_\lambda h^{++\dot{\alpha}+} &= \mathcal{D}^{++} \lambda^{+\dot{\alpha}} - \lambda^{++}\bar{\theta}^{+\dot{\alpha}},
        \\
        \delta_\lambda h^{(+4)} &= \mathcal{D}^{++} \lambda^{++}.
    \end{cases}
\end{equation}

These transformations fully reproduce the linearized gauge freedom of $\mathcal{N}=2$ Weyl multiplet \cite{Galperin:1987ek}.
Note that eqs. \eqref{eq: spin 2 system} specifying rigid symmetries of the hypermultiplet, are just the conditions of vanishing
of the variations \eqref{eq: spin 2 gauge transformations},  $\delta_\lambda h^{++M} = 0$. So one can interpret $\mathcal{N}=2$ rigid superconformal
group as the transformations  preserving the flat $\mathcal{N}=2$ conformal supergravity background $h^{++M}=0$.
This is a consequence of the fact that the multiplet of $\mathcal{N}=2$ conformal supergravity can be obtained through the analytic gauging of rigid $\mathcal{N}=2$ superconformal transformations.

\medskip

Since we did not impose any conditions on the parameters $\lambda^M(\zeta)$ in section \ref{sec: s=2 sc}, the action $S_{free} + S_{int}^{(s=2)}$
is exactly invariant under the transformations $\delta_\lambda + \kappa_2\delta_{diff}$ with arbitrary analytic parameters $\lambda^M(\zeta)$:
\begin{equation}\label{eq: spin 2 nonlinear}
    \delta^\lambda_{nonl} \hat{\mathcal{H}}^{++}_{(s=2)}
    :=
    \left(\delta_\lambda + \kappa_2 \delta_{diff} \right) \hat{\mathcal{H}}^{++}_{(s=2)}
    =
        [\mathcal{D}^{++} + \kappa_2\hat{\mathcal{H}}^{++}_{(s=2)} ,   \hat{\Lambda} ]
        -
        \lambda^{++} \mathcal{D}^0.
\end{equation}
 In this way we recover the \textit{non-linear gauge freedom of $\mathcal{N}=2$ Weyl multiplet} elaborated in \cite{Galperin:1987ek}.
In the full nonlinear case, superconformal transformations become a subgroup of the full gauge supergroup of conformal supergravity. The latter is realized
on the analytic gauge potentials $h^{++M}$ by the same formulas \eqref{eq: spin 2 gauge transformations}, however with the replacement
\bea
\mathcal{D}^{++} \; \Rightarrow \; \mathfrak{D}^{++} := \mathcal{D}^{++}+ \kappa_2 \hat{\mathcal{H}}^{++}_{(s=2)}\,. \lb{ChangeFull}
\eea

\medskip

The superconformal coupling \eqref{eq: 1 12 12 vertex} of the hypermultiplet to the spin ${\bf 1}$ superfield $V^{++}$
is also invariant under the full conformal supergravity gauge supergroup; the $U(1)$ gauge transformations \eqref{eq: spin 1 gauge} are modified just through
the replacement \eqref{ChangeFull} in the gauge transformation of $V^{++}$. The sum of the  $q^{+a}$ action \eqref{Spin2Total} covariantized by ${\cal N}=2$ Weyl multiplet
and the $(\mathbf{1}, \mathbf{\frac{1}{2}}, \mathbf{\frac{1}{2}})$ coupling \eqref{eq: 1 12 12 vertex},
\begin{equation}\label{eq: 1,1/2,1/2 sugra}
    S =- \frac{1}{2} \int d\zeta^{(-4)} q^{+a} \left( \mathfrak{D}^{++} + \kappa_1 V^{++}J \right) q^+_a\,,
\end{equation}
 is invariant under both the full $\mathcal{N}=2$ conformal supergravity gauge supergroup and the modified gauge
$U(1)$ transformations:
\begin{equation}
    \delta_\lambda V^{++} = [\mathfrak{D}^{++}, \lambda] = \mathfrak{D}^{++}\lambda.
\end{equation}
So we have constructed the vertex $(\mathbf{1}, \mathbf{\frac{1}{2}}, \mathbf{\frac{1}{2}})$ in $\mathcal{N}=2$ conformal supergravity background.
Note that the superfield $V^{++}$ in the action \eqref{eq: 1,1/2,1/2 sugra} does not directly interact with $h^{++M}$. At the component level, such an interaction
is induced as a result of elimination  of the auxiliary fields of the hypermultiplet.

\subsection{Wess-Zumino gauge: $\mathcal{N}=2$ Weyl supermultiplet} \label{eq: s=2 WZ gauge}

To specify the physical contents of Weyl multiplet, one needs to gauge away the pure gauge degrees of freedom, thus fixing the Wess-Zumino
gauge for the set of spin $\mathbf{2}$ analytic potentials:
\begin{equation}
    \begin{cases}
        h^{++\alpha\dot{\alpha}}
        =
        -4i \theta^{+\rho}\bar{\theta}^{+\dot{\rho}} \Phi^{\alpha\dot{\alpha}}_{\rho\dot{\rho}}
        -  (\bar{\theta}^+)^2 \theta^{+}_{\rho} \psi^{(\alpha\rho)\dot{\alpha}i} u^-_i
        +
         (\theta^+)^2 \bar{\theta}^{+}_{\dot{\rho}} \bar{\psi}^{\alpha(\dot{\alpha}\dot{\rho})i}_{}u_i^-
        \\\qquad\qquad\qquad\qquad\qquad\qquad\qquad\qquad\qquad  +  (\theta^+)^2 (\bar{\theta}^+)^2 V^{\alpha\dot{\alpha}(ij)}u^-_iu^-_j\,,\\
        h^{++\mu+} =(\theta^+)^2 \bar{\theta}^+_{\dot{\mu}} P^{\mu\dot{\mu}}
        + \left(\bar{\theta}^+\right)^2 \theta^+_\nu T^{(\nu\mu)}
        +  (\theta^+)^2 (\bar{\theta}^+)^2 \chi^{\mu i}u^-_i, \;\;\;
        \\
        h^{++\dot{\mu}+} = \widetilde{h^{++\mu+}}\,,
        \\
        h^{(+4)} \;\;\,= (\theta^+)^2 (\bar{\theta}^+)^2 D\,.
    \end{cases}
\end{equation}
Here we find out the physical content of $\mathcal{N}=2$ Weyl multiplet \cite{Bergshoeff:1980is, Muller:1986ku, Galperin:1987ek}
involving graviton, a doublet of conformal gravitinos, gauge fields for $SU(2)_R$ and $\gamma_5$ transformations; all other fields are auxiliary
(after some redefinition):
\begin{equation}
    \Phi^{\alpha\dot{\alpha}}_{\rho\dot{\rho}},
    \quad
    \psi^{(\alpha\beta)\dot{\alpha}i},
    \quad
    V_{\alpha\dot{\alpha}}^{(ij)},
    \quad
    P^{\mu\dot{\mu}};
    \qquad
    T^{(\mu\nu)},
    \quad
     \chi^{\mu i},
    \quad
    D.
\end{equation}

At the linearized level, the residual gauge freedom of the theory is spanned by the parameters:
\begin{equation}\label{eq: s=2 residual}
    \begin{cases}
        \lambda^{\alpha\dot{\alpha}} \Rightarrow a^{\alpha\dot{\alpha}}(x) - 4i \epsilon^{\alpha i}(x) u^-_i \bar{\theta}^{+\dot{\alpha}} - 4i \theta^{+\alpha} \bar{\epsilon}^{\dot{\alpha}i}(x) u^-_i - 4i \theta^{+\alpha}\bar{\theta}^{+\dot{\alpha}} \lambda^{(ij)}(x)u^-_i u^-_j \,, \\
        \lambda^{\mu+} \Rightarrow \epsilon^{\mu i}(x) u^+_i + \theta^{+\nu} \big[\big\{ \frac12 \big[{a(x)} + i {b(x)}\big] + \lambda^{(ij)}(x) u^+_i u^-_j \big\}  \delta_\nu^\mu
        + l_{(\nu}^{\;\;\;\mu)}(x) \big] - i (\theta^+)^2 \partial^\mu_{\dot{\rho}}\bar{\epsilon}^{\dot{\rho} i}(x) u^-_i \,, \\
        \bar{\lambda}^{\dot{\mu}+} \Rightarrow \bar{\epsilon}^{\mu i}(x) u^+_i + \bar{\theta}^{+\dot{\nu}} \big[\big\{\frac12 \big[{a(x)} - i {b(x)}\big]
        + \lambda^{(ij)}(x) u^+_i u^-_j \big\}  \delta_{\dot{\nu}}^{\dot{\mu}} + l_{(\dot{\nu}}^{\;\;\;\dot{\mu})}(x) \big]
        +
        i (\bar{\theta}^+)^2 \partial^{\dot{\mu}}_\rho \epsilon^{\rho i}(x) u^-_i\,,
        \\
        \lambda^{++} \Rightarrow
        \lambda^{ij}(x) u^+_i u^+_j
        +
        4i \theta^{+\alpha}\bar{\theta}^{+\dot{\alpha}} \partial_{\alpha\dot{\alpha}}a(x)
        +
        2i \left(\theta^{+\alpha} \partial_{\alpha\dot{\rho}} \bar{\epsilon}^{\dot{\rho} i}(x) - \partial_{\dot{\alpha}\rho} \epsilon^{\rho i} (x) \bar{\theta}^{+\dot{\alpha}} \right)u^+_i \,.
    \end{cases}
\end{equation}
These parameters can be identified as follows:
\begin{itemize}
    \item $a^{\alpha\dot{\alpha}}(x)$ are the remnants of the diffeomorphism parameters which  now form the basic gauge freedom of the free spin 2 field;
    \item $\epsilon^{\hat{\mu}i}(x)$ originate from the parameters of local supersymmetry and provide ${\cal N}=2$ counterparts
    of the local $a^{\alpha\dot{\alpha}}$ transformations;
    \item $l^{(\mu\nu)}(x)$ and $l^{(\dot{\mu}\dot{\nu})}(x)$ are the former parameters of local Lorentz transformations which can be used to gauge away
    the antisymmetric part of $\Phi^{\alpha\dot{\alpha}}_{\rho\dot{\rho}}$ and so to leave in the latter only the symmetric part (traceless ``conformal graviton'' and the trace itself);
    \item $a(x)$ is a parameter of Weyl transformation;
    \item $b(x)$ is a parameter of local $U_R(1)$ transformations;
    \item $\lambda^{(ij)}(x)$ are parameters of local $SU(2)_R$ transformations.
\end{itemize}
Rigid parameters of  special conformal transformations $k_{\alpha\dot{\alpha}}$ and  conformal supersymmetry $\eta^{\hat{\mu}i}$ are contained in derivatives of gauge parameters:
\begin{equation}
    k_{\alpha\dot{\alpha}}= \partial_{\alpha\dot{\alpha}} a(x)\vert ,
    \quad
    \eta^{\alpha i} =  \frac{1}{2} \partial^{\alpha}_{\dot{\rho}} \bar{\epsilon}^{\dot{\rho} i}(x)\vert,
    \quad
     \bar{\eta}^{\dot{\alpha}i} =  - \frac{1}{2} \partial^{\dot{\alpha}}_\rho \epsilon^{\rho i}(x)\vert.
\end{equation}
Here the symbol ``|'' means restriction to $x^{\alpha\dot{\alpha}}$-independent parts of the parameters.

To find the residual gauge transformations and their action on the component fields, one needs to require the preservation
of the Wess-Zumino gauge, that is in $\delta_{\lambda} h^{++M} $  there should be no terms which could not be compensated by the
appropriate transformations of  fields in   $h^{++M}_{WZ}$.
From this condition one can determine the parameters $\lambda_{comp}^M$
and the action of these transformations on the fields of $\mathcal{N}=2$ Weyl multiplet.
As a result, the linearized transformation law for graviton is:
\begin{equation}
    \delta_\lambda \Phi^{\alpha\dot{\alpha}\rho\dot{\rho}}_{}
    =
    \partial^{\rho\dot{\rho}} a^{\alpha\dot{\alpha}}
    -
    2
    l^{(\alpha\rho)}_{} \epsilon^{\dot{\alpha}\dot{\rho}}
    -
    2 l^{(\dot{\alpha}\dot{\rho})} \epsilon^{\alpha\rho}
    -
     a \epsilon^{\alpha\rho}\epsilon^{\dot{\alpha}\dot{\rho}}.
\end{equation}
The decomposition of the field $\Phi^{\alpha\dot{\alpha}\rho\dot{\rho}}$ into the irreducible parts is as follows\footnote{The linearized relation with the metric tensor $g_{ab} = \eta_{ab} + h_{ab}$ is given by:
    $$h^{ab} = \sigma^a_{\alpha\dot{\alpha}}\sigma^b_{\beta\dot{\beta}} \Phi^{(\alpha\beta)(\dot{\alpha}\dot{\beta})} + \frac{1}{2}\eta^{ab} \Phi.$$}:
\begin{equation}
    \Phi^{\alpha\dot{\alpha}\rho\dot{\rho}}
    =
    \Phi^{(\alpha\rho)(\dot{\alpha}\dot{\rho})}
    +
    \epsilon^{\dot{\alpha}\dot{\rho}} \Phi^{(\alpha\rho)}
    +
    \epsilon^{\alpha\beta} \Phi^{(\dot{\alpha}\dot{\beta})}
    +
    \epsilon^{\alpha\beta}  \epsilon^{\dot{\alpha}\dot{\rho}}
    \Phi.
\end{equation}
The parameters $l^{(\alpha\rho)}$, $l^{(\dot{\alpha}\dot{\rho})}$ and $a$ can be used to gauge away all the components except for the symmetric part:
\begin{equation}
    \delta\Phi^{(\alpha\rho)(\dot{\alpha}\dot{\rho})} =
      \partial^{(\dot{\rho}(\rho} a^{\alpha)\dot{\alpha})}.
\end{equation}
In this gauge we have $a = \frac{1}{4} \partial_{\rho\dot{\rho}} a^{\rho\dot{\rho}}$, $l^{(\alpha\rho)} = \frac{1}{4} \partial^{(\rho\dot{\rho}} a^{\alpha)}_{\dot{\rho}}$, $l^{(\dot{\alpha}\dot{\rho})} = \frac{1}{4} \partial^{(\dot{\rho}\rho} a^{\dot{\alpha})}_{\rho}$.

Other gauge fields can be worked out in a similar way. Their irreducible form and gauge transformation laws are given by:
\begin{equation}
    \delta \psi^{(\alpha\rho)\dot{\alpha}i}
    =
    \partial^{\dot{\alpha}(\alpha}  \epsilon^{\rho)i},
    \qquad
    \delta V_{\alpha\dot{\alpha}}^{(ij)} = \partial_{\alpha\dot{\alpha}} \lambda^{(ij)},
    \qquad
    \delta P_{\alpha\dot{\beta}} = \frac{1}{2}\partial_{\alpha\dot{\alpha}} b.
\end{equation}
The fields $ T^{(\mu\nu)}, \chi^{\mu i}, D$ are auxiliary and, after the appropriate redefinition, become invariant under gauge transformations.
So $\mathcal{N}=2$ Weyl multiplet collects $\textbf{24}_B+\textbf{24}_F$
off-shell degrees of freedom.

Note that the same form
of WZ gauge for the analytic gauge potentials can be fixed by starting with the full nonlinear ${\cal N}=2$ conformal supergravity
group from the very beginning.

\subsection{$s=2$ superconformal current superfields}\label{s=2 superconformal current superfields}

According to the superfield version of Noether's theorem, the conserved superfield currents are associated with rigid symmetry  transformations.
The parameters $\lambda^M$ that satisfy the relation $[\mathcal{D}^{++}, \,\hat{\Lambda}] = 0$ form rigid symmetry of the free hypermultiplet.
Using the expression \eqref{eq: s=2 variation} for the variation of the action  one can easily generalize, to $\mathcal{N}=2$ superconformal case,
the current superfields found in \cite{Buchbinder:2022vra} for the non-conformal case. Actually, since the variation of the action on shell is vanishing,
we obtain a set of conservation laws for each of the unconstrained parameters $\lambda^M$.
Equivalently, one can obtain these current superfields by varying cubic coupling \eqref{eq: s=2 general coupling} with respect to $\mathcal{N}=2$ Weyl potentials $h^{++M}$.
As a result, we obtain:
\begin{equation}\label{eq: s=2 conservation laws}
    \begin{split}
        M = \alpha\dot{\alpha}
        \qquad
        &\Rightarrow
        \qquad
        J^{++}_{\alpha\dot{\alpha}} = - \frac{1}{2}q^{+a} \partial_{\alpha\dot{\alpha}} q^+_a,
        \qquad
        \mathcal{D}^{++} J^{++}_{\alpha\dot{\alpha}} = 0;
        \\
        M = \alpha
        \qquad
        &\Rightarrow
        \qquad
        J^+_\alpha = - \frac{1}{2} q^{+a} \partial^-_\alpha q^+_a,
        \qquad
        \quad
        \mathcal{D}^{++} J^+_\alpha = 4i \theta^{+\dot{\rho}} J_{\alpha\dot{\rho}};
        \\
        M = \dot{\alpha}
        \qquad
        &\Rightarrow
        \qquad
        J^+_{\dot{\alpha}} = - \frac{1}{2} q^{+a} \partial^-_{\dot{\alpha}} q^+_a,
        \qquad
        \quad
        \mathcal{D}^{++} J^+_{\dot{\alpha}} = -4i \theta^{+\rho} J_{\rho\dot{\alpha}};
        \\
        M = ++
        \qquad
        &\Rightarrow
        \qquad
        J = - \frac{1}{2} q^{+a} \partial^{--} q^+_a,
        \qquad
        \quad
        \mathcal{D}^{++} J  = - \theta^{+\hat{\rho}} J^+_{\hat{\rho}}.
    \end{split}
\end{equation}
As was shown in \cite{Buchbinder:2022vra} for the component expansion of the non-conformal currents superfields, the conservation laws of the superfield currents
\eqref{eq: s=2 conservation laws} lead to the standard $x$-space conservation of the component currents.
All the above current superfields are analytic, but $J^+_{\hat{\alpha}}$ and $J$ are not invariant under $\mathcal{N}=2$ supersymmetry.
The $\mathcal{N}=2$ supersymmetry-invariant  supercurrent is defined by the non-analytic superfield:
\begin{equation}\label{eq: cal J}
    \mathcal{J} := - \frac{1}{2} q^{+a} \mathcal{D}^{--} q^+_a
    =
    J
    +
    \theta^{-\hat{\rho}} J_{\hat{\rho}}
    -
    4i \theta^{-\rho} \bar{\theta}^{-\dot{\rho}} J^{++}_{\rho\dot{\rho}}.
\end{equation}
It embodies all the analytic currents and satisfies the conservation law
\begin{equation}\label{eq: mathcal J s=2}
    \mathcal{D}^{++} \mathcal{J} = 0\,,
\end{equation}
which immediately reproduces the conservation laws \eqref{eq: s=2 conservation laws}.

Note that the supercurrent $\mathcal{J}$ could be obtained from the representation of unconstrained analytic parameters $\lambda^M$ through an unconstrained
non-analytic superfield  parameter $l^{--} (\zeta, \theta^-)$:
\begin{equation}
    \hat{\Lambda } = \left(D^{+}\right)^4 \left( l^{--} \mathcal{D}^{--}  \right).
\end{equation}
The variation \eqref{eq: s=2 variation} of the free hypermultiplet action takes the form:
\begin{equation}
    \delta_{l}^{(s=2)} S_{free} = \int d^4x d^8 \theta du\, \left(\mathcal{D}^{++} l^{--}\right) q^{+a} \mathcal{D}^{--} q^+_a\,,
\end{equation}
which immediately leads to  \eqref{eq: mathcal J s=2}.

One can also obtain $\mathcal{J}$ by varying the coupling \eqref{eq: s=2 general coupling} with respect to an unconstrained non-analytic prepotential
$\Upsilon(\zeta, \theta^-)$ of $\mathcal{N}=2$ conformal supergravity defined as:
\be \label{HUpsilon}
\hat{\mathcal{H}}^{++}_{(s=2)} := (\mathcal{D}^+)^4 \left( \Upsilon \mathcal{D}^{--} \right).
\ee
This way of representing analytic gauge potentials  through the non-analytic Mezincescu-type prepotential was used in refs. \cite{Zupnik:1998td, Kuzenko:1999pi}.
It allows one  to relate harmonic gauge potentials to the prepotentials of non-geometric type used for the superfield description of supergravity
beyond the HSS approach \cite{Howe:1981qj, Gates:1981qq, Rivelles:1981qz, Butter:2010sc}.

So we conclude that the transformations \eqref{eq: s=2 gauge} correspond to the  non-analytic current superfield $\mathcal{J}$ defined in \eqref{eq: cal J} and obeying the appropriate
conservation law \eqref{eq: mathcal J s=2} on shell.
This is the ``master'' current superfield discussed recently in \cite{Buchbinder:2022vra} and originally introduced in \cite{Kuzenko:1999pi} (see also a recent work \cite{Kuzenko:2023vgf}).
As compared with the non-conformal spin $\mathbf{2}$  supercurrent \cite{Buchbinder:2022vra},
 we observe the appearance of a new analytic current $J = -\frac{1}{2} q^{+a} \partial^{-- } q^+_a$ associated with the rigid conformal parameter $\lambda^{++}$
 in  \eqref{eq:superconformal transfromations}.

\medskip

Under the $\mathcal{N}=2$ superconformal transformations of the hypermultiplet,  $\mathcal{J}$ transforms as:
\begin{equation}
    \delta_{sc} \mathcal{J} = -\hat{\Lambda} \mathcal{J}
    -
    \Omega \mathcal{J}
    +
    \frac{1}{2} q^{+a} [\mathcal{D}^{--},\, \hat{\Lambda}] q^+_a.
\end{equation}
The last term implies the presence of inhomogeneities in the current transformation laws. E.g., for dilatations (parameter $a$ in \eqref{eq:superconformal symmetry}) we obtain
$$[\mathcal{D}^{--}, \,\hat{\Lambda}] = -4i a \theta^{-\alpha} \bar{\theta}^{-\dot{\alpha}} \partial_{\alpha\dot{\alpha}}
+ \frac{1}{2} a \theta^{-\hat{\alpha}}\partial^-_{\hat{\alpha}}, $$
so  $\mathcal{J}$ transforms under dilatations as:
\begin{equation}
    \delta_{(a)} \mathcal{J} = - \hat{\Lambda}_{(a)} \mathcal{J}
    -
    \Omega_{(a)}  \mathcal{J}
    +
    4i a \theta^{-\alpha} \bar{\theta}^{-\dot{\alpha}} J_{\alpha\dot{\alpha}}
    -
    \frac{1}{2} a \theta^{-\hat{\alpha}} J_{\hat{\alpha}}.
\end{equation}
Using the relation \eqref{eq: cal J}, one can equivalently rewrite this as
\begin{equation}
    \delta_{(a)} \mathcal{J} = - \hat{\Lambda}_{(a)} \mathcal{J}
    -
    \Omega_{(a)}  \mathcal{J}
    -
    \frac{1}{2} a \theta^{-\hat{\alpha}} \partial^+_{\hat{\alpha}} \mathcal{J}.\lb{deltab}
\end{equation}
The last term appeared due to the dilatation rescaling of non-analytic $\theta^{-\hat{\alpha}}$. So, defining $\hat{\Lambda}_{na} : = \hat{\Lambda} + \lambda^{-\hat{\alpha}}\partial^+_{\hat{\alpha}}$,
the variation \eqref{deltab} can be cast in the more suggestive form:
\begin{equation}
    \delta_{(a)} \mathcal{J} = - \hat{\Lambda}_{(a)} \mathcal{J}
-
\Omega_{(a)}  \mathcal{J}.
\end{equation}

\section{$\mathcal{N}=2$ spin $\mathbf{3}$ superconformal multiplet and superconformal $(\mathbf{3}, \mathbf{\tfrac{1}{2}}, \mathbf{\tfrac{1}{2}})$ coupling}
\label{sec: spin 3}

The spin $\mathbf{3}$ superconformal interaction with hypermultiplet is the first non-trivial case which was never discussed before in the HSS approach.
The most general form of the two-derivative analytic vertex is:
\begin{equation}\label{eq: s=3 most general}
    S_{int}^{(s=3)} = - \frac{\kappa_3}{2}\int d\zeta^{(-4)}\, q^{+a} \, h^{++MN} \partial_N \partial_M J \,q^+_a .
\end{equation}
Here we introduced unconstrained analytic gauge potentials $h^{++MN}(\zeta)$, which satisfy the conditions:
\begin{equation}\label{eq: spin 3 indices}
    h^{++MN} = (-1)^{P(M)P(N)} h^{++NM},
\end{equation}
with
\begin{equation}
    P(M) :=
    \begin{sqcases}
        0 \quad \text{for} \quad M = \alpha\dot{\alpha}, ++ \;\;\;\;\text{(bosonic indices)};
        \\
        1 \quad \text{for} \quad M = \alpha +, \, \dot{\alpha} +
       \quad \text{(fermionic indices)}.
    \end{sqcases}
\end{equation}
The conditions \eqref{eq: spin 3 indices} are necessary in order to avoid ``double counting'' of terms of the same type, for example,
\begin{equation}
    h^{++\alpha+\dot{\beta}+} \partial^-_{\dot{\beta}} \partial^-_\alpha
    =
    h^{++\dot{\beta}+\alpha+}
    \partial^-_\alpha
 \partial^-_{\dot{\beta}}.
\end{equation}

Taking this into account, the complete expansion of the operator with two derivatives has the form:
\begin{equation}
    \begin{split}
        h^{++MN}\partial_N \partial_M
        =&\,
        h^{++\alpha\dot{\alpha}\beta\dot{\beta}} \partial_{\beta\dot{\beta}} \partial_{\alpha\dot{\alpha}}
        \\
        &+
        h^{++[\beta+\gamma]+} \partial^-_\gamma \partial^-_\beta
        +
        h^{++[\dot{\beta}+\dot{\gamma}]+} \partial^-_{\dot{\gamma}} \partial^-_{\dot{\beta}}
        +
        h^{(+6)} \partial^{--} \partial^{--}
        \\
        &+
        2h^{++\beta+\alpha\dot{\alpha}}
        \partial_{\alpha\dot{\alpha}} \partial^-_\beta
        +
        2h^{++\dot{\beta}+\alpha\dot{\alpha}}
        \partial_{\alpha\dot{\alpha}} \partial^-_{\dot{\beta}}
        +
        2 h^{++\alpha\dot{\alpha}++} \partial^{--} \partial_{\alpha\dot{\alpha}}
        \\
        &+
        2 h^{++\beta+\dot{\gamma}+} \partial^-_{\dot{\gamma}} \partial^-_{\beta}
        +
        2 h^{++++\beta+} \partial^-_\beta \partial^{--}
        +
        2 h^{++++\dot{\beta}+} \partial^-_{\dot{\beta}} \partial^{--}.
    \end{split} \label{s3Split}
\end{equation}
We require reality of the action \eqref{eq: s=3 most general}, so the analytic gauge potentials satisfy the following tilde-conjugation rules:
\begin{equation}
    \widetilde{h^{++MN}\partial_N \partial_M}
    =
    h^{++MN}\partial_N \partial_M.
\end{equation}
It then follows that the  analytic potentials $h^{++MN}$ obey the reality conditions:
\begin{subequations}\label{eq: spin 3 reality}
    \begin{equation}
        \widetilde{h^{++\alpha\dot{\alpha}\beta\dot{\beta}}} = h^{++\alpha\dot{\alpha}\beta\dot{\beta}},
        \qquad
        \widetilde{h^{(+6)}} = h^{(+6)},
    \end{equation}
    \begin{equation}
        \widetilde{h^{++[\beta+\gamma]+}} =
       - h^{++[\dot{\beta}+\dot{\gamma}]+},
        \qquad
        \widetilde{ h^{++[\dot{\beta}+\dot{\gamma}]+}}
        =
        -h^{++[\beta+\gamma]+},
    \end{equation}
    \begin{equation}
        \widetilde{h^{++\beta+\alpha\dot{\alpha}}}
        =
        -
        h^{++\dot{\beta}+\alpha\dot{\alpha}},
        \qquad
        \widetilde{h^{++\dot{\beta}+\alpha\dot{\alpha}}}
        =
        h^{++\beta+\alpha\dot{\alpha}},
    \end{equation}
    \begin{equation}
        \widetilde{h^{++\alpha\dot{\alpha}++}} = h^{++\alpha\dot{\alpha}++},
        \qquad
        \widetilde{h^{++\alpha+\dot{\alpha}+}} = -
        h^{++\alpha+\dot{\alpha}+},
    \end{equation}
    \begin{equation}
        \widetilde{h^{++++\beta+}} = - h^{++++\dot{\beta}+},
        \qquad
        \widetilde{h^{++++\dot{\beta}+}}
        =
        h^{++++\beta+}.
    \end{equation}
\end{subequations}
At this step we deal with the most general form of the analytic gauge potentials  $h^{++MN}$,
without assuming in advance any symmetry between the Lorentz spinorial indices hidden in the
multi-indices $M$ and $N$ \footnote{This is an essential difference from the non-conformal case \cite{Buchbinder:2021ite}, where all indices of the
same chirality were assumed to be symmetrized.}.

\medskip
Next we require $\mathcal{N}=2$ superconformal invariance of vertex \eqref{eq: s=3 most general} and determine the minimal
set of potentials $h^{++MN}(\zeta)$ needed to secure this invariance.
After that we will analyze gauge freedom of the  coupling obtained, as well as the irreducible physical field contents of the corresponding superconformal
spin $\mathbf{3}$ supermultiplet.

\subsection{$\mathcal{N}=2$ superconformal symmetry}

The hypermultiplet transformations \eqref{eq:superconformal transfromations} with arbitrary analytic parameters take the superfield Lagrangian
in \eqref{eq: s=3 most general}, up to total derivative, into:
\begin{equation} \lb{tran6.5}
    \begin{split}
        \delta_{diff}
        \left( q^{+a} h^{++MN}  \partial_N  \partial_M  J q^+_a \right)
        =&
        q^{+a} ( \hat{\Lambda} h^{++MN} )
        \partial_N \partial_M
        J q^+_a
        \\&
        -
        2
        q^{+a} h^{++MN} (\partial_{N}\lambda^K) \partial_K \partial_M J q^+_a
        \\&
        +
        \frac{1}{2} (-1)^{P(K)} \partial_K \left( h^{++MN} \partial_N\partial_M \lambda^K\right)
        q^{+a} J q^+_a
        \\&
        +
        \frac{1}{2} (-1)^{P(M)} (\partial_Mh^{++MN}) (\partial_N \Omega) q^{+a} J q^+_a.
    \end{split}
\end{equation}
Note that in the process of calculation of this variation we made use of the property
\be \lb{RedInt}
J_{ab} = J_{ba} \;\Rightarrow \; q^{+a}J_{ab}(\partial_Z q^{+b}) = \frac12 \partial_Z (q^{+a} J_{ab} q^{+b}),
\ee
which ensures reducing all terms with one derivative to those without derivatives by integration by parts.

We observe the presence of two types of terms: those with two derivatives acting on the hypermultiplet and terms without derivatives at all.
To cancel all these terms  one is led to slightly modify the vertex \eqref{eq: s=3 most general} by introducing the spin ${\bf 1}$ superfield  $h^{++}$
and adding the relevant $(\mathbf{1}, \mathbf{\frac{1}{2}}, \mathbf{\frac{1}{2}})$-type vertex:
\begin{equation}\label{eq: s=3 most general-1}
(6.1) \;\Rightarrow \;        S_{int}^{(s=3)} = - \frac{\kappa_3}{2} \int d\zeta^{(-4)}\, q^{+a} h^{++MN} \partial_N \partial_M J q^+_a
        -
        \frac{\kappa_3}{2} \int d\zeta^{(-4)} q^{+a} h^{++} J q^+_a.
\end{equation}
Superfield $h^{++}$ satisfies the reality condition $\widetilde{h^{++}} = h^{++}$.

\medskip

We start our analysis with the two-derivative terms. Requiring local superconformal transformation laws for the analytic potentials,
\begin{equation}\label{eq: s=3 SC}
    \delta_{diff} h^{++MN}
    =
    -\hat{\Lambda}h^{++MN}
    +
    h^{++MK} (\partial_K \lambda^N)
    +
    (-1)^{P(N)\left[P(M)+ P(K) \right]}
    h^{++KN} (\partial_K\lambda^M),
\end{equation}
one can cancel the analogous terms  with two derivatives in \eqref{tran6.5}. The first term is the transport term, while the second and third ones mix up the potentials carrying different indices.
If one chooses as the parameters just the rigid superconformal parameters \eqref{eq:superconformal symmetry} then it is not difficult to make sure that it is necessary
 to include into the game the whole set of potentials $h^{++MN}$.
 For example, under the conformal supersymmetry (parameter $\eta^i_\rho$ in \eqref{eq:superconformal symmetry}) we have:
\begin{equation}\label{eq: h++}
    \delta_{\eta^i_\rho} h^{++\alpha+\dot{\beta}+}
    =
    -
    \hat{\Lambda} h^{++\alpha+\dot{\beta}+}
    +
    h^{++(\alpha\beta)\dot{\beta}+} \left( \eta^i_\beta  u^+_i\right) + \dots\,.
\end{equation}
We observe that  the potential $h^{++\alpha+\dot{\beta}+}$ is mixed with $h^{++(\alpha\beta)\dot{\beta}+}$.

This peculiarity leads to an important difference of the superconformal vertices from the non-conformal ones constructed in \cite{Buchbinder:2022kzl}.
Indeed, to respect the standard Poincar\'e  supersymmetry (parameters $a^{\alpha\dot{\alpha}}$ and $\epsilon^{\hat{\alpha}i}$
in \eqref{eq:superconformal symmetry}) it would be enough to deal only with the restricted set of potentials $h^{++M\alpha\dot{\alpha}}$.

The superdiffeomorphism transformation of the $(\mathbf{1}, \mathbf{\frac{1}{2}}, \mathbf{\frac{1}{2}})$ part of the vertex reads
\begin{equation}
    \delta_{diff} \int d\zeta^{(-4)} q^{+a} h^{++} J q^+_a
    =
    \int d\zeta^{(-4)} q^{+a} \left( \delta_{diff} h^{++} + \hat{\Lambda} h^{++} \right) J q^+_a\,,
\end{equation}
and it is required to cancel the terms without derivatives in \eqref{tran6.5}. This is achieved with
\begin{equation}\label{eq: s=3 s=1 SC}
    \begin{split}
        \delta_{diff} h^{++} =& - \hat{\Lambda} h^{++}
        -
        \frac{1}{2} (-1)^{P(K)} \partial_K \left( h^{++MN} \partial_N\partial_M \lambda^K\right)
        \\&
        -
    \frac{1}{2} (-1)^{P(M)} (\partial_Mh^{++MN}) (\partial_N \Omega)\,.
    \end{split}
\end{equation}
The first term  coincides with the similar term in the superconformal transformation of spin $\mathbf{1}$
multiplet \eqref{eq: superconformal vector} and so it automatically leaves the action invariant.
Then the appropriate parts of the  two-derivative transformations in \eqref{tran6.5} are canceled by the remaining terms in \eqref{eq: s=3 s=1 SC}.

Thus we arrive at the cubic vertices which are {\it invariant under $\mathcal{N}=2$ superdiffeomorphism transformations} with the general analytic parameters $\lambda^M$ (i.e. invariant under
the complete gauge group of $\mathcal{N}=2$ conformal supergravity).
The spin $\mathbf{2}$ gauge transformations act on the spin $\mathbf{3}$ potentials according to \eqref{eq: s=3 SC} and \eqref{eq: s=3 s=1 SC},
so that the vertex  \eqref{eq: s=3 most general-1} is invariant under the sum of these transformations and the  hypermultiplet transformations \eqref{eq:superconformal transfromations}.

Substituting the superconformal  parameters  \eqref{eq:superconformal symmetry} into \eqref{eq: s=3 SC} and \eqref{eq: s=3 s=1 SC}
yields rigid $\mathcal{N}=2$ superconformal transformation laws of the spin $\mathbf{3}$ analytic potentials. The above reasoning indicates
that we need to introduce from the very beginning the most general set of analytic gauge potentials $h^{++MN}$ and $h^{++}$ in order  to realize $\mathcal{N}=2$ superconformal symmetry.
It is useful to combine the total set of gauge potentials  into the spin $\mathbf{3}$  second-order analytic operator as:
\begin{equation}
    \hat{\mathcal{H}}^{++}_{(s=3)} := h^{++MN}\partial_N\partial_M
    +
    h^{++}. \label{Spin3Hat}
\end{equation}

The precise realization of rigid ${\cal N}=2$ superconformal transformations on the analytic gauge potentials in \eqref{Spin3Hat} is given in Appendix C. It
is shown there that all $h^{++MN}$ with antisymmetric combinations of the Lorentz indices $\alpha, \dot\beta$
form a set closed under ${\cal N}=2$ superconformal group, while the remaining ``essential'' ones (with symmetric combinations of indices)
transform through this set and themselves. In other words, $h^{++MN}$
constitute not fully reducible representation. The auxiliary spin ${\bf 1}$ gauge potential $h^{++}$ properly transforms through $h^{++MN}$. The linearized gauge
transformations to be discussed in the next subsection are compatible with this not fully reducible superconformal structure: the conformally
invariant subset just mentioned is transformed by gauge parameters which do not appear in the gauge transformations of the ``essential'' potentials.
Just this notable group-theoretical property allows one to gauge away the irreducible subset of gauge potentials without breaking of superconformal symmetry
and to end up with the essential potentials as carriers of the irreducible ${\bf s}=3$ ${\cal N}=2$ gauge multiplet (in the proper Wess-Zumino gauges).

\subsection{Gauge freedom}

As the following step we analyze the gauge freedom of the action:
\begin{equation}
    S_{free} + S^{(s=3)}_{int} = -\frac{1}{2} \int d\zeta^{(-4)}
    q^{+a} \left(  \mathcal{D}^{++} + \kappa_3 h^{++MN} \partial_N \partial_MJ + \kappa_3 h^{++}J \right)  q^+_a.
\end{equation}

A generalization of the ${\bf s}=2$ gauge transformations \eqref{eq: s=2 gauge} is obtained by inserting one more derivative,
\begin{equation}
    \partial_M = \{\partial_{\alpha\dot{\alpha}}, \partial^-_{\hat{\alpha}}, \partial^{--} \}\,,
\end{equation}
in the hypermultiplet transformation law.
Then the most general ${\bf s}=3$ generalization of ${\bf s}=2$ gauge freedom \eqref{eq: s=2 gauge} is given
by \footnote{Anti-graded bracket is defined as $\{F_1, F_2\}_{AGB} := [F_1, F_2]$ for fermionic objects and $\{B_1, B_2\}_{AGB} := \{B_1, B_2\}$ for bosonic ones. Also, $\{F, B\}_{AGB} := [F,  B]$.}
\begin{equation}\label{eq: s=3 general transfromations-1}
    \begin{split}
    \delta_\lambda^{(s=3)} q^{+a}
    =&
    - \kappa_3 \, \hat{\mathcal{U}}_{(s=3)} J q^{+a}
    \\
    =& -\frac{\kappa_3}{2}\{\hat{\Lambda}^{M}, \partial_{M} \}_{AGB} J q^{+a}
    -
    \frac{\kappa_3}{4} \{ \Omega^{M}, \partial_{M}  \}_{AGB}  J q^{+a}.
    \end{split}
\end{equation}
Here we have introduced the first-order analytic operators\footnote{We define the ordering of the indices $M = \{\alpha\dot{\alpha}, \hat\alpha+, ++\}$ as:
$
\alpha\dot{\alpha} > \hat\alpha+  > ++
$.}
\begin{equation}\hat{\Lambda}^M :=  \sum_{N\leq M} \lambda^{MN}\partial_N,
\end{equation}
with the analytic parameters satisfying the condition $\lambda^{MN} = (-1)^{P(M)P(N)} \lambda^{NM}$,
as well as the analytic weight factor
 \begin{equation}\Omega^M :=  \sum_{N<M} (-1)^{ P(N )}\partial_N \lambda^{NM}.
 \end{equation}
The  transformation law \eqref{eq: s=3 general transfromations-1} is of the second order in the superspace derivatives.

 Gauge parameters satisfy reality conditions, which follows from the requirement of reality of variation \eqref{eq: s=3 general transfromations-1}. These conditions have the same form as those
 for the analytic potentials \eqref{eq: spin 3 reality}:
 \begin{subequations}\label{eq: spin 3 parameters reality}
    \begin{equation}
        \widetilde{\lambda^{\alpha\dot{\alpha}\beta\dot{\beta}}} = \lambda^{\alpha\dot{\alpha}\beta\dot{\beta}},
        \qquad
        \widetilde{\lambda^{(+4)}} = \lambda^{(+4)},
    \end{equation}
    \begin{equation}
        \widetilde{\lambda^{[\beta+\gamma]+}} =
        \lambda^{[\dot{\beta}+\dot{\gamma}]+},
        \qquad
        \widetilde{ \lambda^{[\dot{\beta}+\dot{\gamma}]+}}
        =
        \lambda^{[\beta+\gamma]+},
    \end{equation}
    \begin{equation}
        \widetilde{\lambda^{\beta+\alpha\dot{\alpha}}}
        =
        -
        \lambda^{\dot{\beta}+\alpha\dot{\alpha}},
        \qquad
        \widetilde{\lambda^{\dot{\beta}+\alpha\dot{\alpha}}}
        =
        \lambda^{\beta+\alpha\dot{\alpha}},
    \end{equation}
    \begin{equation}
        \widetilde{\lambda^{\alpha\dot{\alpha}++}} = \lambda^{\alpha\dot{\alpha}++},
        \qquad
        \widetilde{\lambda^{\alpha+\dot{\alpha}+}} = -
        \lambda^{\alpha+\dot{\alpha}+},
    \end{equation}
    \begin{equation}
        \widetilde{\lambda^{++\beta+}} = - \lambda^{++\dot{\beta}+},
        \qquad
        \widetilde{\lambda^{++\dot{\beta}+}}
        =
        \lambda^{++\beta+}.
    \end{equation}
 \end{subequations}

The variation of the free hypermultiplet Lagrangian under the general transformations  \eqref{eq: s=3 general transfromations-1}
with arbitrary analytic parameters $\lambda^{MN}(\zeta)$ has the form (up to a total derivative)\footnote{Using this result one can obtain rigid symmetries
(``higher-spin'' superconformal symmetries) of the free massless hypermultiplet and the corresponding current superfields. We hope to address this issue elsewhere.}:
\begin{equation}\label{eq: s=3 local variation}
    \begin{split}
    \delta_\lambda^{(s=3)} S_{free}
    =&\;
    \frac{\kappa_3}{4} \int d\zeta^{(-4)}\,
    q^{+a} \left[\mathcal{D}^{++}, \{\hat{\Lambda}^M, \partial_M\}_{AGB}\right] J q^+_a
    \\=&\;
    \frac{ \kappa_3}{2} \int d\zeta^{(-4)}\, q^{+a} [\mathcal{D}^{++}, \hat{\Lambda}^M] \partial_M J q^+_a
    \\&+
    \frac{\kappa_3}{4} \int d\zeta^{(-4)}\, q^{+a} \left\{ \hat{\Lambda}^M, [\mathcal{D}^{++}, \partial_M] \right\}_{AGB} J q^+_a.
    \end{split}
\end{equation}
The first line involves terms with two derivatives. The second line, modulo integration by parts, collects terms with two derivatives and those without derivatives.

Requiring gauge invariance
\begin{equation}
    \delta^{(s=3)}_\lambda S_{free} + \delta^{(s=3) }_\lambda S_{int}^{(s=3)} = 0
\end{equation}
to the leading order gives the linearized gauge transformation law for the analytic potentials. It can be formally represented as\footnote{One needs to integrate by parts
the terms with one derivative and to reduce them to terms without derivatives in order to be able to cancel them by a gauge transformation of the $h^{++}$ term in $\hat{\mathcal{H}}^{++}_{(s=3)}$.
In formula \eqref{eq: s=3 operator transformation} we assume that such manipulations have been done.}:
\begin{equation}\label{eq: s=3 operator transformation}
    \begin{split}
    \delta^{(s=3)}_\lambda \hat{\mathcal{H}}^{++}_{(s=3)}
    =&
    \;
    \left[\mathcal{D}^{++}, \hat{\mathcal{U}}_{(s=3)}\right]
    \\
    =&\;
    \frac{1}{2}
     \left[\mathcal{D}^{++}, \{\hat{\Lambda}^M, \partial_M\}_{AGB}\right]
    \\=&\;
    [\mathcal{D}^{++}, \hat{\Lambda}^M] \partial_M
    +
    \frac{1}{2}
    \left\{ \hat{\Lambda}^M, [\mathcal{D}^{++}, \partial_M] \right\}_{AGB}.
     \end{split}
\end{equation}
The action $S_{free} + S_{int}^{(s=3)}$ also respects an additional $U(1)$ gauge freedom:
\begin{equation}
    \delta_\lambda q^{+a} = - \kappa_3 \lambda J q^{+a},
    \qquad \delta_\lambda h^{++} = \mathcal{D}^{++} \lambda.
\end{equation}

\subsection{Wess-Zumino gauge:  $\mathcal{N}=2$ superconformal spin $\mathbf{3}$ multiplet }\label{sec: s=3 sc multiplet}

The linearized gauge transformations of independent analytic potentials can be deduced from \eqref{eq: s=3 operator transformation}:

\begin{subequations}\label{eq: s=3 gauge transformations}
\begin{equation}\label{eq: s=3 gauge transformations vector}
    \begin{cases}
        \delta_\lambda h^{++\alpha\dot{\alpha}\beta\dot{\beta}} &=      \mathcal{D}^{++}\lambda^{\alpha\beta\dot{\alpha}\dot{\beta}}
        + 2i \left( \lambda^{\alpha\dot{\alpha}\beta+} \bar{\theta}^{+\dot{\beta}}
        + \lambda^{\beta\dot{\beta}\alpha+} \bar{\theta}^{+\dot{\alpha}}\right)
       \\&\qquad\qquad\qquad
        - \,2i
        \left( \bar{\lambda}^{\alpha\dot{\alpha}\dot{\beta}+} \theta^{+\beta}
        + \bar{\lambda}^{\beta\dot{\beta}\dot{\alpha}+} \theta^{+\alpha}
        \right),
        \\
       2 \delta_\lambda h^{++\alpha\dot{\alpha}\beta+} &=  \mathcal{D}^{++}\lambda^{\alpha\dot{\alpha}\beta+}
        - 8i \lambda^{[\alpha+\beta]+} \bar{\theta}^{+\dot{\alpha}}
        -
        8i \lambda^{\beta+\dot{\alpha}+} \theta^{+\alpha}
        - \lambda^{\alpha\dot{\alpha}++} \theta^{+\beta},
        \\
        2\delta_\lambda h^{++\alpha\dot{\alpha}++} &= \mathcal{D}^{++} \lambda^{++\alpha\dot{\alpha}}
        + 4 i \lambda^{++\alpha+} \bar{\theta}^{+\dot{\alpha}}
        -
        4i \lambda^{++\dot{\alpha}+}\theta^{+\alpha},
    \end{cases}
\end{equation}

\begin{equation}\label{eq: spin 3 b}
    \begin{cases}
    \delta_\lambda h^{++[\alpha+\beta]+} &= \mathcal{D}^{++} \lambda^{[\alpha+\beta]+}
    -
    \lambda^{++[\alpha+} \theta^{+\beta]},
    \\
    2\delta_\lambda h^{++\alpha+\dot{\alpha}+} &=
    2\mathcal{D}^{++} \lambda^{\alpha+\dot{\alpha}+}
    -  \lambda^{++\alpha+} \bar{\theta}^{+\dot{\alpha}}
    +
    \lambda^{++\dot{\alpha}+} \theta^{+\alpha},
    \\
    2\delta_\lambda h^{++\hat{\alpha}+++} &= \mathcal{D}^{++} \lambda^{\hat{\alpha}+++}
    -2  \theta^{+\hat{\alpha}} \lambda^{(+4)},
    \\
    \delta_\lambda h^{(+6)} &=  \mathcal{D}^{++} \lambda^{(+4)} ,
    \end{cases}
\end{equation}

\begin{equation}\label{eq: spin 3 c}
    \begin{split}
\delta_\lambda h^{++}  = \mathcal{D}^{++} \lambda
&+ 2i \bar{\theta}^{+\dot{\rho}} \left( \partial_{\alpha\dot{\rho}} \partial^-_\beta  \lambda^{[\alpha+\beta]+} \right)
+
2i \theta^{+\rho} \left( \partial_{\rho\dot{\alpha}} \partial^-_{\dot{\beta}} \lambda^{[\dot{\alpha}+\dot{\beta}]+} \right)
\\
&
+ 2i \bar{\theta}^{+\dot{\rho}} \left( \partial_{\alpha\dot{\rho}} \partial^-_{\dot{\beta}} \lambda^{\alpha+\dot{\beta}+} \right)
-
2i \theta^{+\rho} \left( \partial_{\rho\dot{\alpha}} \partial^-_\beta \lambda^{\dot{\alpha}+\beta+} \right)
\\
&
+
2i \bar{\theta}^{+\dot{\rho}} \left( \partial_{\alpha\dot{\rho}} \partial^{--} \lambda^{\alpha+++}   \right)
-
2i \theta^{+\rho} \left( \partial_{\rho\dot{\beta}} \partial^{--} \lambda^{\dot{\beta}+++} \right)
\\
&
- 8i \partial_{\alpha\dot{\alpha}} \lambda^{\alpha+\dot{\alpha}+}
+
\frac{1}{2} \theta^{+\hat{\rho}} \left( \partial^-_{\hat{\rho}} \partial^{--} \lambda^{(+4)} \right)
-
2 (\partial^{--} \lambda^{(+4)}).
\end{split}
\end{equation}
\end{subequations}

Using these transformations, one can impose Wess-Zumino type gauge.
Potentials of the form $h^{++\alpha\dot{\alpha}M}$ span $\mathcal{N}=2$ spin $\mathbf{3}$ superconformal multiplet ($\mathbf{s}=3$ Weyl multiplet):
\begin{equation}\label{eq: spin 3 GF}
    \begin{cases}
        \begin{split}
            h^{++(\alpha\beta)(\dot{\alpha}\dot{\beta})}
            &=
            -4i \theta^{+}_{\rho}\bar{\theta}^{+}_{\dot{\rho}} \Phi^{(\alpha\beta\rho)(\dot{\alpha}\dot{\beta}\dot{\rho})}
            -
            (\bar{\theta}^+)^2 \theta^{+}_{\rho} \psi^{(\alpha\beta\rho)(\dot{\alpha}\dot{\beta})i} u_i^-
            \\&\;\;\;\;\;\;\;\;\;\;\;\;\;\;\;\;
            -
            (\theta^+)^2 \bar{\theta}^{+}_{\dot{\rho}} \bar{\psi}^{(\alpha\beta)(\dot{\alpha}\dot{\beta}\dot{\rho})i} u_i^-
            +
            (\theta^+)^2 (\bar{\theta}^+)^2 V^{(\alpha\beta)(\dot{\alpha}\dot{\beta})ij}u^-_i u^-_j\,,\\
            h^{++(\alpha\beta)\dot{\alpha}+}
            &=
            (\theta^{+})^2 \bar{\theta}^{+}_{\dot{\nu}} \mathcal{P}^{(\alpha\beta)(\dot{\alpha}\dot{\nu})}
            +
            (\bar{\theta}^{+})^2 \theta^{+}_{\nu} T^{(\alpha\beta\nu)\dot{\alpha}}_{}
            +
            (\theta^{+})^4 \chi^{(\alpha\beta)\dot{\alpha}i}u_i^-\,,
            \\
            h^{++\alpha(\dot{\alpha}\dot{\beta})+}&=
            \widetilde{h^{++(\alpha\beta)\dot{\alpha}+}}\,,
            \\h^{++\alpha\dot{\alpha}++} \;\,&=
            (\theta^+)^2 (\bar{\theta}^+)^2 D^{\alpha\dot{\alpha}}\,.
        \end{split}
    \end{cases}
\end{equation}
It is essential that the field $\mathcal{P}^{(\alpha\beta)(\dot{\alpha}\dot{\nu})}$ is {\it real},
$$\widetilde{\mathcal{P}^{(\alpha\beta)(\dot{\alpha}\dot{\nu})}} = \mathcal{P}^{(\alpha\nu)(\dot{\alpha}\dot{\beta})}.$$
The originally present imaginary part of such a field  proves to be  pure gauge.

All other potentials (including those parts of the original potentials which are antisymmetric in the spinorial indices)
can be fully gauged away\footnote{Similar pure gauge field parameters were also used in ref. \cite{Kuzenko:2022hdv} (see sect. 3.4 there).
These fields can also be gauged away.
After eliminating these redundant fields, gauge transformations cease to be linear in fields. The purpose of introducing extra fields in the work \cite{Kuzenko:2022hdv}
was the desire to close the algebra of gauge transformations.
In our case, their introduction is dictated by $\mathcal{N}=2$ superconformal invariance and, since gauge transformations are chosen to have a general form,
we expect that the algebra of gauge transformations will be automatically closed.}
using the gauge freedom \eqref{eq: spin 3 b} and \eqref{eq: spin 3 c} (see also discussion in appendix \ref{app: N2SCDeriv}). The technical details
of this procedure are collected in appendix \ref{app: WZ gauge}. In the physical sector we are left with the following fields\footnote{Some fields require redefinitions,
here we assume that such a procedure has been  performed. Explicitly, these redefinitions are given in Appendix \ref{app: WZ gauge}.
For simplicity and clarity of notation we also use the properly rescaled gauge parameters here. The precise relation between the
gauge parameters $a^{(\alpha\beta)(\dot{\alpha}\dot{\beta})}$, $v^{\alpha\dot{\alpha}(ij)}$, $p^{\beta\dot{\beta}}$, $t^{(\alpha\beta)}$, $c$
used below and the components of the analytic superfield parameters $\lambda^{MN}$ used in Appendix \ref{app: WZ gauge}
can be established by comparing with eqs. \eqref{Gauge3}, \eqref{Triplet},  \eqref{Pgfield},
\eqref{IIc4}, and \eqref{app D alpha}.}:\\
\vspace{0.5cm}

\textbf{\textit{\underline{Bosonic sector}}} :

 $\bullet$\textit{ Conformal spin 3 field }with gauge freedom (7 off-shell d.o.f.):
 \begin{equation}
    \delta \Phi^{(\alpha\beta\rho)(\dot{\alpha}\dot{\beta}\dot{\rho})}
    =
    \partial^{(\dot{\rho} (\rho} a^{\alpha\beta) \dot{\alpha}\dot{\beta})}.
 \end{equation}

$\bullet$ \textit{Triplet of conformal gravitons (spin 2 fields) }(15 off-shell d.o.f.):
\begin{equation}
   \delta V^{(\alpha\beta)(\dot{\alpha}\dot{\beta}) (ij)} = \partial^{(\dot{\alpha}(\alpha}v^{\beta)\dot{\beta})(ij)}.
\end{equation}

$\bullet$ \textit{Conformal graviton} (5 off-shell d.o.f.):
\begin{equation}
    \delta \mathcal{P}^{(\alpha\beta)(\dot{\alpha}\dot{\beta})} =  \partial^{(\dot{\alpha}(\alpha}
    p^{\beta)\dot{\beta})}.
\end{equation}

$\bullet$ \textit{Gauge field for self-dual two-form symmetry} (10 off-shell d.o.f.):
\begin{equation}
    \delta T^{(\alpha\beta\rho)\dot{\alpha}} = \partial^{\dot{\alpha}(\rho} t^{\alpha\beta)}.
\end{equation}
Fields $T^{(\alpha\beta\gamma)\dot{\alpha}}$ and complex conjugated $\bar{T}^{(\dot{\alpha}\dot{\beta}\dot{\gamma})\alpha}$  are in one-to-one correspondence with a real tensor field $T^{[ab]d}$:
\begin{equation}
    T^{[ab]c} = \sigma^{[ab]}_{(\alpha\beta)}\sigma^c_{\gamma\dot{\gamma}} T^{(\alpha\beta\gamma)\dot{\gamma}}
    +
    \tilde{\sigma}^{[ab]}_{(\dot{\alpha}\dot{\beta})} \sigma^c_{\dot{\gamma}\gamma}  \bar{T}^{(\dot{\alpha}\dot{\beta}\dot{\gamma})\gamma}.
\end{equation}
Due to the $\sigma$-matrices properties, the following identity holds:
\begin{equation}
    T^{[abc]} =0 \qquad \Leftrightarrow
    \qquad
    T^{[ab]c} + T^{[bc]a} + T^{[ca]b} =0.
\end{equation}
These symmetry properties correspond to the simple hook Young diagram $\tiny\yng(2,1)$.
Additionally, properties of $\sigma$-matrices imply the traceless condition $T^{[ab]}_{\;\;\;\;\;\;b}=0$. The gauge freedom amounts to:
\begin{equation}
    \delta T^{[ab]c} = 2\partial^{[a}t^{b]c}  - 2 \partial^c t^{[ab]},
    \qquad\qquad
    t^{[ab]}= \sigma^{[ab]}_{(\alpha\beta)} t^{(\alpha\beta)} + \sigma^{[ab]}_{(\dot{\alpha}\dot{\beta})} \bar{t}^{(\dot{\alpha}\dot{\beta})}.
\end{equation}
This field is called  ``hook field'' (or conformal pseudo-graviton field). Hook field was firstly studied in \cite{Curtright:1980yj, Curtright:1980yk} as
a generalized gauge field\footnote{It is not difficult to construct a conformal and gauge-invariant action for the hook field, see for example Appendix C of ref. \cite{Kuzenko:2020jie}.}.
The basic motivation for their consideration was the construction of dual formulations of gauge fields with spin $s\neq 1$. These fields can be viewed as a natural
generalization of the notoph field of Ogievetsky and Polubarinov \cite{Ogievetsky:1966eiu}\footnote{This kind of gauge theories  was later re-discovered by Kalb and Ramond \cite{CaRa}.}
(for review see \cite{Ivanov:2016lha}).

$\bullet$ \textit{Spin 1 gauge field} (3 off-shell d.o.f.):
\begin{equation}
    \delta D^{\alpha\dot{\alpha}} = \partial^{\alpha\dot{\alpha}}\,c.
\end{equation}

\textbf{\textit{\underline{Fermionic sector}}}\footnote{The relations between the gauge parameters $b^{(\alpha\beta)\dot{\beta}i}$, $c^{\beta i}$  and components of
 the superfield parameters $\lambda^{MN}$ can be found by comparing with eqs. \eqref{52psi} and \eqref{32chi}.}:

$\bullet$ \textit{Doublet of conformal spin $\frac{5}{2}$ fields }(24 off-shell d.o.f.):
\begin{equation}
    \delta \psi^{(\alpha\beta\rho)(\dot{\alpha}\dot{\beta})i} =  \partial^{(\dot{\alpha}(\rho} b^{\alpha\beta)\dot{\beta})i}.
\end{equation}

$\bullet$ \textit{Doublet of gauge spin $\frac{3}{2}$ fermions} $\chi^{(\alpha\beta)\dot{\alpha}i}$ (16 off-shell d.o.f.):
\begin{equation}
    \delta \chi^{(\alpha\beta)\dot{\alpha} i}
    =
    \partial^{\dot{\alpha}(\alpha} c^{\beta)i}.
\end{equation}

So $\mathcal{N}=2$ off-shell superconformal spin $\mathbf{3}$ multiplet contains $\mathbf{40}_B+ \mathbf{40}_F$ off-shell degrees of freedom.
Note that the spin 3 and spin 1 fields appear in the same $\mathcal{N}=2$ gauge supermultiplet.
This may simplify the implementation of the assumption of refs. \cite{Grigoriev:2016bzl, Beccaria:2017nco} about the gauge invariance
of a system of conformal spins 1 and 3 in an arbitrary curved background.

The residual gauge transformations and their action on these fields can be analyzed in full analogy with the spin $\mathbf{2}$ case considered in Section \ref{eq: s=2 WZ gauge}.
 We do not give the explicit formulas, because the detailed component considerations are beyond the scope of our study.

\medskip

All the  potentials except $h^{++\alpha\dot{\alpha}M}$ can be put equal to zero using the original large  gauge freedom.
One can choose such a gauge from the very beginning to bring the vertex to the simpler form:
\begin{equation}\label{eq: s=3 vertex in fixed gauge}
    S^{(s=3)}_{int|fixed}
    =
    -\frac{\kappa_3}{2} \int d\zeta^{(-4)}\; q^{+a} h^{++\alpha\dot{\alpha}M} \partial_M \partial_{\alpha\dot{\alpha}} J q^+_a,
\end{equation}
where, like in the non-conformal case, the spinorial indices of the same chirality in $h^{++\alpha\dot\alpha M}$ are assumed to be symmetrized. In such a form the vertex,
 up to terms involving harmonic derivative $\partial^{--}$, fully matches the  non-conformal $(\mathbf{3}, \mathbf{\tfrac{1}{2}},  \mathbf{\tfrac{1}{2}})$ vertex.
However in such a gauge one is led to accompany the local superconformal transformations \eqref{eq: s=3 SC} by the proper compensating gauge transformations in order to preserve the gauge:
\begin{equation}\label{eq: delta WZ}
    \delta_{dif|mod} h^{++MN}_{WZ} =   \delta_{diff}  h_{WZ}^{++MN}
    +
    \delta_{\lambda|WZ} h^{++MN}
     \sim h_{WZ}^{++MN}.
\end{equation}
This condition fixes the parameters $\lambda^{MN}_{WZ}$ to have the proper explicit  dependence on gauge potentials $h_{WZ}^{++MN}$, so that the variation \eqref{eq: delta WZ}
proves to be linear in the components of $h_{WZ}^{++MN}$. So the vertex \eqref{eq: s=3 vertex in fixed gauge}
is invariant under the modified local superconformal transformations
\begin{equation}
    \delta_{dif|mod} q^{+a} = \delta_{dif} q^{+a}  + \delta_{\lambda|WZ} q^{+a}.
\end{equation}
These transformations generically involve the spin $\mathbf{3}$ potentials and so are essentially nonlinear.

As an example, we quote the explicit form of such a transformation in the sector of conformal supersymmetry (parameter $\eta^i_\alpha$).
In WZ gage $h_{WZ}^{++\alpha\dot{\beta}++}=0$ that amounts to the condition:
\begin{equation}
    \delta_{dif|WZ} h^{++\alpha\dot{\beta}++}_{WZ}
    =
    \mathcal{D}^{++}\lambda_{WZ}^{+\alpha\dot{\beta}+}
    -
    \lambda_{WZ}^{+\alpha++}\bar{\theta}^{+\dot{\beta}}
    +
    \lambda_{WZ}^{+\dot{\beta}++}\theta^{+\alpha}
    +
    h_{WZ}^{++(\alpha\beta)\dot{\beta}+} \eta^i_\beta u^+_i
    =0.
\end{equation}
Using the explicit form of WZ gauge for  $h_{WZ}^{++(\alpha\beta)\dot{\beta}+}$, one has:
\begin{equation}
    \lambda_{WZ}^{+\alpha\dot{\beta}+} =
    -
   h^{++(\alpha\beta)\dot{\beta}+}_{WZ} \eta^i_{\beta} u^-_i + \dots.
\end{equation}
Here ellipses  stand for possible contributions from $\lambda^{+\hat{\alpha}++}$. The resulting modified hypermultiplet superconformal transformation
with the parameter $\eta^i_\alpha$ (conformal supersymmetry) is thus found to be:
\begin{equation}\label{eq: 640}
    \begin{split}
    \delta^\eta_{dif|mod} q^{+a}
    =&
    \delta^\eta_{dif} q^{+a}
    +
    \frac{\kappa_3}{2} \left\{ h^{++(\alpha\beta)\dot{\beta}+}_{WZ} \eta^i_{\beta} u^-_i \partial^{-}_{\dot{\beta}}, \partial^-_\beta\right\}_{AGB} J q^{+a}
    \\&+
        \frac{\kappa_3}{4} \left\{ \partial^{-}_{\dot{\beta}}h^{++(\alpha\beta)\dot{\beta}+}_{WZ} \eta^i_{\beta} u^-_i , \partial^-_\beta\right\}_{AGB} J q^{+a}
        +
        \dots
        \end{split}
\end{equation}

\medskip

To summarize, it was necessary to start with the most general form of the vertex \eqref{eq: s=3 most general-1}
in order to realize (local) $\mathcal{N}=2$ superconformal symmetry
linearly on the hypermultiplet. The elimination of the auxiliary analytic potentials leads to the minimal set of the gauge potentials on which rigid ${\cal N}=2$ superconformal
group generically acts by nonlinear transformations explicitly involving the spin $\mathbf{3}$ gauge potentials.

\subsection{${\bf s}=3$ superconformal current superfields}\label{s=3 superconformal current superfields}

Putting $\delta_\lambda^{(s=3)} S_{free} = 0$ in \eqref{eq: s=3 local variation}, we recover rigid symmetries of the free hypermultiplet action.
There exist two ways to derive the corresponding ${\bf s}=3$ Noether current superfields. One can either study the variation \eqref{eq: s=3 local variation} of the action,
or, equivalently, vary the cubic coupling \eqref{eq: s=3 most general} with respect to the analytic potentials $h^{++MN}$. The relevant current superfields are given by the expressions:
\begin{equation}\label{eq: s=3 analytic}
    J^{++}_{MN} = - \frac{1}{2} q^{+a}
    \partial_N \partial_M J q^+_a,
    \qquad
    \mathcal{D}^{++} J^{++}_{MN} = - \frac{1}{2} q^{+a}
    [\mathcal{D}^{++} ,\partial_N \partial_M] J q^+_a.
\end{equation}
When deducing the current conservation condition in \eqref{eq: s=3 analytic}, we made use of the free hypermultiplet
equations of motion \eqref{eq: hyper eom}. The current superfields obtained in this way are sources of the equations of motion for  the spin $\mathbf{3}$ gauge potentials.
In this article we do not discuss the issue of constructing an $\mathcal{N}=2$ spin $\mathbf{3}$ superconformal  action and the corresponding equations of motion.

In the more detailed notation, we are left with  nine independent current superfields:
\begin{equation}
    J^{++}_{\alpha\beta\dot{\alpha}\dot{\beta}},
    \quad
    J^{+}_{\alpha\beta\dot{\alpha}},
    \quad
    J^{+}_{\dot{\alpha}\dot{\beta}\alpha},
    \quad
    J^{}_{\alpha\beta},
    \quad
    J^{}_{\dot{\alpha}\dot{\beta}},
    \quad
    J^{}_{\alpha\dot{\alpha}},
    \quad
    J^-_{\alpha},
    \quad
    J^-_{\dot{\alpha}},
    \quad
    J^{--}. \label{MoreDet}
\end{equation}
The current superfields \eqref{eq: s=3 analytic} (or \eqref{MoreDet})  are analytic but they are not invariant under $\mathcal{N}=2$ supersymmetry.
Like in the ${\bf s}=2$ case, one can introduce non-analytic current superfields which are invariants of $\mathcal{N}=2$ supersymmetry.
In contrast to the ${\bf s}=2$ case, here we deal with few different ``master'' currents.

The simplest option corresponds to the choice $M=\alpha\dot{\alpha}$. The non-analytic current superfield has the following form:
\begin{equation}\label{eq: spin 3 CSF}
    \mathcal{J}_{\alpha\dot{\alpha}} = -\frac{1}{2} q^{+a} \mathcal{D}^{--} \partial_{\alpha\dot{\alpha}} J q^+_a,
    \qquad
    \mathcal{D}^{++} \mathcal{J}_{\alpha\dot{\alpha}}
    =
    -\frac{1}{2} q^{+a}  \partial_{\alpha\dot{\alpha}} J q^+_a.
\end{equation}
This expression satisfies various conservation laws, for example:
\begin{equation}
    \mathcal{D}^{++} \left( D^+_{\hat{\beta}} \mathcal{J}_{\alpha\dot{\alpha}}  \right)
    =
    0,
    \qquad
    \mathcal{D}^{++} \left( D^+_{\beta} D^+_{\dot{\beta}} \mathcal{J}_{\alpha\dot{\alpha}}  \right)
    =
    0.
\end{equation}
For the choice $M = \hat{\alpha}$ we analogously obtain:
\begin{equation}
    \mathcal{J}^-_{\hat{\alpha}} = - \frac{1}{2} q^{+a} \mathcal{D}^{--} \mathcal{D}^-_{\hat{\alpha}} J q^+_a,
    \qquad
    \mathcal{D}^{++}    \mathcal{J}^-_{\hat{\alpha}} = 0.
\end{equation}
At last, choosing $M = ++$ yields:
\begin{equation}
    \mathcal{J}^{--} = - \frac{1}{2} q^{+a} \mathcal{D}^{--} \mathcal{D}^{--} J q^+_a,
    \qquad
    \mathcal{D}^{++}    \mathcal{J}^{--} = 0.
\end{equation}

The set of the ${\bf s}=3$ superconformal current superfields  $\{\mathcal{J}_{\alpha\dot{\alpha}}, \mathcal{J}^-_{\hat{\alpha}},     \mathcal{J}^{--} \}$
incorporate all the analytic supercurrents \eqref{eq: s=3 analytic}, \eqref{MoreDet}  in their $\theta^-$ expansions. For example,
\begin{equation}
    \mathcal{J}_{\alpha\dot{\alpha}}
    =
    -4i \theta^{-\beta}\bar{\theta}^{-\dot{\beta}}
    J^{++}_{\alpha\beta\dot{\alpha}\dot{\beta}}
    +
    \theta^{-\beta} J^{+}_{\alpha\beta\dot{\alpha}}
    +
    J_{\alpha\dot{\alpha}}.
\end{equation}

An alternative way to derive these superconformal currents is through varying cubic couplings with respect to the unconstrained non-analytic
prepotentials $\{\Upsilon^{\alpha\dot{\alpha}}, \Upsilon^{+\hat{\alpha}}$, $\Upsilon^{++}, \Upsilon^{--}\}$ defined as:
\begin{equation}\label{eq: s=3 prepotental}
    \hat{\mathcal{H}}^{++}_{(s=3)} := (D^+)^4 \left( \Upsilon^{\alpha \dot{\alpha}} \partial_{\alpha\dot{\alpha}} \mathcal{D}^{--}
    + \Upsilon^{+ \hat{\alpha}} \mathcal{D}^-_{\hat{\alpha}} \mathcal{D}^{--} +
    \Upsilon^{ ++} \mathcal{D}^{--} \mathcal{D}^{--}  + \Upsilon^{--} \right).
\end{equation}
From this definition one can deduce the transformation laws of non-analytic prepotentials. In the next section we will show that it is possible
to select a gauge $\Upsilon^{+\hat{\alpha}}=0$, $\Upsilon^{++}=0$ and $\Upsilon^{--}=0$. In this gauge we can describe the spin $\mathbf{3}$ supermultiplet
in terms of unconstrained non-analytic prepotential $\Upsilon^{\alpha\dot{\alpha}}$. Such a prepotential (in the gauge where it does not depend on harmonics)
can presumably be identified with the one introduced in ref. \cite{Kuzenko:2021pqm}.
Thus the relation \eqref{eq: s=3 prepotental}  gives a hint of how the prepotentials of ref. \cite{Kuzenko:2021pqm} could appear within
the harmonic superspace approach. It should be pointed out that all these prepotentials and their gauge freedom are of non-geometric character, like the original
Mezincescu potential for ${\cal N}=2$ Maxwell theory. In contrast, the analytic gauge potentials have the clear geometric meaning as the objects covariantizing
the analyticity-preserving harmonic derivative ${\cal D}^{++}$.

\subsection{Superconformal $(\mathbf{3}, \mathbf{\frac{1}{2}},\mathbf{\frac{1}{2}})$ vertex in conformal supergravity background}

In this subsection we generalize the previous results to $\mathcal{N}=2$ conformal supergravity background.

Let us start with the action containing the spin $\mathbf{2}$ and the spin $\mathbf{3}$ couplings to hypermultiplet,
\begin{equation}\label{eq: s=2 + s=3}
    S = - \frac{1}{2} \int d\zeta^{(-4)} \, q^{+a} \left( \mathcal{D}^{++} + \kappa_2 \hat{\mathcal{H}}^{++}_{(s=2)}
    + \kappa_3 \hat{\mathcal{H}}^{++}_{(s=3)}J \right) q^+_a.
\end{equation}
Here the operators $\hat{\mathcal{H}}^{++}_{(s=2)}$ and $\hat{\mathcal{H}}^{++}_{(s=3)}$ were defined in \eqref{eq: spin 2 operator} and \eqref{Spin3Hat}.
This action is exactly invariant under the nonlinear spin $\mathbf{2}$ gauge
transformations \eqref{eq:superconformal transfromations}, \eqref{eq: spin 2 nonlinear}, \eqref{eq: s=3 SC}
(and so is also invariant under rigid $\mathcal{N}=2$ superconformal transformations), as well as under the linearized spin $\mathbf{3}$
gauge transformations \eqref{eq: s=3 general transfromations-1}, \eqref{eq: s=3 operator transformation}  ({\it i.e.}, to the leading order in $\kappa_2$, $\kappa_3$).

Under the spin $\mathbf{3}$ gauge transformations \eqref{eq: s=3 general transfromations-1} of the hypermultiplet
the vertex $(\mathbf{2}, \mathbf{\tfrac{1}{2}}, \mathbf{\tfrac{1}{2}})$ transforms as:
\begin{equation}
    \begin{split}
    \delta^{(s=3)}_\lambda \left( q^{+a} \hat{\mathcal{H}}^{++}_{(s=2)} q^+_a \right)
    = &
    - \kappa_3\, q^{+a} \left[  \hat{\mathcal{H}}^{++}_{(s=2)}, \hat{\mathcal{U}}_{(s=3)}  \right] J q^+_a
    \\
    =
    &-
    \frac{\kappa_3}{2} q^{+a} \left[  \hat{\mathcal{H}}^{++}_{(s=2)}, \{\hat{\Lambda}^{M}, \partial_{M} \}_{AGB}  \right] J q^+_a
\\  &   -
    \frac{\kappa_3}{4} q^{+a} \left[  \hat{\mathcal{H}}^{++}_{(s=2)}, \{\Omega^{M}, \partial_{M} \}_{AGB}  \right] J q^+_a.
    \end{split}
\end{equation}
One can cancel these terms (using integrations by parts) by introducing the additional spin $\mathbf{2}$-dependent
terms in the gauge transformations \eqref{eq: s=3 operator transformation} of the spin $\mathbf{3}$ multiplet:
\begin{equation}
    \begin{split}
    \delta^{ad}_\lambda \hat{\mathcal{H}}^{++}_{(s=3)}
        =
    &
    \frac{\kappa_2}{2}  \left[  \hat{\mathcal{H}}^{++}_{(s=2)}, \{\hat{\Lambda}^{M}, \partial_{M} \}_{AGB}  \right]
    +
    \frac{\kappa_2}{4}  \left[  \hat{\mathcal{H}}^{++}_{(s=2)}, \{\Omega^{M}, \partial_{M} \}_{AGB}  \right].
    \end{split}
\end{equation}
These terms deform the transformations law \eqref{eq: s=3 operator transformation} by the general $\mathcal{N}=2$ conformal supergravity background:
\begin{equation}
    \begin{split}
    \delta_{\lambda|full}^{(s=3)} \hat{\mathcal{H}}^{++}_{(s=3)}
    =&
    \left[\mathfrak{D}^{++}, \hat{\mathcal{U}}_{(s=3)} \right]
    \\
    =&
   \; \frac{1}{2} \left[\mathfrak{D}^{++}, \{ \hat{\Lambda}^M, \partial_M\}_{AGB}\right]
     +
    \frac{\kappa_2}{4}  \left[  \hat{\mathcal{H}}^{++}_{(s=2)}, \{\Omega^{M}, \partial_{M} \}_{AGB}  \right].
    \end{split}
\end{equation}
The last term acts only on the $h^{++}$ part of $\hat{\mathcal{H}}^{++}_{(s=3)}$.

\medskip

As a result, we have found that the action \eqref{eq: s=2 + s=3} is invariant with respect to the spin $\mathbf{3}$ transformations to the leading order in  $\kappa_3$
and to any order in $\kappa_2$. This means that we have constructed a cubic vertex $(\mathbf{3}, \mathbf{\tfrac{1}{2}}, \mathbf{\tfrac{1}{2}})$ which is invariant under
the gauge transformations of conformal $\mathcal{N}=2$ supergravity.  In the component approach this amounts to the property  that, after elimination of the auxiliary fields,
one will  recover the superconformal action of the spin $\mathbf{3}$ supermultiplet coupling $(\mathbf{3}, \mathbf{\frac{1}{2}}, \mathbf{\frac{1}{2}})$
in {\it generic}  ${\cal N}=2$  Weyl  supergravity background.
Note that the spin $\mathbf{3}$ multiplet fields in the action \eqref{eq: s=2 + s=3} do not  directly interact with the supergravity fields; the interaction is mediated
by the auxiliary fields of hypermultiplet.

\section{Generalization to arbitrary spin $\mathbf{s}$}
\label{sec: spin s}

In this section we generalize the results for the superconformal spin $\mathbf{3}$ hypermultiplet coupling  to the general spin $\mathbf{s}$ case.
We follow the general strategy of section \ref{sec: The general strategy of construction of superconformal couplings and multiplets}.

The relevant cubic superconformal $(\mathbf{s}, \mathbf{\tfrac{1}{2}}, \mathbf{\tfrac{1}{2}})$ vertex has the  form:

\begin{equation}\label{eq: spin s vertex}
    S_{int}^{(s)}
    =
    -
    \frac{\kappa_s}{2}
    \int d\zeta^{(-4)}\, q^{+a} \hat{\mathcal{H}}^{++}_{(s)} \left( J \right)^{P(s)} q^+_a,
    \qquad
    P(s) :=
    \begin{sqcases}
        0\;\;(even\; s),\\
        1\;\; (odd\; s).
    \end{sqcases}
\end{equation}
Here $\hat{\mathcal{H}}^{++}_{(s)}$ is analytic differential operator including general terms with $s-1$, $s-3$, $\dots$ $1/0$ (for even $s$/odd $s$) derivatives:
\begin{equation}\label{eq: operator H s}
    \hat{\mathcal{H}}^{++}_{(s)}
    :=
    h^{++M_1\dots M_{s-1}} \partial_{M_{s-1}}\dots \partial_{M_1}
    +
        h^{++M_1\dots M_{s-3}} \partial_{M_{s-3}}\dots  \partial_{M_1}
        +
        \dots
        +
            \begin{sqcases}
            h^{++M}\partial_M \;\;(even\; s)\\
            h^{++} \quad\quad\;\;\; (odd\; s).
        \end{sqcases}
\end{equation}
Like in the ${\bf s}=2$ and ${\bf s}=3$ cases, the necessity to include the derivatives of general type  follows from the requirement of $\mathcal{N}=2$ superconformal invariance.
The analytic superfields $h^{++\dots}(\zeta)$  for any pair of adjacent  indices satisfy the symmetry conditions:
\begin{equation}\label{eq: spin s permutation}
    h^{++M_1\dots M_n  M_k \dots M_{s-1}}
    =
    (-1)^{P(M_k)P(M_n)}
    h^{++M_1\dots M_k M_n \dots M_{s-1}}.
\end{equation}
From here one can deduce how to permute any 2 indices. Also, the operator $\hat{\mathcal{H}}^{++}_{(s)}$ satisfies the reality condition:
\begin{equation}
    \widetilde{\hat{\mathcal{H}}^{++}_{(s)}} = \hat{\mathcal{H}}^{++}_{(s)}.
\end{equation}

\subsection{$\mathcal{N}=2$ superconformal symmetry}
The analytic superdiffeomorphism transformation \eqref{eq: superconformal hyper 0} of the hypermultiplet  generates the following transformation of the vertex:
\begin{equation}\label{eq: general s variation}
    \begin{split}
        \delta_{diff} S_{int}^{(s)}
    =&
    \frac{\kappa_s}{2}
    \int d\zeta^{(-4)}\, q^{+a} [\hat{\mathcal{H}}^{++}_{(s)}, \hat{\Lambda}] \left( J \right)^{P(s)} q^+_a
    +
    \frac{\kappa_s}{4}
    \int d\zeta^{(-4)}\, q^{+a} [\hat{\mathcal{H}}^{++}_{(s)}, \Omega] \left( J \right)^{P(s)} q^+_a
            \\&-
    \frac{\kappa_s}{2}
    \int d\zeta^{(-4)}\, q^{+a} \delta_{diff} \hat{\mathcal{H}}^{++}_{(s)} \left( J \right)^{P(s)} q^+_a.
    \end{split}
\end{equation}
Calculating the commutators in the first line, we get terms with various numbers of derivatives acting on the hypermultiplet, analogously to the spin $\mathbf{3}$
case (recall eq. \eqref{tran6.5}).

$\bullet$ For even spin $\mathbf{s}$ one can always reduce the terms with even number of derivatives to those with odd number.
In this case, they are entirely compensated by the corresponding transformations of the gauge potentials in \eqref{eq: operator H s}.

\medskip

For example, the contribution of the  two-derivative term in the spin $\mathbf{4}$ case is:
\begin{equation}
    q^{+a} h^{++MNK} \left(  \partial_K \partial_N \lambda^R  \right) \partial_R \partial_M q^+_a.
\end{equation}
The expression $T^{++MR} := h^{++MNK} \left(  \partial_K \partial_N \lambda^R  \right) $ has the proper symmetry under permutation of the indices $R$ and $M$:
\begin{equation}\label{eq: T symm}
    T^{++MR}
    =
    (-1)^{P(R)P(M)}
    T^{++RM}.
\end{equation}
because it is a coefficient of $\partial_R \partial_M$.
After integration by parts and omitting total derivatives we obtain:
\begin{equation}
    T^{++MR} q^{+a} \partial_R \partial_M q^+_a
    \Rightarrow
    -(-1)^{(P(M)+P(R))P(R)}
    \left(\partial_R T^{++MR}\right) q^{+a}  \partial_M q^+_a
    -
    T^{++MR} \partial_R q^{+a}  \partial_M q^+_a.
\end{equation}
Due to the symmetry \eqref{eq: T symm} the second term  vanishes. So we have reduced the term with two derivatives acting on $q^+_a$ to a
term with one derivative.

\medskip

In the general case, one should integrate by parts and bring all the terms either to an odd number of derivatives acting on $q^+_a$
(and those without derivatives), which can be canceled by the proper transformation of gauge potentials, or
to a term with equal number of derivatives acting on $q^{+a}$ and $q^+_a$, and then use the identities of the type:
\begin{equation}
    h^{++\dots N_1\dots N_n M_1 \dots M_n\dots} \partial_{N_1} \dots \partial_{N_n} q^{+a} \partial_{M_1} \dots \partial_{M_n} q^+_a = 0,
\end{equation}
which are a direct generalization of the identity $q^{+a}q^+_a=0$.

\medskip

$\bullet$ For odd $\mathbf{s}$ one can also transform the terms with an odd number of derivatives to those with the even number,
integrating by parts and making use of the relation:
\begin{multline}
    h^{++N_1\dots N_n M_1 \dots M_n K}  \partial_{N_1} \dots \partial_{N_n} q^{+a} \partial_K \partial_{M_1} \dots \partial_{M_n}  J q^+_a
    \\=
    \frac{1}{2}     h^{++N_1 \dots N_n M_1 \dots M_n K}  \partial_K \left( \partial_{N_1} \dots \partial_{N_n} q^{+a}  \partial_{M_1} \dots \partial_{M_n}  J q^+_a \right).
\end{multline}

As a result,  for any $\mathbf{s}$ we are able to cancel terms coming from $[\hat{\mathcal{H}}^{++}_{(s)}, \hat{\Lambda}]$ and $[\hat{\mathcal{H}}^{++}_{(s)}, \Omega]$
by the proper transformations of the set of gauge potentials \eqref{eq: operator H s} and thereby to ensure the
diffeomorphism (and so superconformal) invariance of the cubic interaction \eqref{eq: spin s vertex}. Once again,
since we have not used the explicit form of $\mathcal{N}=2$ superconformal parameters anywhere,
this vertex is covariant under the complete gauge group of $\mathcal{N}=2$ conformal supergravity.

Based upon this reasoning, from \eqref{eq: general s variation}
 we can figure out the transformation law of the analytic spin ${\bf s}$  operator $\hat{\mathcal{H}}^{++}_{(s)}$, which can be symbolically written as:
\begin{equation}\label{eq: sc spin s}
    \delta_{diff} \hat{\mathcal{H}}^{++}_{(s)}
    =
    [\hat{\mathcal{H}}^{++}_{(s)}, \hat{\Lambda}]
    +
    \frac{1}{2}
    [\hat{\mathcal{H}}^{++}_{(s)}, \Omega].
\end{equation}
Here we assumed that the various terms in the right hand side  must be re-organized as was explained  above. The auxiliary gauge potentials of the lower spins
play the same role as in the spin $\mathbf{3}$ case: they cancel terms with a lesser number of derivatives, which result from the
commutators $[\hat{\mathcal{H}}^{++}_{(s)}, \hat{\Lambda}]$ and $[\hat{\mathcal{H}}^{++}_{(s)}, \Omega]$.

\subsection{Gauge freedom}

The action
\begin{equation}\label{eq: s 12 12}
    S_{free} + S_{int}^{(s)}
    =
    -
    \frac{1}{2}
    \int d\zeta^{(-4)} \, q^{+a} \left( \mathcal{D}^{++} + \kappa_s \hat{\mathcal{H}}^{++}_{(s)} \left( J \right)^{P(s)}   \right) q^+_a
\end{equation}
is invariant to the leading order in $\kappa_s$ under the hypermultiplet gauge transformations of the form ($k={s, s-2, s-4 \dots}$):
\begin{equation}\label{eq: spin s gauge transformations - k}
    \begin{split}
    \delta_\lambda^{(k)} q^{+a}
    =&
    - \kappa_s \, \hat{\mathcal{U}}^{(k)}_{(s)}  \left( J \right)^{P(s)} q^{+a}
    \\
    =&
    - \frac{\kappa_s}{2} \left\{\hat{\Lambda}^{M_1 \dots M_{k-2}}, \partial_{M_{k-2}} \dots  \partial_{M_1} \right\}_{AGB} \left( J \right)^{P(s)} q^{+a}
    \\
    &
    - \frac{\kappa_s}{4} \left\{\Omega^{M_1 \dots M_{k-2}}, \partial_{M_{k-2}} \dots  \partial_{M_1} \right\}_{AGB} \left( J \right)^{P(s)} q^{+a}
    \end{split}
\end{equation}
accompanied by the gauge transformations of the gauge potentials:
\begin{equation}\label{eq: spin s gauge transformations}
    \begin{split}
    \delta^{(k)}_\lambda \hat{\mathcal{H}}^{++}_{(s)}
    =&
    \left [\mathcal{D}^{++}, \hat{\mathcal{U}}^{(k)}_{s}\right]
    \\
    =&
    \left[\mathcal{D}^{++}, \hat{\Lambda}^{M_1 \dots M_{k-2}}\right]  \partial_{M_{k-2}} \dots  \partial_{M_1}
    \\&+
    \frac{1}{2}
    \left\{  \hat{\Lambda}^{M_1 \dots M_{k-2}}, [\mathcal{D}^{++}, \partial_{M_{k-2}} \dots  \partial_{M_1} ] \right\}_{AGB}.
    \end{split}
\end{equation}
These formulas are a direct generalization of the spin $\mathbf{3}$ transformations \eqref{eq: s=3 operator transformation}. The spin $\mathbf{s}$ gauge transformation of the hypermultiplet
contains  $s-1$ superspace derivatives.

The formula \eqref{eq: spin s gauge transformations} is symbolic like \eqref{eq: sc spin s}, and it makes sense only when its r.h.s. acts on the hypermultiplet, i.e.
when it is sandwiched between two hypermultiplet superfields.
One needs to reorganize the terms in the last line as was explained above for the diffeomorphisms invariance, using the fact
that they act on the hypermultiplet and then integrating by parts. Analogously to the spin $\mathbf{3}$ case \eqref{eq: s=3 gauge transformations},
it is straightforward  to determine the gauge transformations of the analytic spin $\mathbf{s}$  prepotentials from  \eqref{eq: spin s gauge transformations},
but the resulting formulas will be rather cumbersome. The transformation $\delta^{(k)}_\lambda$ corresponds to the gauge freedom for the spin $k$ part
included in the operator $\hat{\mathcal{H}}^{++}_{s}$ \eqref{eq: operator H s}.

Here we used the following notations for the first-order analytic operator:
\begin{equation}
    \hat{\Lambda}^{M_1\dots M_k}:=
    \sum_{N\leq M_k \dots \leq M_1}
    \lambda^{M_1\dots M_k N} \partial_N
\end{equation}
and for the analytic weight factor:
\begin{equation}
    \Omega^{M_1\dots M_k}:=
        \sum_{N\leq M_k \dots \leq M_1}
    (-1)^{P(N)}\partial_N \lambda^{N M_1\dots M_k }.
\end{equation}
Analytic parameters satisfy the conditions
\begin{equation}
    \lambda^{\dots MN\dots }
    =
    (-1)^{P(M)P(N)}
    \lambda^{\dots NM\dots }
\end{equation}
for any pair of adjacent indices. These conditions have the same form as those for analytic gauge potentials in \eqref{eq: spin s permutation}.
The reality of the variation \eqref{eq: spin s gauge transformations - k} implies the appropriate reality conditions for the transformation parameters.

These transformations constitute the gauge freedom of the spin $\mathbf{s}$, spin $\mathbf{s}-2$, $\dots$ parts of the differential
operator $\hat{\mathcal{H}}^{++}_{(s)}$ (i.e. those entering with $\mathbf{s}-1$, $\mathbf{s}-3$, $\dots$ derivatives).

It is worth noting an important property that the operator $\hat{\mathcal{U}}^{(k)}_{s} (J)^{P(s)}$ satisfies the remarkable
relation\footnote{We are grateful to the referee, who suggested to explicitly mention this property.
It can also be interpreted as $\left(\hat{\mathcal{U}}^{(k)}_{s} (J)^{P(s)} \right)^T = - \hat{\mathcal{U}}^{(k)}_{s} (J)^{P(s)}$, where $\hat{\mathcal{O}}^T$ is defined as $  \int d \zeta^{(-4)} \psi^{(q)a} \, \hat{\mathcal{O}}  \, \phi^{(4-q)}_a
    =
    \int d \zeta^{(-4)} \, \hat{\mathcal{O}}^T \, \psi^{(q)a}  \phi^{(4-q)}_a$ . }:
\begin{equation}\label{eq: transp}
    \int d \zeta^{(-4)} \psi^{(q)a} \, \hat{\mathcal{U}}^{(k)}_{s} (J)^{P(s)} \, \phi^{(4-q)}_a
    =
    -\int d \zeta^{(-4)} \, \hat{\mathcal{U}}^{(k)}_{s} (J)^{P(s)} \, \psi^{(q)a}  \phi^{(4-q)}_a.
\end{equation}
Here $\psi^{(q)a} = (\psi^{(q)}, \tilde{\psi}^{(q)})$ and $\phi^{(4-q)a} = (\phi^{(4-q)}, \tilde{\phi}^{(4-q)})$ are arbitrary analytic superfields.

\subsection{Wess-Zumino gauge: $\mathcal{N}=2$ superconformal spin $\mathbf{s}$ multiplet}

 The gauge freedom \eqref{eq: spin s gauge transformations} enables one to eliminate a large number of fields.
The Wess-Zumino gauge can be imposed quite analogously to the spin $\mathbf{3}$ case (as described in detail in Appendix \ref{app: WZ gauge}). The field contents
of this gauge completely repeats the form of the corresponding Wess-Zumino gauge in the case of spin $\mathbf{3}$:

\begin{equation}
    \begin{cases}
        \begin{split}
            h^{++\alpha(s-1)\dot{\alpha}(s-1)}
            &=
            -4i \theta^{+}_{\rho}\bar{\theta}^{+}_{\dot{\rho}} \Phi^{(\rho\alpha(s-1))(\dot{\rho}\dot{\alpha}(s-1))}
            -
            (\bar{\theta}^+)^2 \theta^{+}_{\rho} \psi_{}^{(\rho\alpha(s-1))\dot{\alpha}(s-1)i} u_i^-
            \\&\;\;\;\;-
            (\theta^+)^2 \bar{\theta}^{+}_{\dot{\rho}} \bar{\psi}^{\alpha(s-1)(\dot{\alpha}(s-1)\dot{\rho})i} u_i^-
            +
            (\theta^+)^2 (\bar{\theta}^+)^2 V^{\alpha(s-1)\dot{\alpha}(s-1)ij}u^-_i u^-_j\,,\\
            h^{++\alpha(s-1)\dot{\alpha}(s-2)+}
            &=
            (\theta^{+})^2 \bar{\theta}^{+}_{\dot{\nu}} P^{\alpha(s-1)(\dot{\alpha}(s-2)\dot{\nu})}
            +
            (\bar{\theta}^{+})^2 \theta^{+}_{\nu} T^{(\alpha(s-1)\nu)\dot{\alpha}(s-2)}_{}
            \\&\qquad\qquad\qquad\qquad\qquad\quad\;\;\,+
            (\theta^{+})^4 \chi^{\alpha(s-1)\dot{\alpha}(s-2)i}u_i^-\,,
            \\
            h^{++\alpha(s-2)\dot{\alpha}(s-1)+}&=
            \widetilde{h^{++\alpha(s-1)\dot{\alpha}(s-2)+}}\,,
            \\h^{(+4)\alpha(s-2)\dot{\alpha}(s-2)} &=
            (\theta^+)^2 (\bar{\theta}^+)^2 D^{\alpha(s-2)\dot{\alpha}(s-2)}\,.
        \end{split}
    \end{cases}
\end{equation}
All the remaining analytic potentials in \eqref{eq: operator H s} are pure gauge and can be entirely gauged away. Also, as in the spin $\mathbf{3}$ case, one can consider a special
gauge in which only the potentials $h^{++\alpha(s-2)\dot{\alpha}(s-2)M}$ survive \footnote{In such a gauge one can rewrite the analytic differential operator as
$$
\mathcal{H}^{++}_{s} = h^{++\alpha(s-2)\dot{\alpha}(s-2)M} \partial_M \partial^{(s-2)}_{\alpha(s-2)\dot{\alpha}(s-2)}
=
(D^+)^4 \left(\Upsilon^{\alpha(s-2)\dot{\alpha}(s-2)}\mathcal{D}^{--} \right) \partial^{(s-2)}_{\alpha(s-2)\dot{\alpha}(s-2)}.
$$
This gives a direct connection with the Mezincescu-type prepotentials studied in \cite{Kuzenko:2021pqm}.}. Generically, in such a gauge the superconformal transformations
must be accompanied  by gauge transformations with the composite parameters  involving gauge potentials (recall, e.g., the spin $\mathbf{3}$ example \eqref{eq: 640}).

So we have obtained $\mathcal{N}=2$ spin $\mathbf{s}$ superconformal off-shell gauge multiplet as the set of surviving fields in W-Z gauge.
It consists of the fields with gauge transformations:

\textbf{\textit{\underline{Bosonic sector}}}:

$\bullet$ \textit{Conformal spin $s$ gauge field }($2s+1$ off-shell d.o.f.):
\begin{equation}
    \delta \Phi^{\alpha(s)\dot{\alpha}(s)} = \partial^{(\alpha(\dot{\alpha}} a^{\alpha(s-1))\dot{\alpha}(s-1))}.
\end{equation}
Such fields are also known as Fradkin-Tseytlin fields \cite{Fradkin:1985am}.

$\bullet$ \textit{Triplet of the spin $s-1$  conformal gauge fields} [$3(2s-1)$ off-shell d.o.f.]:
\begin{equation}
    \delta V^{\alpha(s-1)\dot{\alpha}(s-1)ij}
    =
     \partial^{(\alpha(\dot{\alpha}} v^{\alpha(s-2))\dot{\alpha}(s-2))ij}.
\end{equation}

$\bullet$ \textit{Conformal spin $s-1$ gauge field} [$2s-1$ off-shell d.o.f.]:
\begin{equation}
    \delta P^{\alpha(s-1)\dot{\alpha}(s-1)} = \partial^{(\alpha(\dot{\alpha}} p^{\alpha(s-2))\dot{\alpha}(s-2))}.
\end{equation}

$\bullet$  \textit{Generalized conformal ``hook-type'' gauge field} [$2(2s-1)$ off-shell d.o.f.]:
\begin{equation}
    \delta T^{\alpha(s)\dot{\alpha}(s-2)}
    =
    \partial^{(\alpha(\dot{\alpha}} t^{\alpha(s-1)) \dot{\alpha}(s-3))}.
\end{equation}
Such a complex gauge field was already considered in the context of $\mathcal{N}=1$ superconformal multiplets in \cite{Kuzenko:2017ujh} (see eq. (3.16) there)
and in \cite{Kuzenko:2020opc}. Gauge invariant field strengths and conformal actions for such fields were also presented in  \cite{Kuzenko:2017ujh}.
In  refs. \cite{Kuzenko:2019ill, Kuzenko:2019eni, Kuzenko:2020jie} the gauge invariant actions for such fields were constructed in conformally flat spaces.

$\bullet$ \textit{Spin $s-2$ conformal gauge field} [$2s-3$ off-shell d.o.f.]:
\begin{equation}
    \delta D^{\alpha(s-2)\dot{\alpha}(s-2)}
    =
    \partial^{(\alpha(\dot{\alpha}}\Omega^{\alpha(s-3))\dot{\alpha}(s-3))}.
\end{equation}

\textbf{\textit{\underline{Fermionic sector}}}:

$\bullet$ \textit{Doublet of the fermionic spin $s - \frac{1}{2}$ gauge field} [$8s$ off-shell d.o.f.]:
\begin{equation}
    \delta \psi^{\alpha(s)\dot{\alpha}(s-1)i}
    =
    \partial^{(\alpha(\dot{\alpha}}b^{\alpha(s-1))\dot{\alpha}(s-2))i}.
\end{equation}

$\bullet$ \textit{Doublet of the spin $s-\frac{3}{2}$ fermionic gauge} fields [$8(s-1)$ off-shell d.o.f.]:
\begin{equation}
    \delta \chi^{\alpha(s-1)\dot{\alpha}(s-2)i}
    =
    \partial^{(\alpha(\dot{\alpha}} c^{\alpha(s-2)) \dot{\alpha}(s-3))i}.
\end{equation}

So the general integer-spin $\mathbf{s}$ \, $\mathcal{N}=2$ superconformal multiplet encompasses $\mathbf{8(2s-1)}_B+\mathbf{8(2s-1)}_F$
off-shell degrees of
freedom\footnote{In the case of non-conformal $\mathcal{N}=2$ spin $\mathbf{s}$ supermultiplet one deals with $ \mathbf{8 [s^2 + (s-1)^2]}_B + \mathbf{8 [s^2 + (s-1)^2]}_F$
off-shell degrees of freedom. These multiplets were constructed in \cite{Buchbinder:2021ite} as a generalization of off-shell multiplet of $\mathcal{N}=2$ Einstein supergravity.
The superconformal multiplets described here naturally generalize the Weyl multiplet of conformal $\mathcal{N}=2$ supergravity \cite{Galperin:1987ek, Ivanov:2022vwc}
to arbitrary integer higher spins. It is interesting to note that the number of d.o.f. in $\mathcal{N}=2$ superconformal multiplet
can be parametrized as $ \mathbf{8 [s^2 - (s-1)^2]}_B + \mathbf{8 [s^2 - (s-1)^2]}_F$. This leads to the conjecture on the structure of
superconformal compensators for the general spin ${\bf s}$: they should be composed of two towers of all integer $\mathcal{N}=2$ superconformal higher spins.}.
Interestingly, all fields in the $\mathcal{N}=2$ superconformal higher-spin multiplets are gauge fields: no non-gauge auxiliary fields are present
(the cases of  $\mathcal{N}=2$ spin $\mathbf{1}$ theory and $\mathcal{N}=2$ conformal supergravity are an exception). It is the significant difference from the case
of non-conformal $\mathcal{N}=2$ higher spins \cite{Buchbinder:2021ite}.
It is worth noting that there appear no auxiliary fields in the superconformal $\mathcal{N}=1$ higher spin multiplets as well \cite{Kuzenko:2017ujh}
\footnote{Based on these affinities, it is reasonable to assume that an arbitrary $\mathcal{N}$-extended superconformal multiplet also does not contain auxiliary fields.
(We thank the referee for pointing out for us ref. \cite{Raptakis:2023gyu},
    where $\mathcal{N}$ extended supermultiplets and their component contents were sketched.)}.
In this connection we mention that the  $\mathcal{N}=2$ higher-spin superconformal multiplets constructed can be decomposed into the sum of three $\mathcal{N}=1$ supermultiplets:
higher-spin $\mathbf{s}$ multiplet ($\mathbf{4s}_B+ \mathbf{4s}_F$ off-shell d.o.f), higher-spin $\mathbf{s-1}$ multiplet ($\mathbf{4(s-1)}_B+ \mathbf{4(s-1)}_F$ off-shell d.o.f)
 and higher-spin $\mathbf{s-\frac{1}{2}}$ multiplet ($\mathbf{4(2s-1)}_B + \mathbf{4(2s-1)}_F$ off-shell d.o.f.).

\medskip

Generalizing the spin $\mathbf{3}$ superconformal current superfields of Section \ref{s=3 superconformal current superfields}
to the general spin $\mathbf{s}$ is straightforward. For example, for the special case of vector indices we find:
\begin{equation}
    \begin{split}
        &\mathcal{J}_{\alpha(s-2)\dot{\alpha}(s-2)} = -\frac{1}{2} q^{+a} \mathcal{D}^{--} \partial^{s-2}_{\alpha(s-2)\dot{\alpha}(s-2)} J q^+_a,
        \\
        &\mathcal{D}^{++} \mathcal{J}_{\alpha(s-2)\dot{\alpha}(s-2)}
        =
        -\frac{1}{2} q^{+a}  \partial^{s-2}_{\alpha(s-2)\dot{\alpha}(s-2)} J q^+_a.
    \end{split}
\end{equation}
These expressions satisfy various conservation laws, e.g.,
\begin{equation}
    \mathcal{D}^{++} \left( D^+_{\hat{\beta}} \mathcal{J}_{\alpha(s-2)\dot{\alpha}(s-2)}   \right)
    =
    0,
    \qquad
    \mathcal{D}^{++} \left( D^+_{\beta} D^+_{\dot{\beta}} \mathcal{J}_{\alpha(s-2)\dot{\alpha}(s-2)}   \right)
    =
    0.
\end{equation}
Other supercurrents can be constructed in a similar way. We leave the general case for the future work.

\subsection{Summary of the superconformal spin $\mathbf{s}$}

The action \eqref{eq: s 12 12} admits the natural generalization to an arbitrary $\mathcal{N}=2$ conformal supergravity background:
\begin{equation}\label{eq: s 12 12 sugra}
    S = - \frac{1}{2}\int d\zeta^{(-4)} q^{+a} \left(\mathfrak{D}^{++}
    +
    \kappa_s \hat{\mathcal{H}}^{++}_{(s)} (J)^{P(s)} \right)q^+_a\,.
\end{equation}

The generalized action \eqref{eq: s 12 12 sugra} is  invariant under:
\begin{enumerate}
    \item \textit{Nonlinear spin $\mathbf{2}$ gauge transformations} (i.e. $\mathcal{N}=2$ conformal supergravity group). The action of these transformations
    on the spin $\mathbf{s}$ analytic potentials is given in \eqref{eq: sc spin s}.
    \item \textit{Spin $\mathbf{s}$ gauge transformations} to the leading order in $\kappa_s$ (like in the spin $\mathbf{3}$ case, one needs
    to add the proper $\hat{\mathcal{H}}^{++}_{(s=2)}$ terms to the $\hat{\mathcal{H}}^{++}_{(s)}$ gauge transformation law).
    The full form  of such transformations, with the proper spin $\mathbf{2}$ part added, is given by:
    \begin{subequations}
    \begin{equation}
\label{eq: spin s gauge transformations-full}
            \begin{split}
                \delta^{(k)}_\lambda \hat{\mathcal{H}}^{++}_{(s)}
                =&
                \left [\mathfrak{D}^{++}, \hat{\mathcal{U}}^{(k)}_{s} \right]
                \\
                =&
                 \frac{1}{2} \left[ \mathfrak{D}^{++},
                 \left\{\hat{\Lambda}^{M_1 \dots M_{k-2}}, \partial_{M_{k-2}} \dots  \partial_{M_1} \right\}_{AGB} \right]
                 \\&
                 +
                 \frac{\kappa_2}{4}
                 \left[ \hat{\mathcal{H}}^{++}_{(s=2)},
                 \left\{\Omega^{M_1 \dots M_{k-2}}, \partial_{M_{k-2}} \dots  \partial_{M_1} \right\}_{AGB} \right],
            \end{split}
    \end{equation}
        \begin{equation}\label{eq: spin s gauge transformations - k-1}
            \begin{split}
                \delta_\lambda^{(k)} q^{+a}
                =& - \kappa_s \,
                 \hat{\mathcal{U}}^{(k)}_{s}  \left( J \right)^{P(s)}  q^{+a}
                 \\
                =&
                - \frac{\kappa_s}{2} \left\{\hat{\Lambda}^{M_1 \dots M_{k-2}}, \partial_{M_{k-2}} \dots  \partial_{M_1} \right\}_{AGB} \left( J \right)^{P(s)} q^{+a}
                \\
                &
                - \frac{\kappa_s}{4} \left\{\Omega^{M_1 \dots M_{k-2}}, \partial_{M_{k-2}} \dots  \partial_{M_1} \right\}_{AGB} \left( J \right)^{P(s)} q^{+a},
            \end{split}
        \end{equation}
        \end{subequations}
where $k= s, s-2, s-4 \dots$. The gauge transformation $\delta_\lambda^{(k)}$ corresponds to the transformation of the spin $k$ part
of the operator $\hat{\mathcal{H}}^{++}_{(s)}$ (see eq. \eqref{eq: operator H s}).
\end{enumerate}
As in the cases of interaction of the spin $\mathbf{1}$ and $\mathbf{3}$ multiplets with $\mathcal{N}=2$ conformal supergravity fields, the interaction of
$\mathcal{N}=2$ spin $\mathbf{s}$ multiplet with $\mathcal{N}=2$ conformal supergravity multiplet is mediated by auxiliary hypermultiplet fields.

Thus eq. \eqref{eq: s 12 12 sugra} provides the covariant superconformal vertex  $(\mathbf{s}, \mathbf{\tfrac{1}{2}}, \mathbf{\tfrac{1}{2}})$
in an arbitrary $\mathcal{N}=2$ conformal supergravity background.

\section{Fully consistent higher-spin hypermultiplet coupling}
\label{sec: 8 fully consistent}

In the previous sections we have constructed the superconformal cubic vertices $(\mathbf{s}, \mathbf{\frac{1}{2}}, \mathbf{\frac{1}{2}})$ which are consistent
to the leading order in the higher-spin analogs of Einstein constant.
In this section, we will consider the possibility of making the resulting cubic vertices invariant with respect to gauge transformations in the next orders in these
coupling constants.

For example, consider the simplest case of the spin $\mathbf{3}$ in curved superspace:
\begin{equation}
    S_{(s=3)} = -\frac{1}{2} \int d\zeta^{(-4)} \,
    q^{+a} \left(\mathfrak{D}^{++} + \kappa_3 \hat{\mathcal{H}}^{++}_{(s=3)} J \right) q^+_a.
\end{equation}
This action is gauge invariant to the leading order in $\kappa_3$. In the next order
we have the following gauge transformation of cubic vertex under the spin $\mathbf{3}$ gauge transformations \eqref{eq: s=3 general transfromations-1}
of the hypermultiplet:
\begin{equation}\label{eq: kappa 3 variation}
    \begin{split}
    \delta_\lambda^{(s=3)} \left( -\frac{\kappa_3}{2} q^{+a} \hat{\mathcal{H}}^{++}_{(s=3)} J  q^+_a \right)
    =&
    - \frac{\kappa^2_3}{2}
    q^{+a}  \left[\hat{\mathcal{H}}^{++}_{(s=3)}, \hat{\mathcal{U}}_{(s=3)} \right] q^+_a
    \\
    =&
    -\frac{\kappa_3^2}{4} q^{+a}  \left[\hat{\mathcal{H}}^{++}_{(s=3)}, \left\{\hat{\Lambda}^M + \frac{1}{2} \Omega^M, \partial_M\right\}_{AGB}\right] q^+_a.
    \end{split}
\end{equation}
So we arrived at the differential operator of the third order  in superspace derivatives.
Making use of the spin $\mathbf{4}$ superconformal multiplet described in the previous section (modulo integrations by parts),
one can compensate this term by  deforming the spin $\mathbf{4}$ differential operator $\hat{\mathcal{H}}^{++}_{s=4}$ transformation law as:
\begin{equation}
    \begin{split}
    \kappa_4 \delta_\lambda^{(s=3)}
    \hat{\mathcal{H}}^{++}_{s=4}
    =&
    -\frac{\kappa^2_3}{2}
    \left[  \hat{\mathcal{H}}^{++}_{(s=3)}, \hat{\mathcal{U}}_{(s=3)} \right]
    \\
    =&
    - \frac{\kappa_3^2}{4}  \left[\hat{\mathcal{H}}^{++}_{(s=3)}, \left\{\hat{\Lambda}^M + \frac{1}{2} \Omega^M, \partial_M\right\}_{AGB}\right].
    \end{split}
\end{equation}
Here we assumed that the appropriate integration by parts has been performed, like  in the previous sections.
Such a modified transformation law mixes different $\mathcal{N}=2$ superconformal multiplets, i.e. it is a nonabelian-type gauge symmetry.
So the action
\begin{equation}\label{eq: s=3,4}
    S_{s=3,4} = -\frac{1}{2} \int d\zeta^{(-4)} \,
    q^{+a} \left(\mathfrak{D}^{++} + \kappa_3 \hat{\mathcal{H}}^{++}_{(s=3)} J
    +
    \kappa_4 \hat{\mathcal{H}}^{++}_{(s=4)}  \right) q^+_a
\end{equation}
respects the spin ${\bf s}=3$ gauge invariance to $\kappa_3^2$ order.
However, the action \eqref{eq: s=3,4} is not invariant in the $\kappa_3\kappa_4$ order. Then the procedure just described
can be continued step by step.

To summarize this procedure, we introduce an analytic differential operator that includes all integer higher spins:
\begin{equation}
    \hat{\mathcal{H}}^{++} := \sum_{s=1}^{\infty}  \kappa_s \hat{\mathcal{H}}^{++}_{(s)} (J)^{P(s)}.
\end{equation}
The action of the infinite tower of integer $\mathcal{N}=2$ superconformal higher spins interacting with the hypermultiplet
in an arbitrary $\mathcal{N}=2$ conformal supergravity background reads:
\begin{equation}\label{full vertex}
    S_{full} = - \frac{1}{2} \int d\zeta^{(-4)} \,q^{+a}
    \left(\mathcal{D}^{++} + \hat{\mathcal{H}}^{++} \right) q^+_a.
\end{equation}
Then, assuming the proper gauge transformation of $\hat{\mathcal{H}}^{++}$, one can achieve gauge invariance to any order in couplings constants.
Collecting the hypermultiplet gauge transformations \eqref{eq: spin s gauge transformations - k} for all spins, we obtain
\begin{equation}
    \delta_\lambda q^{+a} = - \hat{\mathcal{U}}_{hyp} q^{+a} = - \sum_{s=1}^{\infty} \kappa_s \, \hat{\mathcal{U}}_{s} \left( J\right)^{P(s)} q^{+a},
\end{equation}
where we used the notation:
\begin{equation}
    \hat{\mathcal{U}}_{s} q^{+a}: = \sum_{k ={s, s-2, \dots}}   \hat{\mathcal{U}}_{s} ^{(k)} q^{+a},
\end{equation}
with $ \hat{\mathcal{U}}_{s} ^{(k)}$ being defined in \eqref{eq: spin s gauge transformations - k}. This transformation acts linearly on the hypermultiplet superfield. As a consequence of \eqref{eq: transp}, the operator $\hat{U}_{hyp}$ satisfies
the condition $\hat{U}^T_{hyp}= - \hat{U}_{hyp}$.  For the set of gauge  superfields we obtain the transformation law:
\begin{equation}
    \delta_\lambda \hat{\mathcal{H}}^{++}
    =
    \left[ \mathcal{D}^{++} +  \hat{\mathcal{H}}^{++},   \hat{\mathcal{U}}_{gauge}\right],
    \qquad
    \hat{\mathcal{U}}_{gauge} : = \sum_{s=1}^\infty \kappa_s \, \hat{\mathcal{U}}_{s}.
\end{equation}
Here we also assumed the proper integration by parts, as in the previous sections.
This transformation law mixes different spins, so this is a non-Abelian deformation of the spin $\mathbf{s}$ transformation laws.
In the lowest order, the transformation becomes Abelian and reproduces the sum of transformations \eqref{eq: spin s gauge transformations-full} over all
integer spins ${\bf s}\geq 1$.

\medskip

The invariance of \eqref{full vertex} under $\mathcal{N}=2$ conformal supergravity transformations is automatic for the reasons
expounded in the previous section.  So we have constructed the fully consistent gauge-invariant and conformally invariant
interaction of hypermultiplet with an infinite tower of $\mathcal{N}=2$ higher spins
in an arbitrary $\mathcal{N}=2$ conformal supergravity background. To spot some possible hidden subtleties of the general construction,
it seems necessary to perform  a further deeper inspection of this procedure and, in particular,
to make a detailed comparison with the known couplings among higher-spin gauge fields and scalar fields.

\section{Conclusions and outlook}
\label{sec: conclusion}

In this paper we have derived and discussed in detail the structure of the off-shell
manifestly $\mathcal{N}=2$ superconformal cubic interaction of
$\mathcal{N}=2,\, 4D$ hypermultiplet theory with an arbitrary
superconformal higher spin ${\bf s}$ gauge superfield. The basic results can be summarized as:

\begin{itemize}
    \item We considered the off-shell
    hypermultiplet model in $\mathcal{N}=2,\, 4D$
    harmonic superspace and described its rigid and local superconformal
    symmetries. For invariance of the cubic higher-spin vertices $(\mathbf{s}, \mathbf{\frac{1}{2}}, \mathbf{\frac{1}{2}})$  under these symmetries it proved necessary
    to properly modify the superconformal transformations of the hypermultiplet by  the corresponding superconformal gauge superfields;

    \item
    To this end, we introduced the complete set of $\mathcal{N}=2$, $4D$ unconstrained analytic
    spin $\mathbf{s}$ superconformal higher-spin potentials, defined their superconformal and gauge
    transformations and revealed the physical field contents of the corresponding higher-spin Weyl
    supermultiplets in Wess-Zumino gauges. Their most notable features are: (i) all fields in the multiplets starting
    from ${\bf s}=3$ are gauge; (ii) the sets of bosonic fields necessarily contain ``hook-type'' generalized gauge fields;

    \item
    As a result, we have derived the 
     $\mathcal{N}=2$
    superconformal cubic vertex of the hypermultiplet coupled to superconformal higher spin external
    gauge superfields. Generically, the vertex has the structure: {\it higher spin
    superconformal gauge superfields $\times$ superconformal
    hypermultiplet supercurrents}. So the corresponding supercurrents
    can be explicitly constructed in terms of the hypermultiplet superfields, like it has been done for the spin $\mathbf{2}$ and $\mathbf{3}$ cases;

    \item As particular cases, we have constructed and discussed in detail the off-shell
    $(\mathbf{s}, \mathbf{\tfrac{1}{2}}, \mathbf{\tfrac{1}{2}})$ vertices in the background
    of $\mathcal{N}=2$ conformal supergravity for ${\bf s} = 2, 3$.

\end{itemize}

It should be specially pointed out that the geometric basis of the superconformal ${\cal N}=2, 4D$ off-shell gauge supermultiplets and their couplings to $q^+$ hypermultiplets, like
in the previously discussed non-conformal case, proved to be the preservation of ${\cal N}=2$ Grassmann harmonic analyticity. First of all, the fundamental gauge
potentials encompassing superconformal gauge multiplets are unconstrained ${\cal N}=2$ analytic harmonic superfields. Secondly,  they are naturally recovered from the demand of analyticity
of the $q^{+a}$ Lagrangian and requiring them to be closed under the analyticity-preserving coordinate realization of rigid ${\cal N}=2$ superconformal symmetry.\\

\noindent Finally, let us list possible directions of the future study:
\begin{itemize}
    \item \textit{Dynamical actions for higher-spin $\mathcal{N}=2$ superconformal multiplets}

    The natural foremost task is to construct $\mathcal{N}=2$ Fradkin-Tseytlin superconformal actions for the superconformal multiplets presented, at least at the linearized level.
    In components, these actions should be reducible to higher-spin generalizations of the square of the linearized generalized Weyl tensors
    which were firstly introduced in \cite{Fradkin:1985am}.
   For $\mathcal{N}$-extended superconformal higher-spin multiplets, the linearized actions were provided in \cite{Kuzenko:2021pqm}.
    In the HSS approach, such actions were not considered even for the standard $\mathcal{N}=2$ Weyl ($\mathbf{s}={2}$) multiplet.
    It is expected that the actions constructed in \cite{Kuzenko:2021pqm} can be obtained from the actions in the harmonic superspace by fixing the harmonic-independent gauge.

    \item \textit{Superconformal current superfield and rigid higher-spin superconformal symmetries}

    In this paper, we have addressed the important issue of the rigid symmetries of the free hypermultiplet and of
    the corresponding superfield currents only in passing. In fact, like in \cite{Buchbinder:2022vra}, one can easily identify the corresponding
    rigid symmetries by imposing the obvious conditions on the parameters \eqref{eq: spin s gauge transformations}:
    \begin{equation}
            \left[\mathcal{D}^{++}, \hat{\Lambda}^{M_1 \dots M_{k-2}}\right]  \partial_{M_{k-2}} \dots  \partial_{M_1}
        +
        \frac{1}{2}
        \left\{  \hat{\Lambda}^{M_1 \dots M_{k-2}}, [\mathcal{D}^{++}, \partial_{M_{k-2}} \dots  \partial_{M_1} ] \right\}_{AGB}
        =
        0
    \end{equation}
The solutions of these equations (modulo possible terms vanishing after integrations by parts) yield rigid higher-order conformal symmetries of the free hypermultiplet.
It would be of significant interest  to study the algebra of the corresponding group variations and compare
the result with the consideration in \cite{Kuzenko:2023vgf}, where rigid symmetries of the on-shell hypermultiplet were described (and were sketched also for the off-shell hypermultiplet).

Using $\mathcal{N}=2$ superfield Noether theorem or directly varying the
cubic interactions with respect to the superfield gauge potentials, one can derive the
conserved supercurrents for these symmetries. As one of the
instructive examples one could construct the ``master'' current
superfields, discussed in \cite{Buchbinder:2022vra} and briefly
sketched here. An interesting task is to study the component current
expansion of the supercurrents obtained.
Another important problem is the study of the superconformal transformation laws of the current
superfields (see also \cite{Kuzenko:2023vgf} for an alternative approach to conformal supercurrents).

    \item \textit{Induced actions}

Finding out the 
$\mathcal{N}=2$ superconformal
interaction vertex for the hypermultiplet coupled to external gauge
higher spin superfields opens a principal possibility to study the
higher spin quantum effects in such a theory. One of the topical
problems in this area is the one-loop effective action of a
higher-spin gauge field induced by its interaction with a lower-spin
quantum field. For the explicit construction of such an effective
action, there exists a general procedure going back to Schwinger and
DeWitt  and based upon  the representation of the effective action
as an integral over the proper time (see, e.g., \cite{BS}). In
general, the induced effective action is essentially non-local.
However, it can be perturbatively calculated as a series in the
background field derivatives, which makes it possible to obtain
various local invariants as functionals of the background gauge
fields. In the context of the theory of higher spin fields, this
opens up the possibility to find out, by direct algorithmic
calculations, new invariants depending on the higher spin gauge
fields. To the best of our knowledge, the study of the induced
effective action in the conformal theory of  higher spin fields was
initiated in refs \cite{Segal:2002gd} in the world-line
approach (see also the later paper \cite{Bonezzi:2017mwr}). In the
theory of the lower-spin fields cubically coupled to conformal gauge
higher spin fields, some approaches to the problem of calculating
the induced effective actions were worked out  in refs.
\cite{Bekaert:2010ky}, \cite{Bonora1}, \cite{Bonora2},
\cite{Kuzenko:2022hdv}, \cite{KuLaPo}, including the superfield
approaches  for $\mathcal{N}=1$ and $\mathcal{N}=2$ supersymmetric
higher-spin theories formulated in $\mathcal{N}=1$ superspace
\cite{Kuzenko:2022hdv}, \cite{KuLaPo}. In the present paper, we have
constructed the manifestly $\mathcal{N}=2$ superconformally
invariant cubic interaction vertex for a hypermultiplet coupled to
$\mathcal{N}=2$ higher spin gauge superfields. This makes it
possible to develop the manifestly $\mathcal{N}=2$ supersymmetric
proper time technique and use it to calculate the induced effective
action depending on $\mathcal{N}=2$ higher-spin gauge superfields
treated as classical external superfields. In other words, knowing
the explicit expressions for the general superfield coupling of the
hypermultiplet to the $\mathcal{N}=2$ higher-spin superconformal
gauge potentials could help to find out the invariant Lagrangians of
the latter.

    \item \textit{Higher-spin conformal compensators}

$\mathcal{N}=2$ supersymmetric extension of Fronsdal theory
constructed in \cite{Buchbinder:2021ite} generalized merely one of
the available versions of $\mathcal{N}=2$ Einstein supergravity.
An important question is how to construct the higher-spin
generalization of other versions of $\mathcal{N}=2$ supergravity. It
is well known that the most general set of distinct versions of
Einstein (super)gravity can be obtained by making use of the method
of (super)conformal compensators.
It is of primary interest to learn
what is a generalization of this compensator mechanism  to
higher ${\cal N}=2$ spins\footnote{Few earlier ideas regarding conformal compensators for higher spins were adduced in \cite{Segal:2002gd}.}.
To answer this question  it is
necessary, first of all, to explore the issue of quartic interacting
conformal vertices. The severe restrictions imposed by extended
supersymmetry and harmonic superspace methods could greatly simplify
the  problem of constructing such vertices\footnote{One of the possible sources of such vertices
in the HSS approach was addressed in a recent paper \cite{Ivanov23}.}.  On the other hand, the generic matter conformal compensator for
$\mathcal{N}=2$ supergravity is just the massless hypermultiplet with the wrong sign of kinetic term (plus vector $\mathcal{N}=2$
compensator with the analogous ``wrong'' sign of the kinetic term) \cite{Galperin:1987ek,18}\footnote{The relevant off-shell version
of Einstein $\mathcal{N}=2$ supergravity was dubbed ``principal version'' in \cite{Galperin:1987ek,18};
it is the only one which admits the most general $\mathcal{N}=2$ matter off-shell couplings.}. So there naturally emerges
the problem of extending this picture to higher-spin $\mathcal{N}=2$ supergravity. It is obvious in advance that, in order
to recover the hypermultiplet coupling of ref. \cite{Buchbinder:2022kzl},  one needs to start with a conformal system
involving at least two independent hypermultiplet superfields, one being a compensator.

    \item \textit{AdS background }

One more actual problem is to develop a similar formalism for ${\cal N}=2$ higher spins in the AdS and other conformally flat backgrounds. Since the super AdS group
is a subgroup of ${\cal N}=2$ superconformal group, we hope that such a problem can be attacked, based largely on the results of the present work.

    \item \textit{Construction of more general interactions}

    An important task is to generalize supercurrents and cubic vertices constructed here for hypermultiplets to the more general cases of interaction
    with other matter ${\cal N}=2$ multiplets, e.g., with $\mathcal{N}=2$ Maxwell multiplet (massless or massive). We hope to tackle
    this task (closely related also to the issue of conformal compensators) elsewhere.
\end{itemize}

\acknowledgments

The authors are grateful to M. Tsulaia, M.A. Vasiliev and Yu.M.
Zinoviev for valuable discussion of some aspects of the paper. The authors also thank K. Koutrolikos for useful correspondence. Work of N.Z. was partially supported by the grant 22-1-1-42-2 from
the Foundation for the Advancement of Theoretical Physics and
Mathematics ``BASIS''.
The authors are grateful to the anonymous referee for useful and suggestive comments.

\appendix

\section{Wess-Zumino gauge for superconformal spin $\mathbf{3}$}
\label{app: WZ gauge}

In this appendix, we expound how to fix the Wess-Zumino gauges for the spin $\mathbf{3}$ analytic potentials $h^{++MN}$.
The relevant linearized gauge transformations are collected in \eqref{eq: s=3 gauge transformations}.
We show that one can fix gauge in such a way that all superfields, except their subset $h^{++M\alpha\dot{\alpha}}$, are gauged away. Then we deduce the Wess-Zumino form
of the residual gauge potentials and find out  the irreducible off-shell component content of the ${\bf s}=3$ ${\cal N}=2$ gauge multiplet.

\subsection{Fixing ``harmonic'' freedom}

\medskip
As the first important step, consider the analytic superfield $h^{(+n)K}$ with the following gauge freedom:
\begin{equation}\label{eq: gauge law}
    \delta_\lambda h^{(+n)K} = \mathcal{D}^{++} \lambda^{(+(n-2))K}.
\end{equation}
Here $K$ is an arbitrary multi-index, $\lambda^{(+(n-2))K}$ is an unconstrained analytic superfield parameter.
Terms of just this type appear in the transformation laws of all gauge potentials, see \eqref{eq: s=3 gauge transformations}.
Once this gauge freedom is fixed, we can inspect contributions of other terms.

The generic component expansions of $h^{(+n)K}(\zeta)$ and $\lambda^{+(n-2)K}$ read, respectively,
\begin{equation}
    \begin{split}
        h^{(+n)K}(\zeta)
        =&
        A^{(+n)K}
        +
        \theta^{+\hat{\rho}} B^{(+(n-1))K}_{\hat{\rho}}
        \\&+
        (\theta^+)^2 C^{(+(n-2))K}_1
        +
        (\bar{\theta}^+)^2 C^{(+(n-2))K}_2
        +
        \theta^{+\alpha} \bar{\theta}^{+\dot{\alpha}} C^{(+(n-2))K}_{\alpha\dot{\alpha}}
        \\&+
        (\theta^+)^2 \bar{\theta}^{+\dot{\alpha}} D^{(+(n-3))K}_{\dot{\alpha}}
        +
        (\bar{\theta}^+)^2 \theta^{+\alpha} D^{(+(n-3))K}_{\alpha}
        +
        (\theta^+)^4 E^{(+(n-4))K}\,,\label{A1}
    \end{split}
\end{equation}
\begin{equation}
    \begin{split}
        \lambda^{(+(n-2))K}(\zeta)
        =&
        a^{(+(n-2))K}
        +
        \theta^{+\hat{\rho}} b^{(+(n-3))K}_{\hat{\rho}}
        \\&+
        (\theta^+)^2 c^{(+(n-4))K}_1
        +
        (\bar{\theta}^+)^2 c^{(+(n-4))K}_2
        +
        \theta^{+\alpha} \bar{\theta}^{+\dot{\alpha}} c^{(+(n-4))K}_{\alpha\dot{\alpha}}
        \\&+
        (\theta^+)^2 \bar{\theta}^{+\dot{\alpha}} d^{(+(n-5))K}_{\dot{\alpha}}
        +
        (\bar{\theta}^+)^2 \theta^{+\alpha} d^{(+(n-5))K}_{\alpha}
        +
        (\theta^+)^4 e^{(+(n-6))K}.
    \end{split} \label{A2}
\end{equation}
The coefficients $A,B \dots$ and $a,b \dots$ are arbitrary $x$-dependent harmonic functions with the properly fixed harmonic charges.

The result of action of the partial harmonic derivative $\partial^{++}$ on \eqref{A2} is as follows:
\begin{equation}
    \begin{split}
        \partial^{++}   \lambda^{(+(n-2))K}(\zeta)
        =&
        \partial^{++} a^{(+(n-2))K}
        +
        \theta^{+\hat{\rho}}    \partial^{++} b^{(+(n-3))K}_{\hat{\rho}}
        \\&+
        (\theta^+)^2    \partial^{++} c^{(+(n-4))K}_1
        +
        (\bar{\theta}^+)^2  \partial^{++} c^{(+(n-4))K}_2
        \\&+
        \theta^{+\alpha} \bar{\theta}^{+\dot{\alpha}}   \partial^{++} c^{(+(n-4))K}_{\alpha\dot{\alpha}}
        \\&+
        (\theta^+)^2 \bar{\theta}^{+\dot{\alpha}}   \partial^{++} d^{(+(n-5))K}_{\dot{\alpha}}
        +
        (\bar{\theta}^+)^2 \theta^{+\alpha}     \partial^{++} d^{(+(n-5))K}_{\alpha}
        \\&+
        (\theta^+)^4    \partial^{++}  e^{(+(n-6))K}.
    \end{split} \label{A22}
\end{equation}
In this expression, the harmonic derivative produces general harmonic functions if the charge of the corresponding function $\geq 0$. Then,
for the harmonic charges with $n\geq 5$, one can gauge away all the components by the gauge transformations \eqref{eq: gauge law}. For $n=4$,
one cannot gauge away by this mechanism the highest component in the harmonic expansion of $E^K$ , for $n=3$ those in the expansion of $D_{\hat{\alpha}}^K, E^K$,
and so forth.

The corresponding residual gauge freedom is specified by the lowest components of the $  \lambda^{(+(n-2))K}(\zeta)$ coefficients  with the positive harmonic charge.
For example, in $n=3$ case these parameters are $a^{i K}, b^{K}$. Due to the presence of the term with $x$-derivative in ${\cal D}^{++}$, these surviving parameters
(with derivatives on them) can appear in the transformations of some other non-vanishing components. Also, the appropriate contributions from the terms with explicit
$\theta$ s  in \eqref{eq: s=3 gauge transformations} can modify the residual gauge transformations and ensure some additional gauge conditions.
All these subtleties can be uniquely fixed from the condition of preserving the final Wess-Zumino type gauges.

\medskip

 Now we can proceed to the precise discussion of the gauge-fixing procedure for the superconformal spin $\mathbf{3}$ potentials.
Using merely terms with  harmonic derivatives, and based on the reasoning around eqs. \eqref{A2} - \eqref{A22}, we can partially fix the gauge as:
\begin{equation}
    \begin{split}
        &h^{++\alpha\dot{\alpha}\beta\dot{\beta}}
        =
        i (\theta^+)^2 C^{\alpha\dot{\alpha}\beta\dot{\beta}}
        -
        i (\bar{\theta}^+)^2 \bar{C}^{\alpha\dot{\alpha}\beta\dot{\beta}}
        -
        4i \theta^{+\rho} \bar{\theta}^{+\dot{\rho}} \Phi_{\rho\dot{\rho}}^{\alpha\dot{\alpha}\beta\dot{\beta}}
        \\&
        \qquad
        \qquad
        +
        (\bar{\theta}^+)^2 \theta^{+\rho} \psi_\rho^{\alpha\dot{\alpha}\beta\dot{\beta}i}u^-_i
        +
        (\theta^+)^2 \bar{\theta}^{+\dot{\rho}} \bar{\psi}_{\dot{\rho}}^{\alpha\dot{\alpha}\beta\dot{\beta}i}u^-_i
        +(\theta^+)^4 V^{\alpha\dot{\alpha}\beta\dot{\beta}ij}u^-_i u^-_j,
        \\
        & h^{++\alpha\dot{\alpha}\beta+}
        =
        (\theta^+)^2 \bar{\theta}^{+\dot{\rho}} P_{\dot{\rho}}^{\alpha\dot{\alpha}\beta}
        +
        (\bar{\theta}^+)^2 \theta^{+\rho} T_{\rho}^{\alpha\dot{\alpha}\beta}
        +
        (\theta^+)^4 \chi^{\alpha\dot{\alpha}\beta i}u^-_i,
        \\
        &
        h^{++\alpha\dot{\alpha}++}
        =
        (\theta^+)^4 D^{\alpha\dot{\alpha}},
        \\
        &
        h^{++[\alpha+\beta]+} = (\theta^+)^4 K^{[\alpha\beta]},
        \\
        &
        h^{++\alpha+\dot{\alpha}+} = i (\theta^+)^4 K^{\alpha\dot{\alpha}},
        \\
        &
        h^{++\hat\alpha+++} = 0,
        \\
        &
        h^{(+6)} = 0.
    \end{split}\label{A23}
\end{equation}
The reality conditions for the involved fields can be figured out from the generalized reality conditions for the analytic gauge potentials \eqref{eq: spin 3 reality}.
The ultimate effect of the shift transformations \eqref{eq: s=3 gauge transformations} on the component fields in \eqref{A23} can be determined
by considering separately various sectors. Using these transformations, one can find the transformation laws of the remaining fields and learn which fields survive
after the WZ gauges have been completely fixed (up to the residual gauge transformations involving only $x$-derivatives of the relevant parameters).

\subsection{Further gauge-fixing}
Inspecting the transformations \eqref{eq: s=3 gauge transformations} more carefully, we found that, besides the primary gauge-fixing \eqref{A23},
based on the general properties of the harmonic expansions, the further steps of the gauge fixing can be effected, which become possible due to the presence
of the explicit $\theta$s in \eqref{eq: s=3 gauge transformations}. Namely, it is self-consistent to put
\be
\hat{h}^{(+4)} =0\,, \quad \hat{h}^{++} = 0\,, \quad \hat{h}^{+3 \hat\alpha} = 0\,, \quad h^{++\alpha+\dot\beta+} = 0 \;\; ({\rm and \,\,c.c.})\,, \label{Antysymm}
\ee
where
\bea
&& \hat{h}^{++}:= \epsilon_{\alpha\beta} \epsilon_{\dot\alpha\dot\beta} h^{++\alpha\dot\alpha\, \beta\dot\beta}\,, \; \hat{h}^{+3\dot\alpha}
:= \epsilon_{\alpha\beta} h^{++\alpha+\beta\dot\alpha}, \nn
&&  \bar{\hat{h}}^{+3 \alpha} = \epsilon_{\dot\alpha\dot\beta} h^{++\dot\alpha+\alpha\dot\beta}, \; \hat{h}^{(+4)} :=\frac12\epsilon_{\beta\alpha}h^{++[\beta + \alpha]+}
\quad  ({\rm and \,\, c.c.})\,.\label{DefAnty}
\eea

Then all physical fields are contained  in the remaining parts of the original gauge  potentials
\bea
h^{++(\alpha\,\beta)(\dot\alpha\, \dot\beta)}\,, \quad h^{++(\alpha+\beta)\dot\alpha}\,, \quad  h^{++(\dot\alpha+\dot\beta)\alpha}\,,\quad h^{++\alpha\dot\alpha ++}\,. \label{Phys}
\eea
Requiring the gauges \eqref{Antysymm} and the last two gauges in \eqref{A23} to be preserved under the general
linearized gauge transformations \eqref{eq: s=3 gauge transformations} imposes the following constraints on the relevant residual analytic gauge parameters:
\bea
&&h^{(+6)} = 0\quad \Longrightarrow  \quad  {\cal D}^{++}\lambda^{(+4)} = 0\,, \label{a1} \\
&&h^{+5\hat\beta} = 0\quad \Longrightarrow  \quad {\cal D}^{++}\lambda^{+3\hat\beta} - 2\lambda^{(+4)}\theta^{+\hat\beta} = 0 \,, \label{a2} \\
&&\hat{h}^{(+ 4)} = 0 \quad \Longrightarrow  \quad {\cal D}^{++}\hat{\lambda}^{+ 2} - \lambda^{+3 \alpha}\theta^+_\alpha = 0\,, \quad {\rm and \,\,c.c.}\,, \label{a5} \\
&& h^{++\beta+\dot\gamma+} = 0 \quad \Longrightarrow  \quad {\cal D}^{++}\lambda^{\beta + \dot\gamma +} - \frac12 \big(\lambda^{+3\beta}\bar\theta^{+\dot\gamma}
- \lambda^{+3 \dot\gamma}\theta^{+\beta} \big) = 0\,,  \label{BasConstrlamb} \\
&&\hat{h}^{+3\dot\alpha} = 0 \quad \Longrightarrow  \quad
{\cal D}^{++}\hat{\lambda}^{+\dot\alpha}  + 8i \hat{\lambda}^{+2} \bar\theta^{+\dot\alpha}+ \hat{\lambda}^{++\alpha\dot\alpha}\theta^{+}_\alpha = 0\,, \quad {\rm and \,\,c.c.}\,, \label{a4}\\
&&\hat{h}^{++} = 0 \quad \Longrightarrow  \quad {\cal D}^{++}\hat{\lambda}  - 4i \hat{\lambda}^{+\dot\beta}\bar\theta^+_{\dot\beta}
+ 4i \hat{\lambda}^{+\beta}\theta^+_{\beta} = 0\,, \quad {\rm and \,\,c.c.}\,, \label{a3}
\eea
where
\bea
&&\hat{\lambda} := \epsilon_{\alpha\beta} \epsilon_{\dot\alpha\dot\beta} \lambda^{\alpha\beta\dot\alpha\dot\beta}, \quad
\hat{\lambda}^{+\dot\beta} := \epsilon_{\alpha\beta}\lambda^{\alpha+ \beta\dot\beta}\, {\rm and \;c.c.}\,, \quad \hat{\lambda}^{+2}
:= \epsilon_{\alpha\beta}\lambda^{[\alpha +\beta]+} \quad  {\rm and \;c.c.}\,, \nn
&&  \hat{\lambda}^{++\alpha\dot\alpha} := \lambda^{++\alpha\dot\alpha} - 8i \lambda^{\alpha+\dot\alpha+}\,. \label{Restrlambda}
\eea

The gauge transformations of the ``physical'' set \eqref{Phys} are given by
\bea
&& \delta h^{++(\alpha\,\beta)(\dot\alpha\, \dot\beta)} = {\cal D}^{++} \lambda^{(\alpha\beta)(\dot\alpha\dot\beta)} +
4i \big(\lambda^{(\beta+\alpha)(\dot\alpha}\bar\theta^{+\dot\beta)} - \lambda^{(\dot\alpha+\dot\beta)(\beta} \theta^{+\alpha)}\big), \label{b1} \\
&& 2 \delta h^{++(\beta+\alpha)\dot\alpha} ={\cal D}^{++} \lambda^{(\alpha+\beta)\dot\alpha} - \theta^{+(\alpha}\big(\lambda^{\beta)\dot\alpha ++} + 8i\lambda^{\beta) +\dot\alpha +}\big)
\,, \quad {\rm and \,\,c.c.}\,,\label{b2} \\
&& 2 \delta h^{++\alpha\dot\alpha++} = {\cal D}^{++} \lambda^{++\alpha\dot\alpha} + 4i\big(\lambda^{+3\alpha}\bar\theta^{+\dot\alpha} - \lambda^{+3\dot\alpha}\theta^{+\alpha}\big). \label{b3}
\eea

We observe that the eqs. \eqref{a1} - \eqref{a4} involve some gauge parameters which appear also in \eqref{b1} - \eqref{b3}. So we need
first to fully exhibit the consequences of \eqref{a1} - \eqref{BasConstrlamb}.

\subsection{Bosonic sector}

We will start from the bosonic sector.

For what follows we will need the component structure of the analytic gauge parameters. Firstly we present it for the gauge parameters
associated with the pure gauge potentials appearing in (A.9) - (A.14):
\bea
&& \lambda^{(+4)} = \ell^{+4} + \ell^{+2}(\theta^+)^2  + \bar\ell^{+2}(\bar\theta^+)^2 + \ell^{+2\alpha\dot\alpha}\theta^+_\alpha\bar{\theta}^{+}_{\dot\alpha} + \ell (\theta^{+})^4, \nn
&&\lambda^{+3\alpha} = \mu^{+2 \alpha}_\beta \theta^{+\beta} + \mu^{+2 \alpha}_{\dot\beta} \bar\theta^{+\dot\beta}  +
\gamma^\alpha_{\beta} (\bar\theta^+)^2 \theta^{+\beta} + \gamma^\alpha_{\dot\beta} (\theta^+)^2 \bar\theta^{+\dot\beta}\,, \nn
&& \hat\lambda^{+2} = \sigma^{+2} + \sigma(\theta^+)^2  + \sigma' (\bar\theta^+)^2 + \sigma_1^{\alpha\dot\alpha}\theta^+_\alpha\bar\theta^+_{\dot\alpha} + \sigma^{-2}_2(\theta^{+})^4, \nn
&& \lambda^{\alpha+\dot\beta+} =  \psi^{++\alpha\dot\beta} + \psi_1^{\alpha\dot\beta}(\theta^+)^2  - \bar\psi_1^{\alpha\dot\beta}(\bar\theta^+)^2
+ \psi_2^{\alpha\dot\beta \gamma\dot\gamma}\theta^+_\gamma\bar{\theta}^+_{\dot\gamma} + \psi^{-2\alpha\dot\beta}_3(\theta^{+})^4. \label{VspomExp}\\
&& \hat{\lambda}^{+\alpha} = \nu^{\alpha}_\gamma \theta^{+\gamma} + \nu^{\alpha}_{\dot\gamma} \bar\theta^{+\dot\gamma}  +
\phi^{-2\alpha}_{\gamma} (\bar\theta^+)^2 \theta^{+\gamma} + \phi^{-2\alpha}_{\dot\gamma} (\theta^+)^2 \bar\theta^{+\dot\gamma}\,, \nn
&& \hat{\lambda} = \beta + \beta^{-2} (\theta^+)^2  + \bar\beta^{-2}(\bar\theta^+)^2 + \beta^{-2\gamma\dot\gamma}\theta^+_\gamma\bar{\theta}^{+}_{\dot\gamma} + \beta^{-4}(\theta^{+})^4\,.
\label{hatlam}
\eea

The analogous expansions for the remaining superfield gauge parameters read
\bea
&& \lambda^{(\alpha\beta)(\dot\alpha\dot\beta)} = \rho^{(\alpha\beta)(\dot\alpha\dot\beta)}
+ \big[\rho^{-2(\alpha\beta)(\dot\alpha\dot\beta)}_1(\theta^+)^2 + {\rm c.c.}\big] + \rho^{-2(\alpha\beta)(\dot\alpha\dot\beta)\gamma\dot\gamma}_2
\theta^+_\gamma\bar\theta^+_{\dot\gamma} \nn
&& \quad\quad\quad\quad \quad +\,\rho^{-4(\alpha\beta)(\dot\alpha\dot\beta)}_3(\theta^+)^4, \label{EssExp1} \\
&& \lambda^{(\beta +\alpha)\dot\beta} = \omega^{(\alpha\beta) \dot\beta\gamma}\theta^{+}_{\gamma} +
\omega^{(\alpha\beta) \dot\beta\dot\gamma}\bar\theta^{+}_{\dot\gamma} +
\omega_1^{-2(\alpha\beta)\dot\beta\gamma}(\bar\theta^+)^2 \theta^{+}_{\gamma} + \omega_2^{-2(\alpha\beta)\dot\beta\dot\gamma}(\theta^+)^2 \bar\theta^{+}_{\dot\gamma}, \label{EssSpin} \\
&&\lambda^{+2\alpha\dot\beta} =  \chi^{+2\alpha\dot\beta} + \chi_1^{\alpha\dot\beta}(\theta^+)^2  + \bar\chi_1^{\alpha\dot\beta}(\bar\theta^+)^2
+ \chi_2^{\alpha\dot\beta \gamma\dot\gamma}\theta^+_\gamma\bar{\theta}^+_{\dot\gamma} + \chi^{-2\alpha\dot\beta}_3(\theta^{+})^4. \label{EssHarm}
\eea
The conjugation rules for the component gauge parameters follow from the superfield ones listed earlier.

Eqs. (A.9) - (A.12) yield the following restrictions on the four sets of the component parameters in \eqref{VspomExp}:
\bea
&&(a)\;\,\ell^{+4} = \ell^{+4}_{(0)}, \; \ell^{+2} = \ell^{+2}_{(0)},\;\ell^{+2\alpha\dot\alpha} = \ell^{+2\alpha\dot\alpha}_{(0)} + 4i\,\partial^{\alpha\dot\alpha}\ell^{+3-}_{(0)}, \nn
&&(b)\;\, \ell = \ell_{(0)} + i\partial_{\alpha\dot\alpha}\ell^{+-\alpha\dot\alpha}_{(0)} - 2\Box \ell^{+2-2}_{(0)}, \label{A92} \\
&& {}\nonumber \\
&&(a)\;\,\mu^{+2\beta}_\alpha = \mu^{+2\beta}_{\alpha}{}_{(0)} +2\delta^\beta_\alpha \ell^{+3-}_{(0)}\,, \;\mu^{+2\beta}_{\dot\alpha} = \mu^{+2\beta}_{\dot\alpha}{}_{(0)}\,, \nn
&&\gamma^{\alpha}_{\beta} = \gamma^{\alpha}_{\beta}{}_{(0)} - 2i\,\partial^{\dot\alpha}_\alpha\, \mu^{+-\beta}_{\dot\alpha}{}_{(0)} + 2\,\delta^\alpha_\beta \ell^{+-}_{(0)}\,, \nn
&&(b)\;\,\gamma^{\beta}_{\dot\alpha} = \gamma^{\beta}_{\dot\alpha}{}_{(0)} - 2i\,\partial^{\alpha}_{\dot\alpha}\, \mu^{+-\beta}_{\alpha}{}_{(0)} - \ell^{+-\beta}_{\dot\alpha}{}_{(0)}
-4i\,\partial^{\beta}_{\dot\alpha}\,\ell_{(0)}^{+2-2}\label{A102}\\
&&{}\nonumber \\
&&(a)\;\, \sigma^{+2} = \sigma^{+2}_{(0)}\,, \; \sigma' = \sigma'_{(0)}\,, \; \sigma = \sigma_{(0)} + \frac12\mu^{+- \beta}_{\beta (0)} + \ell^{+2-2}_{(0)}\,,\nn
&&(b)\;\,\sigma_1^{\gamma\dot\gamma} =
\sigma_{1(0)}^{\gamma\dot\gamma} + 4i\,\partial^{\gamma\dot\gamma}\,\sigma_{(0)}^{+-} + \mu^{+-\gamma\dot\gamma}_{(0)} \nn
&&(c)\; \sigma^{-2}_2 = -2\Box \sigma^{-2}_{(0)} + \bar\ell_{(0)}^{-2} + i\partial_{\alpha\dot\alpha}\mu^{-2\alpha\dot\alpha}_{(0)}\,, \quad
(d) \; \gamma^\alpha_{\alpha\,(0)} = -2i \partial_{\gamma\dot\gamma}\,\sigma_{1(0)}^{\gamma\dot\gamma}\,, \label{A112} \\
&&{}\nonumber \\
&&(a)\;\,\psi^{+2\alpha\dot\alpha} = \psi^{+2\alpha\dot\alpha}_{(0)}\,, \; \psi^{\alpha\dot\alpha}_1 = \psi^{\alpha\dot\alpha}_{1(0)}+ \frac14\bar\mu^{+-\alpha\dot\alpha}_{(0)}\,, \nn
&&(b)\;\,\psi^{\alpha\dot\beta \gamma\dot\gamma} = \psi^{\alpha\dot\beta \gamma\dot\gamma}_{(0)} + 4i\partial^{\gamma\dot\gamma}\,\psi^{+-\alpha\dot\beta}_{(0)} +
\frac12\varepsilon^{\rho\gamma}\varepsilon^{\dot\beta\dot\gamma}[ \mu^{+-\alpha}_{\rho (0)} + \ell^{+2-2}_{(0)} \delta^\alpha_{\rho}] \nn
&& \quad\quad\quad\quad\quad +\,\frac12\varepsilon^{\alpha\gamma}\varepsilon^{\dot\alpha\dot\gamma}[\bar\mu^{+-\dot\beta}_{\dot\alpha (0)} + \ell^{+2-2}_{(0)} \delta^{\dot\beta}_{\dot\alpha}], \nn
&& (c)\;\,\psi^{-2\alpha\dot\alpha}_3 = -2 \Box \psi^{-2\alpha\dot\alpha}_{(0)} + \frac{i}{2} [ \partial^{\rho\dot\alpha}\,\mu^{-2\alpha}_{\rho(0)}
+ \partial^{\alpha\dot\beta} \bar\mu^{-2\dot\alpha}_{\dot\beta (0)}] + \frac14 \ell^{-2\alpha\dot\alpha}_{(0)}  + i \partial^{\alpha\dot\alpha}\ell^{-3+}_{(0)}\,,\nn
&& (d)\;\,\gamma^{\alpha\dot\beta}_{(0)} - \bar\gamma^{\alpha\dot\beta}_{(0)} = 4i \partial_{\gamma\dot\gamma}\psi^{\alpha\dot\beta \gamma\dot\gamma}_{2 (0)}\,.\label{A122}
\eea
Analogously, eqs. (A.13), (A.14) yield, for the component parameters in \eqref{hatlam},
\bea
&&(a)\; \nu^\alpha_\gamma = \nu^\alpha_{\gamma (0)} + 8i \delta^\alpha_\gamma\,\bar{\sigma}^{+-}_{(0)}\,, \;\; \nu^\alpha_{\dot\gamma} = \nu^\alpha_{\dot\gamma (0)} + \big[\chi^{+-\alpha} _{\dot\gamma (0)} -
8i \,\psi^{+-\alpha}_{\dot\gamma (0)}\big]\,, \nn
&&(b)\; \phi^{-2\alpha}_\gamma = -2i\partial^{\dot\gamma}_\gamma \,\big[\chi^{-2\alpha} _{\dot\gamma (0)} -
8i \,\psi^{-2\alpha}_{\dot\gamma (0)}\big] - 4i\mu^{-2\alpha}_{\gamma (0)} - \frac{8i}{3}\,\delta^\alpha_\gamma\,\ell^{-3+}_{(0)}, \nn
&& \quad \phi^{-2\alpha\dot\gamma} =
16\,\partial^{\alpha\dot\gamma}\,\bar{\sigma}^{-2}_{(0)}\,, \nn
&& (c)\; \chi_{1 (0)}^{\alpha\dot\gamma} - 8i\, \psi_{1 (0)}^{\alpha\dot\gamma} = 2i\,\partial^{\gamma\dot\gamma}\,\nu^\alpha_{\gamma (0)} - 4i\,\bar\sigma_{1(0)}^{\alpha\dot\gamma}\,, \label{A132} \\
&&{} \nonumber \\
&& (a) \; \beta =\beta_{(0)}\,, \quad \beta^{-2} = 16 \bar{\sigma}^{-2}_{(0)}\,, \quad \nu^\alpha_{\alpha (0)} = \bar\nu^{\dot\alpha}_{\dot\alpha (0)} = 0\,, \nn
&& (b) \; \beta^{-2\gamma\dot\gamma} = -4i\,\big[\chi^{-2\gamma\dot\gamma}_{(0)} - 8i\,\psi^{-2\gamma\dot\gamma}_{(0)}\big]\,, \quad
\nu^{\gamma\dot\gamma}_{(0)} + \bar{\nu}^{\gamma\dot\gamma}_{(0)} = \partial^{\gamma\dot\gamma}\,\beta_{(0)}\,, \nn
&& (c)\; \beta^{-4} = -\frac{16}{3} \ell^{-4}_{(0)}\,, \quad \mu^{-2\alpha}_{\alpha (0)} + \bar{\mu}^{-2\dot\alpha}_{\dot\alpha (0)}  =
\frac{3}{2}\, \partial_{\gamma\dot\gamma}\,\big[\chi^{-2 \gamma\dot\gamma}_{(0)} -8i\,\psi^{-2 \gamma\dot\gamma}_{(0)}\big]. \label{A142}
\eea
Hereafter, the suffix $(0)$ denotes the lowest-order terms in the relevant harmonic expansions.

Thus we have shown that plenty of the superconformal gauge potentials, including $h^{(+6)}$ and $h^{++\hat{\alpha}+++}$ in \eqref{A23} and those in \eqref{Antysymm},
can be completely gauged away and we are left with the restricted set of the analytic superfield potentials which eventually encompass the irreducible ${\bf s}=3$ multiplet.
In the bosonic sector, these basic gauge potentials have the following component expansions (before passing to the partly gauge-fixed form \eqref{A23}):
\bea
&& h^{++(\alpha\beta)(\dot\alpha\dot\beta)} = h_0^{++(\alpha\beta)(\dot\alpha\dot\beta)} +\big[ h_1^{(\alpha\beta)(\dot\alpha\dot\beta)}(\theta^+)^2 + {\rm c.c.}\big]
 + h_2^{(\alpha\beta)(\dot\alpha\dot\beta)\gamma\dot\gamma}\theta^+_\gamma\bar\theta^+_{\dot\gamma} \nn
&& \quad\quad\quad\quad\quad\;\; +\, h_3^{-2 (\alpha\beta)(\dot\alpha\dot\beta)} (\theta^+)^4,  \label{hBas1}\\
&& h^{++(\beta+\alpha)\dot\alpha} = h^{+2(\beta\alpha)\dot\alpha}_{1\,\gamma}\theta^{+\gamma} + h^{+2(\beta\alpha)\dot\alpha}_{2\,\dot\gamma}\bar{\theta}^{+\dot\gamma}
+ h^{(\beta\alpha)\dot\alpha}_{3\,\gamma} (\bar\theta^+)^2\theta^{+\gamma} \nn
&& \quad\quad\quad\quad\quad\;\;+\, h^{(\beta\alpha)\dot\alpha}_{4\,\dot\gamma}(\theta^+)^2 \bar\theta^{+\dot\gamma}, \label{hBas2} \\
&& h^{+4\alpha\dot\alpha} = h_0^{+4\alpha\dot\alpha} + h_1^{+2\alpha\dot\alpha} (\theta^+)^2 + \bar{h}_1^{+2\alpha\dot\alpha} (\bar\theta^+)^2 +
h_2^{+2\alpha\dot\alpha\gamma\dot\gamma}\theta^{+}_\gamma\bar\theta^+_{\dot\gamma} + h_3^{\alpha\dot\alpha} (\theta^+)^4. \label{hBas3}
\eea

The component fields of $h^{++(\alpha\beta)(\dot\alpha\dot\beta)}$ have the following gauge transformation laws:
\bea
&& \delta h_0^{++(\alpha\beta)(\dot\alpha\dot\beta)} = \partial^{++}\rho^{(\alpha\beta)(\dot\alpha\dot\beta)}\,,  \quad
\nn
&& \delta h_1^{(\alpha\beta)(\dot\alpha\dot\beta)} = \partial^{++}\rho^{-2(\alpha\beta)(\dot\alpha\dot\beta)}_1 - 2i\, \bar{\omega}_{(0)}^{(\alpha\beta)(\dot\alpha\dot\beta)}\,,
\,({\rm and\,\,c.c.})\,,   \label{Tra1} \\
&& \delta h_2^{(\alpha\beta)(\dot\alpha\dot\beta)\gamma\dot\gamma} = \partial^{++}\rho^{-2(\alpha\beta)(\dot\alpha\dot\beta)\gamma\dot\gamma}_2
- 4i\,\partial^{\gamma\dot\gamma}\,\rho^{(\alpha\beta)(\dot\alpha\dot\beta)} \nn
&& \qquad\qquad\qquad\quad +\,4i\big[ \omega^{(\alpha\beta)\gamma(\dot\alpha}\varepsilon^{\dot\beta)\dot\gamma}
+ \bar{\omega}^{(\dot\alpha\dot\beta)\dot\gamma(\alpha}\varepsilon^{\beta)\gamma}\big] \,,\label{Tra2} \\
&& \delta h_3^{-2(\alpha\beta)(\dot\alpha\dot\beta)} =\partial^{++}\rho^{-4(\alpha\beta)(\dot\alpha\dot\beta)}_3
-i\,\partial_{\gamma\dot\gamma}\,\rho_{2}^{-2(\alpha\beta)(\dot\alpha\dot\beta)\gamma\dot\gamma} \nn
&& \qquad\qquad\qquad\quad +\,2i\big[\omega_2^{-2(\alpha\beta)(\dot\alpha\dot\beta)} -
\bar\omega_2^{-2(\alpha\beta)(\dot\alpha\dot\beta)}\big]. \label{Tra3}
\eea
Using eqs.\eqref{Tra1}, one can impose the gauges
\bea
&& h_0^{++(\alpha\beta)(\dot\alpha\dot\beta)} = 0 \;\Rightarrow\;  \rho^{(\alpha\beta)(\dot\alpha\dot\beta)} = \rho^{(\alpha\beta)(\dot\alpha\dot\beta)}_{(0)}\,, \nn
&& h_1^{(\alpha\beta)(\dot\alpha\dot\beta)} = 0\,, \;\Rightarrow\; \rho^{-2(\alpha\beta)(\dot\alpha\dot\beta)}_1 =0\,, \,{\omega}_{(0)}^{(\alpha\beta)(\dot\alpha\dot\beta)} =0\,, \,\Rightarrow \,
{\omega}^{(\alpha\beta)\dot\alpha}_{{(0)}\,\,\,\dot\beta} = \frac12 \,\delta^{\dot\alpha}_{\dot\beta}\, \omega^{(\alpha\beta)}_{(0)}\,. \label{GaugeAAA}
\eea

The analysis of consequences of eqs. \eqref{Tra2} and \eqref{Tra3} requires more effort. First of all, we need the gauge transformations of the component fields
in the other two superfield potentials \eqref{hBas3} and \eqref{hBas2}
\bea
&& 2\delta h_0^{+4\alpha\dot\alpha} = \partial^{++} \chi^{+2\alpha\dot\alpha}\,,\label{DeltaIa} \\
&& 2\delta h_1^{+2\alpha\dot\alpha} = \partial^{++} \chi_1^{\alpha\dot\alpha} + 2i\,\bar{\mu}^{+2\alpha\dot\alpha}, \label{DeltaIb} \\
&& 2\delta \bar{h}_1^{+2\alpha\dot\alpha} = \partial^{++} \bar\chi_1^{\alpha\dot\alpha} - 2i\,{\mu}^{+2\alpha\dot\alpha},\label{DeltaIc} \\
&& 2\delta h_2^{+2\alpha\dot\alpha\gamma\dot\gamma} = \partial^{++} \chi_2^{\alpha\dot\alpha\gamma\dot\gamma} - 4i \big[\partial^{\gamma\dot\gamma}\chi^{++\alpha\dot\alpha}
- \varepsilon^{\beta\gamma}\varepsilon^{\dot\alpha\dot\gamma}\mu^{+2\alpha}_\beta
- \varepsilon^{\alpha\gamma}\varepsilon^{\dot\beta\dot\gamma}\bar\mu^{+2\dot\alpha}_{\dot\beta} \big],  \label{DeltaId}\\
&& 2\delta h_3^{\alpha\dot\alpha} = \partial^{++} \chi_3^{-2\alpha\dot\alpha} -i\partial_{\gamma\dot\gamma}\chi_2^{\alpha\dot\alpha \gamma\dot\gamma} -2i\big( \gamma^{\alpha\dot\alpha}
- \bar{\gamma}^{\alpha\dot\alpha}\big), \label{DeltaI} \\
&& \nn
&& 2\delta h^{+2(\beta\alpha)\dot\alpha}_{1\,\gamma} =-\big[ \partial^{++}\omega^{(\alpha\beta)\dot\alpha}_{\,\gamma} + \Sigma^{+2(\alpha\dot\alpha}\delta^{\beta)}_{\gamma}\big]\,,  \label{DeltaIIa} \\
&& 2\delta h^{+2(\beta\alpha)\dot\alpha}_{2\,\dot\gamma} = - \partial^{++}\omega^{(\alpha\beta)\dot\alpha}_{\,\dot\gamma}\,,\label{DeltaIIb} \\
&& 2\delta h^{(\beta\alpha)\dot\alpha\gamma}_{3}  = - \big[\partial^{++}\omega_1^{-2(\alpha\beta)\dot\alpha\gamma} +
2i\partial^{\gamma\dot\gamma}\omega^{(\alpha\beta)\dot\alpha}_{\dot\gamma} - \bar{\Sigma}_1^{(\alpha\dot\alpha}\varepsilon^{\beta)\gamma}\big], \label{DeltaIIc} \\
&&  2\delta h^{(\beta\alpha)\dot\alpha\dot\gamma}_{4} = -\big[\partial^{++}\omega_2^{-2(\alpha\beta)\dot\alpha\dot\gamma}
+ 2i\partial^{\gamma\dot\gamma}\omega^{(\beta\alpha)\dot\alpha}_{\gamma}
-\frac12 \Sigma_2^{(\alpha\beta)\dot\alpha\dot\gamma}\big],  \label{DeltaII}
\eea
where
\bea
&&\Sigma^{++\alpha\dot\alpha} := \chi^{++\alpha\dot\alpha} + 8i \psi^{++\alpha\dot\alpha}\,, \;
\Sigma^{\alpha\dot\alpha}_1 := \chi_1^{\alpha\dot\alpha} + 8i \,\psi_1^{\alpha\dot\alpha}\,,\; \bar\Sigma^{\alpha\dot\alpha}_1 = \bar\chi_1^{\alpha\dot\alpha} - 8i \,\bar\psi_1^{\alpha\dot\alpha}\nn
&&\Sigma^{\alpha\dot\alpha\gamma\dot\gamma}_2 := \chi^{\alpha\dot\alpha\gamma\dot\gamma}_2  + 8i\,\psi_2^{\alpha\dot\alpha\gamma\dot\gamma} +
4i\partial^{\gamma\dot\gamma}\big[ \chi^{+-\alpha\dot\alpha} + 8i\psi^{+-\alpha\dot\alpha}\big]. \label{DefSigm}
\eea

We start with the analysis of \eqref{DeltaIa} - \eqref{DeltaI}. One observes that the fields
$h_0^{+4\alpha\dot\alpha}, h_1^{+2\alpha\dot\alpha}$ and $h_2^{+2\alpha\dot\alpha\gamma\dot\gamma}$ can be
completely gauged away:
\bea
&&h_0^{+4\alpha\dot\alpha} = 0 \; \Rightarrow \; \chi^{+2\alpha\dot\alpha} = \chi^{+2\alpha\dot\alpha}_{(0)}\,, \;h_1^{+2\alpha\dot\alpha} = 0 \; \Rightarrow \;
\chi_1^{\alpha\dot\alpha} = \chi^{\alpha\dot\alpha}_{1 (0)} - 2i\bar\mu^{+1\alpha\dot\alpha}_{(0)}\,, \nn
&& h_2^{+2\alpha\dot\alpha\gamma\dot\gamma} =0 \;\Rightarrow \; \chi_2^{\alpha\dot\alpha\gamma\dot\gamma} = \chi^{\alpha\dot\alpha\gamma\dot\gamma}_{2 (0)}
  + 4i\big[\partial^{\gamma\dot\gamma}\chi^{+-\alpha\dot\alpha}_{(0)}- \varepsilon^{\beta\gamma}\varepsilon^{\dot\alpha\dot\gamma}\mu^{+-\alpha}_{\beta (0)}
- \varepsilon^{\alpha\gamma}\varepsilon^{\dot\beta\dot\gamma}\bar\mu^{+-\dot\alpha}_{\dot\beta (0)}\nn
&& \qquad\qquad\qquad\qquad\qquad\qquad\qquad\qquad  -2\varepsilon^{\alpha\gamma}\varepsilon^{\dot\alpha\dot\gamma} \ell^{+2-2}_{(0)}\big]. \label{h0h1h2}
\eea
In this gauge we also have
\bea
&&\Sigma^{++\alpha\dot\alpha} = \Sigma^{++\alpha\dot\alpha}_{(0)} = \chi^{++\alpha\dot\alpha}_{(0)} + 8i \psi^{++\alpha\dot\alpha}_{(0)}\,, \;
\Sigma^{\alpha\dot\alpha}_1 = \Sigma^{\alpha\dot\alpha}_{1 (0)} = \chi_{1(0)}^{\alpha\dot\alpha} + 8i \,\psi_{1(0)}^{\alpha\dot\alpha}\,,\; \nn
&&\Sigma^{\alpha\dot\alpha\gamma\dot\gamma}_2 = \Sigma^{\alpha\dot\alpha\gamma\dot\gamma}_{2 (0)} + 4i\,\partial^{\gamma\dot\gamma}\Sigma^{+-\alpha\dot\alpha}_{(0)}\,, \quad
\Sigma^{\alpha\dot\alpha\gamma\dot\gamma}_{2 (0)} = \chi^{\alpha\dot\alpha\gamma\dot\gamma}_{2 (0)}  + 8i\,\psi_{2 (0)}^{\alpha\dot\alpha\gamma\dot\gamma},  \label{DefSigmMod}
\eea
where we made use of eq. (\ref{A122}b).

Finally, looking at $\delta h_3^{\alpha\dot\alpha}$ we find that it is possible to impose one more gauge,
\bea
&& h_3^{\alpha\dot\alpha} = h_{3 (0)}^{\alpha\dot\alpha} = D^{\alpha\dot\alpha}\, \Rightarrow \,
2\,\delta D^{\alpha\dot\alpha} = -i\partial_{\gamma\dot\gamma}\chi_{2(0)}^{\alpha\dot\alpha\gamma\dot\gamma}
-2i \big[\gamma^{\alpha\dot\alpha}_{(0)} - \bar\gamma^{\alpha\dot\alpha}_{(0)}\big], \nn
&& 2\,\delta D^{\alpha\dot\alpha} = -i\partial_{\gamma\dot\gamma}\Omega^{\alpha\dot\alpha\gamma\dot\gamma}\,, \quad
\Omega^{\alpha\dot\alpha\gamma\dot\gamma} :=\Sigma^{\alpha\dot\alpha\gamma\dot\gamma}_{2 (0)}\,, \label{DefOmegaB}
\eea
where we used eq. (\ref{A122}d). Below we show that \eqref{DefOmegaB} amounts to the standard Maxwell gauge transformation
for the properly redefined vector field $\tilde{D}^{\alpha\dot\alpha}$.

The next steps towards the eventual WZ gauge are based on the transformations \eqref{DeltaIIa} - \eqref{DeltaII}.  Eqs. \eqref{DeltaIIa}, \eqref{DeltaIIb} imply the possibility to choose the gauge
\bea
&& h^{+2(\beta\alpha)\dot\alpha \gamma} = 0\; \Rightarrow \; \omega^{(\beta\alpha)\dot\alpha \gamma} =
\omega^{(\beta\alpha)\dot\alpha \gamma}_{(0)} + \Sigma_{(0)}^{+-(\alpha\dot\alpha}\varepsilon^{\beta)\gamma}\,, \label{IIa} \\
&& h^{+2(\beta\alpha)\dot\alpha \dot\gamma} = 0 \;\Rightarrow \; \omega^{(\beta\alpha)\dot\alpha \dot\gamma} =
\omega^{(\beta\alpha)\dot\alpha \dot\gamma}_{(0)} = \frac12 \varepsilon^{\dot\gamma\dot\alpha}\,\omega^{(\alpha\beta)}_{(0)}\,, \label{IIb}
\eea
where we used eq. \eqref{GaugeAAA}. Eq. \eqref{DeltaIIc} permits the gauge choice
\bea
&&h_3^{(\alpha\beta)\dot\alpha \gamma} = h_{3 (0)}^{(\alpha\beta)\dot\alpha \gamma} \; \Rightarrow \; \omega_1^{-2(\alpha\beta)\dot\alpha \gamma} = 0\,, \label{IIc1} \\
&& \delta h_{3 (0)}^{(\alpha\beta)\dot\alpha \gamma} = -\frac12 \big(i\,\partial^{\gamma\dot\alpha}\,\omega^{(\alpha\beta)}_{(0)}
+ \varepsilon^{\gamma(\beta}\,\bar\Sigma_{1 (0)}^{\alpha)\dot\alpha}\big). \label{IIc2}
\eea
It is clear from \eqref{IIc2} that the further gauge-fixing is possible,
\bea
&& h_{3 (0)}^{(\alpha\beta)\dot\alpha \gamma} = T^{(\alpha\beta\gamma)\dot\alpha}\; \Rightarrow \;
\bar\Sigma^{\beta\dot\alpha}_{1(0)} = -\frac{2i}{3}\,\partial^{\dot\alpha}_\gamma \,\omega^{(\beta\gamma)}_{(0)}\,, \label{IIc3} \\
&& \delta  T^{(\alpha\beta\gamma)\dot\alpha} = i\partial_{\dot\gamma}^{(\gamma}\, \omega^{\beta\alpha)\dot\alpha\dot\gamma} =
-\frac{i}{2}\,\partial^{\dot\alpha(\gamma} \omega^{\beta\alpha)}_{(0)}\,, \label{IIc4}
\eea
where we used the notations introduced in \eqref{A23}. The field $T^{(\alpha\beta\gamma)\dot\alpha}$ is just the ''hook''  gauge field. It involves ${\bf 16} - {\bf 6} = {\bf 10}$
essential off-shell degrees of freedom.

It remains to reveal the consequences of  the gauge freedom \eqref{DeltaII}. First, it allows for the gauge choice
\bea
&& h^{(\beta\alpha)\dot\alpha\dot\gamma}_{4} = h^{(\beta\alpha)\dot\alpha\dot\gamma}_{4(0)}\, \; \Rightarrow \;
\omega^{-2(\alpha\beta)\dot\alpha\dot\gamma}_2 = 2i\,\partial^{\dot\gamma(\beta}\Sigma^{-2\alpha)\dot\alpha}_{(0)}\,,   \label{ExprOmega} \\
&& \delta h^{(\beta\alpha)\dot\alpha\dot\gamma}_{4(0)} = \frac14\Omega^{(\alpha\dot\alpha\beta)\dot\gamma}
- i\partial^{\gamma\dot\gamma}\omega^{(\alpha\beta)\dot\alpha}_{\gamma(0)}\,. \label{II1}
\eea
Now, let us decompose
\bea
&&\Omega^{\alpha\dot\alpha \beta\dot\beta} = \Omega^{(\alpha\beta)(\dot\alpha\dot\beta)} + \varepsilon^{\alpha\beta}\Omega^{(\dot\alpha\dot\beta)}
-\varepsilon^{\dot\alpha\dot\beta}\Omega^{(\alpha\beta)} -i\varepsilon^{\alpha\beta}\varepsilon^{\dot\alpha\dot\beta} \Omega\,, \label{ExpOm} \\
&& \overline{(\Omega^{(\alpha\beta)(\dot\alpha\dot\beta)})} = - \Omega^{(\alpha\beta)(\dot\alpha\dot\beta)}\,, \; \overline{(\Omega^{(\alpha\gamma)})} = \Omega^{(\dot\alpha\dot\gamma)}\,, \;
\bar{\Omega} = \Omega\,.\label{ConjOm}
\eea
Then, we can impose the further gauge
\bea
&& h^{(\beta\alpha)\dot\alpha\dot\gamma}_{4 (0)} = h^{(\beta\alpha)(\dot\alpha\dot\gamma)}_{4 (0)} \; \Rightarrow \nn
&& \Omega^{(\alpha\beta)} = -2i\partial_{\gamma\dot\alpha}\, \omega^{(\alpha\beta)\dot\alpha\gamma}_{(0)}\,, \quad
\Omega^{(\dot\alpha\dot\beta)} = 2i\partial_{\alpha\dot\gamma}\, \bar\omega^{(\dot\alpha\dot\beta)\alpha\dot\gamma}_{(0)}\,, \label{OmGau1}
\eea
and, using the complex conjugation rules \eqref{ConjOm}, make $h^{(\beta\alpha)(\dot\alpha\dot\gamma)}_{4 (0)}$ real
\bea
&& h^{(\beta\alpha)(\dot\alpha\dot\gamma)}_{4 (0)} = \bar{h}^{(\beta\alpha)(\dot\alpha\dot\gamma)}_{4 (0)} = P^{(\beta\alpha)(\dot\alpha\dot\gamma)}\, \;\Rightarrow  \nn
&& \Omega^{(\alpha\beta)(\dot\alpha\dot\beta)} = -i\big[ \partial_\gamma^{\dot\gamma}\, \omega^{(\alpha\beta)\dot\alpha\gamma}_{(0)} +
\partial_\gamma^{\dot\alpha}\, \omega^{(\alpha\beta)\dot\gamma\gamma}_{(0)} + \partial_{\dot\rho}^\alpha\, \bar\omega^{(\dot\alpha\dot\gamma)\beta\dot\rho}_{(0)}
+\partial_{\dot\rho}^\alpha\, \bar\omega^{(\dot\alpha\dot\gamma)\beta\dot\rho}_{(0)}\big]\,. \label{Real}
\eea
At this step, we are left with the following gauge transformation of real $P^{(\beta\alpha)(\dot\alpha\dot\gamma)}$
\bea
\delta P^{(\beta\alpha)(\dot\alpha\dot\gamma)} = \frac{i}4\big[\partial_\gamma^{\dot\gamma}\, \omega^{(\alpha\beta)\dot\alpha\gamma}_{(0)}
+\partial_\gamma^{\dot\alpha}\, \omega^{(\alpha\beta)\dot\gamma\gamma}_{(0)} - \partial_{\dot\rho}^\alpha\, \bar\omega^{(\dot\alpha\dot\gamma)\beta\dot\rho}_{(0)}
- \partial_{\dot\rho}^\beta\, \bar\omega^{(\dot\alpha\dot\gamma)\alpha\dot\rho}_{(0)}\big]\,. \label{Pprom}
\eea

Now we should be back to the discussion of the structure of gauge potential $h^{++(\alpha\beta)(\dot\alpha\dot\beta)}$. The gauge transformation \eqref{Tra2} implies that
the whole harmonic-dependent part of $h^{(\alpha\beta)(\dot\alpha\dot\beta)\gamma\dot\gamma}$ can be gauged away, in agreement with the general structure \eqref{A23},
\bea
&& h^{(\alpha\beta)(\dot\alpha\dot\beta)\gamma\dot\gamma}_2  = h^{(\alpha\beta)(\dot\alpha\dot\beta)\gamma\dot\gamma}_{2(0)} \; \Rightarrow \nn
&& \rho^{-2(\alpha\beta)(\dot\alpha\dot\beta)\gamma\dot\gamma} =
- 2i\big[\varepsilon^{\beta\gamma}\Sigma^{-2\alpha(\dot\alpha}_{(0)}\varepsilon^{\dot\beta)\dot\gamma} + (\alpha \leftrightarrow \beta)\big],  \label{rho-2} \\
&& \delta h^{(\alpha\beta)(\dot\alpha\dot\beta)\gamma\dot\gamma}_{2(0)} = -4i\partial^{\gamma\dot\gamma}\rho^{(\alpha\beta)(\dot\alpha\dot\beta)}_{(0)}
+ 4i\big[\omega_{(0)}^{(\alpha\beta)\gamma(\dot\alpha}\varepsilon^{\dot\beta)\dot\gamma} + \bar\omega_{(0)}^{(\dot\alpha\dot\beta)\dot\gamma(\alpha}\varepsilon^{\beta)\gamma}\big].
\label{Varh2}
\eea
From \eqref{Varh2} we also observe that all parts of $h^{(\alpha\beta)(\dot\alpha\dot\beta)\gamma\dot\gamma}_{2(0)}$, excepting the totally symmetric one,
can be gauged away, leading to the following gauge transformation for the conformal spin $3$ gauge field (in the notation of \eqref{A23}):
\bea
\delta \Phi^{(\alpha\beta\gamma)(\dot\alpha\dot\beta\dot\gamma)}  = \partial^{(\gamma\dot\gamma} \rho_{(0)}^{(\alpha\beta)(\dot\alpha\dot\beta))} \label{Gauge3}
\eea
(where total symmetrizations with respect to dotted and undotted indices are assumed). Thus we are left with ${\bf 16} - {\bf 9} = {\bf 7}$ off-shell degrees of freedom in
$\Phi^{(\alpha\beta\gamma)(\dot\alpha\dot\beta\dot\gamma)}$.

Preserving the gauge for $h^{(\alpha\beta)(\dot\alpha\dot\beta)\gamma\dot\gamma}_{2(0)}$ just mentioned yields the following restrictions on the gauge $\omega$-parameters
\bea
&& \omega_{(0)}^{(\alpha\beta\gamma)\dot\beta} = -\frac23\,\partial_{\dot\alpha}^{(\gamma}\;\rho_{(0)}^{\alpha\beta) (\dot\alpha\dot\beta)}\,, \quad
\bar\omega_{(0)}^{(\dot\alpha\dot\beta\dot\gamma)\beta} = -\frac23\,\partial_{\alpha}^{(\dot\gamma}\;\rho_{(0)}^{\dot\alpha\dot\beta) (\alpha\beta)}\,, \label{TripOm} \\
&& \omega_{(0)}^{\beta\dot\beta} + \bar{\omega}_{(0)}^{\beta\dot\beta} = -\frac23 \partial_{\alpha\dot\alpha}\rho_{(0)}^{(\alpha\beta)(\dot\alpha\dot\beta)}\,, \quad
\omega_{(0)}^{\beta\dot\beta} := \omega_{(0) \;\; \alpha}^{(\alpha\beta)\dot\beta}\,, \;\bar\omega_{(0)}^{\beta\dot\beta}
:= \bar{\omega}_{\dot\alpha\,(0)}^{(\dot\beta\dot\alpha)\beta}\,. \label{DoubOm}
\eea
Defining new independent gauge parameter
\bea
p^{\beta\dot\beta} := i\big(\omega_{(0)}^{\beta\dot\beta} - \bar{\omega}_{(0)}^{\beta\dot\beta}\big)
\eea
and substituting \eqref{TripOm}, \eqref{DoubOm} in \eqref{Pprom}, we obtain
\bea
\delta P^{(\beta\alpha)(\dot\alpha\dot\gamma)}  = -\frac{1}{3} \partial^{(\beta(\dot\gamma} p^{\alpha)\dot\alpha)}\,. \label{Pgfield}
\eea
So $P^{(\beta\alpha)(\dot\alpha\dot\gamma)}$ is ``conformal graviton'': it carries ${\bf 9} - {\bf 4} = {\bf 5}$ off-shell degrees of freedom.

At this step let us come back to the transformation law \eqref{DefOmegaB}. One can check that
\bea
\delta D^{\alpha\dot\alpha} = -\frac{1}{2}\partial^{\alpha\dot\alpha}\big[\Omega + \frac{2}{9}\, \partial_{\beta\dot\beta}\partial_{\gamma\dot\gamma}\rho_{(0)}^{(\beta\gamma)(\dot\beta\dot\gamma)}\big]
-\frac29 \Box \partial_{\gamma\dot\gamma}\rho_{(0)}^{(\alpha\gamma)(\dot\alpha\dot\gamma)}
\eea
In order to pass to the gauge field with the standard gradient transformation law, let us define
\be
Z^{\alpha\dot\alpha} := \partial_{\beta\dot\beta}\partial_{\gamma\dot\gamma} \Phi^{(\alpha\beta\gamma)(\dot\alpha\dot\beta\dot\gamma)}. \label{DefZ}
\ee
Under the spin ${\bf s}=3$ gauge transformations:
\be
\delta Z^{\alpha\dot\alpha} = \frac{1}{27} \big[\partial^{\alpha\dot\alpha} (\partial_{\beta\dot\beta}\partial_{\gamma\dot\gamma}\rho_{(0)}^{(\beta\gamma)(\dot\beta\dot\gamma)})
+ 5\Box \partial_{\gamma\dot\gamma}\rho_{(0)}^{(\alpha\gamma)(\dot\alpha\dot\gamma)}\big].
\ee
Then it is easy to check that
\be
\tilde{D}^{\alpha\dot\alpha} =    {D}^{\alpha\dot\alpha} +\frac{6}{5} Z^{\alpha\dot\alpha} \label{DefDtilde}
\ee
has the correct spin ${\bf s}=1$ gauge transformation with the properly redefined gauge parameter
\bea
\delta \tilde{D}^{\alpha\dot\alpha} = - \frac{1}{2}\partial^{\alpha\dot\alpha}\tilde{\Omega}\,, \quad \tilde{\Omega} :=
\Omega  + \frac{2}{15}\,(\partial_{\beta\dot\beta}\partial_{\gamma\dot\gamma}\rho_{(0)}^{(\beta\gamma)(\dot\beta\dot\gamma)}).
\label{app D alpha}
\eea

The final step is to reveal the role of the gauge transformation \eqref{Tra3}. It admits imposing the gauge
\bea
&&h_3^{-2(\alpha\beta)(\dot\alpha\dot\beta)} = h_{3 (0)}^{-2(\alpha\beta)(\dot\alpha\dot\beta)} \quad \Rightarrow  \quad
\rho_3^{-4(\alpha\beta)(\dot\alpha\dot\beta)} = 0\,, \nn
&& \delta h_{3 (0)}^{-2(\alpha\beta)(\dot\alpha\dot\beta)} = -12 \partial^{(\dot\alpha (\beta} \Sigma_{(0)}^{-2\alpha)\dot\beta)}, \label{Triplet}
\eea
where we used the expressions \eqref{rho-2} for $\rho^{-2(\alpha\beta)(\dot\alpha\dot\beta)\gamma\dot\gamma}$ and \eqref{ExprOmega} for $\omega_2^{-2(\alpha\beta)(\dot\alpha\dot\beta)}$.
The triplet gauge field $h_{3 (0)}^{-2(\alpha\beta)(\dot\alpha\dot\beta)} := V^{(\alpha\beta)(\dot\alpha\dot\beta)(ij)}u^-_iu^-_j$
carries ${\bf 27} - {\bf 12}  =  {\bf 15}$ off-shell degrees of freedom.

To summarize,  the whole set of bosonic gauge fields carries just total of ${\bf 15} + {\bf 3} + {\bf 7} + {\bf 5} + {\bf 10} = {\bf 40}$ essential off-shell degrees of freedom. All essential bosonic
fields are gauge, in contradistinction to the lower spin (${\bf s} =1$ and ${\bf s}=2$) multiplets containing also auxiliary fields in the bosonic sector.

\subsection{Fermionic sector}
The analysis of the component structure of the conformal ${\cal N}=2$, ${\bf s}=3$ gauge supermultiplet in the fermionic sector basically follows
the same pattern as in the bosonic one, so we will concentrate on the final answers rather than on the intermediate computations.

We will need the following fermionic terms in the general analytic gauge parameters:
\bea
&&\lambda^{+4} \;\Rightarrow \;  (\bar\theta^+)^2\theta^{+\alpha} \ell^{+}_{\alpha} -  (\theta^+)^2 \bar\theta^{+ \dot\alpha}\bar{\ell}^{+}_{\dot\alpha}\,, \nn
&& \lambda^{\alpha+\dot\beta+} \;\Rightarrow \; \theta^{+\gamma} p^{+\alpha\dot\beta}_{\;\;\gamma} + \bar\theta^{+\dot\gamma}\bar{p}^{+\alpha\dot\beta}_{\;\;\dot\gamma}, \nn
&& \lambda^{++\alpha\dot\beta} \;\Rightarrow \;\theta^{+\gamma} k^{+\alpha\dot\beta}_{\;\;\gamma} - \bar\theta^{+\dot\gamma}\bar{k}^{+\alpha\dot\beta}_{\;\;\dot\gamma}, \nn
&& \lambda^{(\beta+\alpha)\dot\alpha}\;\Rightarrow \; \tau^{+(\beta\alpha)\dot\alpha} + (\theta^+)^2 \tau_1^{-(\beta\alpha)\dot\alpha}
+ (\bar\theta^+)^2 \hat{\tau}_1^{-(\beta\alpha)\dot\alpha} + \theta^{+\rho}\bar\theta^{+\dot\rho} \tau^{-(\beta\alpha)\dot\alpha}_{1\;\;\rho\dot\rho} \nn
&& \quad\quad\quad\quad\quad\quad +\, (\theta^+)^4 \tau_2^{-3(\beta\alpha)\dot\alpha}\,, \nn
&& \lambda^{(\beta\alpha)(\dot\alpha\dot\beta)}\;\Rightarrow \;\theta^{+\rho}\lambda^{-(\alpha\beta)(\dot\alpha\dot\beta)}_\rho
 - \bar\theta^{+\dot\rho}\bar\lambda^{-(\alpha\beta)(\dot\alpha\dot\beta)}_{\dot\rho} + (\bar\theta^+)^2\theta^{+\rho}\lambda^{-3(\alpha\beta)(\dot\alpha\dot\beta)}_\rho \nn
&&  \quad\quad\quad\quad\quad\quad -\, (\theta^+)^2\bar\theta^{+\dot\rho}\bar\lambda^{-3(\alpha\beta)(\dot\alpha\dot\beta)}_{\dot\rho}\,. \label{FermParam}
\eea

We can impose the same preliminary gauge conditions as in the bosonic case, in particular, fully gauge away the set of the gauge
potentials $h^{(+6)}$,  $h^{++\hat{\alpha}+++}$, $\hat{h}^{+4}$, $\hat{h}^{++}$,
$\hat{h}^{+3\hat\alpha}$, $h^{++\alpha+\dot\beta+}$ also in the fermionic sector. The whole fermionic part of $h^{+4 \alpha\dot\alpha}$ can also be gauged away.
These gaugings imply certain conditions on the residual
component gauge parameters. In particular, the conditions
\bea
h^{++\alpha+\dot\beta+}_F = 0\,, \qquad h^{++\alpha+\dot\beta+}_F = 0 \eea give rise to \bea p^{+\alpha\dot\beta \gamma} = p^{+\alpha\dot\beta \gamma}_{(0)} =
p^{\alpha\dot\beta \gamma \,i} u^+_i\,, \quad k^{+\alpha\dot\beta \gamma} = k^{+\alpha\dot\beta \gamma}_{(0)} = k^{\alpha\dot\beta \gamma \,i} u^+_i  \;\; ({\rm and\, c.c.}).
\eea
The rest of constraints can also be
straightforwardly solved. The corresponding reduced gauge parameters  are of no interest for our purposes.

We end up with the symmetrized vielbeins $h^{+3 (\alpha\beta)\dot\alpha}_F$ and $h^{+2 (\alpha\beta)(\dot\alpha\dot\beta)}_F$:
\bea
&& h^{+3 (\alpha\beta)\dot\alpha}_F = h^{+3 (\alpha\beta)\dot\alpha}_1 + (\theta^+)^2 h^{+ (\alpha\beta)\dot\alpha}_2 + (\bar\theta^+)^2 \hat{h}^{+ (\alpha\beta)\dot\alpha}_2 +
\theta^+_\rho\bar\theta_{\dot\rho}h^{+ (\alpha\beta)\dot\alpha\rho\dot\rho}_3 \nn
&&\quad\quad\quad\quad\quad\ +\, (\theta^+)^4 h^{-(\beta\alpha)\dot\alpha}_4\,, \label{h+3F} \\
&&h^{+2 (\alpha\beta)(\dot\alpha\dot\beta)}_F = \theta^{+\gamma}h^{+(\alpha\beta)(\dot\alpha\dot\beta)}_{\gamma}
- \bar\theta^{+\dot\rho}\bar{h}^{+(\dot\alpha\dot\beta)(\alpha\beta)}_{\dot\gamma} +(\bar\theta^+)^2 \theta^{+\gamma}\hat{h}^{-(\alpha\beta)(\dot\alpha\dot\beta)}_{\gamma} \nn
&&\quad\quad\quad\quad\quad- (\theta^+)^2 \bar\theta^{+\dot\gamma}\hat{\bar{h}}^{-(\dot\alpha\dot\beta)(\alpha\beta)}_{\dot\gamma}\,. \label{h+2F}
\eea

Firstly we elaborate on \eqref{h+3F}. From the gauge conditions,
\bea
h^{+3 (\alpha\beta)\dot\alpha}_1 = 0\,,\;h^{+ (\alpha\beta)\dot\alpha}_2= 0\,, \;\hat{h}^{+ (\alpha\beta)\dot\alpha}_2 = 0\,, \;
h^{+ (\alpha\beta)\dot\alpha\rho\dot\rho}_3  = 0
\eea
we find the constraints on the residual gauge parameters
\bea
&& \tau^{+(\beta\alpha)\dot\alpha}= \tau^{+(\beta\alpha)\dot\alpha}_{(0)} = \tau^{(\beta\alpha)\dot\alpha i}u^+_i\,, \;\hat{\tau}_1^{-(\beta\alpha)\dot\alpha} = 0\,, \nn
&&\tau_1^{-(\beta\alpha)\dot\alpha} =
\frac12 \varepsilon^{\rho(\alpha} R^{-\beta)\dot\alpha}_\rho\,, \quad R^{-\beta\dot\alpha}_\rho := k^{-\beta\dot\alpha}_{\rho (0)} + 8i p^{-\beta\dot\alpha}_{\rho (0)}
= R^{\beta\dot\alpha i}_\rho u^-_i\,, \nn
&&\tau^{-(\beta\alpha)\dot\alpha}_{1\,\rho\dot\rho} = 4i\partial_{\rho\dot\rho}\tau^{-(\beta\alpha)\dot\alpha}_{(0)}
- \delta^{(\beta}_\rho\,\bar{R}^{-\alpha)\dot\alpha}_{\dot\rho}. \label{Constrh+3F}
\eea
For the surviving part of the gauge potential $h^{-(\beta\alpha)\dot\alpha}_4$, that is $h^{-(\beta\alpha)\dot\alpha}_{4(0)}:= \chi^{-(\beta\alpha)\dot\alpha} = \chi^{(\beta\alpha)\dot\alpha i}u^-_i$
(recall ``master gauge''  \eqref{A23}), at this step we obtain the following gauge transformation
\bea
\delta \chi^{-(\beta\alpha)\dot\alpha}  =  i\partial^{(\beta\dot\rho} \bar{R}^{-\alpha)\dot\alpha}_{\dot\rho} + 2\Box\, \tau^{-(\alpha\beta)\dot\alpha}_{(0)}\,. \label{deltchi1}
\eea

In order to find the final form of this gauge transformation, we need first to work out \eqref{h+2F}.  The choice of the gauge
\bea
&& h^{+(\alpha\beta)(\dot\alpha\dot\beta)}_{\gamma} = 0 \;\;  ({\rm and\,\,c.c.}), \qquad \hat{h}^{-(\alpha\beta)(\dot\alpha\dot\beta)}_{\gamma}
=\hat{h}^{-(\alpha\beta)(\dot\alpha\dot\beta)}_{\gamma (0)}, \nn
\eea
implies the relations
\bea
&& \lambda^{-(\alpha\beta)(\dot\alpha\dot\beta)}_\rho = -4i \delta^{(\alpha}_{\rho} \bar{\tau}^{-(\dot\alpha\dot\beta)\beta)}\,, \quad
\bar\lambda^{-(\alpha\beta)(\dot\alpha\dot\beta)}_{\dot\rho} = -4i \delta^{(\dot\alpha}_{\dot\rho} {\tau}^{-(\alpha\beta)\dot\beta)}\,, \nn
&& \lambda^{-3(\alpha\beta)(\dot\alpha\dot\beta)}_\rho = 0\, \quad ({\rm and\,\,c.c.}), \nonumber
\eea
and the following gauge transformation of $\hat{h}^{-(\alpha\beta)(\dot\alpha\dot\beta)}_{\gamma (0)} := \psi^{(\alpha\beta)(\dot\alpha\dot\beta) i}_{\gamma}u^-_i$
\bea
\delta \psi^{-(\alpha\beta)(\dot\alpha\dot\beta)}_\rho = -8\partial^{(\dot\alpha}_\rho\, \tau_{(0)}^{-(\beta\alpha)\dot\beta} +
4i\,\delta^{(\alpha}_\rho\, \bar\tau_1^{-(\dot\beta\dot\alpha)\beta} + 2i\,\tau^{-(\beta\alpha)(\dot\alpha\dot\beta)}_{1\rho}.
\eea
Using the expressions \eqref{Constrh+3F}, this variation can be rewritten as
\bea
\delta \psi^{-(\alpha\beta)(\dot\alpha\dot\beta)}_\rho = -16\, \partial_\rho^{(\dot\alpha}\, \tau_{(0)}^{-(\alpha\beta)\dot\beta)}  +
4i\,\delta^{(\alpha}_\rho\,\varepsilon^{\dot\gamma(\dot\beta}\bar{R}^{-\beta)\dot\alpha)}_{\dot\gamma}\,. \nonumber
\eea

This transformation law implies
\bea
\delta \psi^{-(\alpha\beta\rho)(\dot\alpha\dot\beta)}= -16\, \partial^{(\dot\alpha(\rho}\, \tau_{(0)}^{-\alpha\beta)\dot\beta)}\,.  \label{52psi}
\eea
while the rest of components in $\psi^{-(\alpha\beta)(\dot\alpha\dot\beta)}_\rho$ can be gauged away
\bea
\psi^{-(\alpha\beta)(\dot\alpha\dot\beta)}_\alpha = 0\; \Rightarrow \;\bar{R}^{-(\dot\alpha\dot\beta)\alpha}
= \frac{8i}{3}\,\partial^{(\dot\alpha}_\rho \tau^{-(\beta\rho) \dot\beta)}_{(0)}\, \label{Rtrace}
\eea
The transformation law \eqref{52psi} means that the complex field $\psi^{-(\alpha\beta\rho)(\dot\alpha\dot\beta)} = \psi^{(\alpha\beta\rho)(\dot\alpha\dot\beta)i}u^-_i$
encompasses the $SU(2)$ doublet of the spin ${\bf 5/2}$  gauge fields with ${\bf 48} - {\bf 24} = {\bf 24}$ essential degrees of freedom off shell.

As the next step, one can define
\bea
&& w^{-(\alpha\beta)\dot\beta} := \partial_{\gamma\dot\alpha}\psi^{-(\alpha\beta\gamma)(\dot\beta\dot\alpha)}\,, \label{Defw} \\
&& \delta w^{-(\alpha\beta)\dot\beta} = -\frac{4}{3}\,\big[ \partial^{(\alpha \dot\beta}\, b^{-\beta)} + 5 \,\Box\,\tau_{(0)}^{-(\alpha\beta)\dot\beta} \big],
\quad b^{-\beta} := \partial_{\gamma \dot\gamma} \tau^{-(\beta\gamma)\dot\gamma}_{(0)}\,.  \label{Tranb}
\eea
Now, coming back to eq. \eqref{deltchi1} and redefining
\bea
\hat{\chi}^{-(\beta\alpha)\dot\alpha} := {\chi}^{-(\beta\alpha)\dot\alpha} + \frac{1}{10}w^{-(\alpha\beta)\dot\alpha}\,,
\eea
we find that $\hat{\chi}^{-(\beta\alpha)\dot\alpha}$ is transformed as
\bea
\delta\hat{\chi}^{-(\beta\alpha)\dot\alpha} = \frac{i}{2} \partial^{(\alpha\dot\alpha}\, c^{-\beta)}\,, \quad c^{-\beta} := \bar{R}^{-\beta \dot\gamma}_{\dot\gamma}
- \frac{12i}{5}\, b^{-\beta}\,. \label{32chi}
\eea
Since $c^{-\beta} = c^{\beta i} u^-_i$ involves just 8 independent real gauge parameters $c^{\beta i}$, the field
$\hat{\chi}^{-(\beta\alpha)\dot\alpha} = \hat{\chi}^{(\beta\alpha)\dot\alpha\,i}u^-_i$ describes $SU(2)$ doublet of the spin ${\bf 3/2}$  gauge fields
with ${\bf 24} - {\bf 8} = {\bf 16}$ essential degrees of freedom off shell.

Thus in the fermionic sector we end up with the spin ${\bf 5/2}$ and spin ${\bf 3/2}$ conformal gauge fields with the total of ${\bf 40}$ off-shell essential
degrees of freedom. This number precisely matches the number of essential degrees of freedom in the bosonic sector, and it remains
to show that the last gauge potential $h^{++}$ does not contribute any degree of freedom in the full WZ gauge.


\subsection{$h^{++}$ gauge potential}

Using the $\mathcal{D}^{++}\lambda$ gauge freedom, one can fix the gauge:
\begin{equation} \lb{WZappr}
    \begin{split}
        h^{++} =& \theta^{+\alpha}\bar{\theta}^{+\dot{\alpha}} A_{\alpha\dot{\alpha}}
        + (\theta^+)^2 {\phi}
        + (\bar{\theta}^+)^2  \bar\phi
        \\
        &
        +
        4 (\bar{\theta}^+)^2 \theta^{+\alpha} \xi^i_\alpha u^-_i
        +
        4 (\theta^+)^2 \bar{\theta}^{+\dot{\alpha}} \bar{\xi}_{\dot{\alpha }}^{i} u^-_i
        +
        (\theta^+)^4 D^{ij}u^-_i u^-_j.
    \end{split}
\end{equation}
This is the standard WZ gauge for the spin $\mathbf{1}$ multiplet. However, the full $h^{++}$ gauge transformation law \eqref{eq: spin 3 c}
contain additional terms which can be used to gauge away all fields in \eqref{WZappr}. In the process, only those
terms in the $\theta$ and $u$-expansions of \eqref{eq: spin 3 c} are of interest, which have the form \eqref{WZappr}.
All other terms can be absorbed into the redefinition of the gauge parameters which were used to ensure \eqref{WZappr}.

After some straightforward algebra, using eqs. \eqref{A92} - \eqref{A122}, we obtain (up to some $U(1)$ gauge transformation of $A^{\gamma\dot{\gamma}}$):
\bea
&&\delta \phi = -i\partial_{\alpha\dot\alpha}\bar{\sigma}_{1(0)}^{\alpha\dot\alpha}, \qquad \delta \bar\phi = i\partial_{\alpha\dot\alpha}{\sigma}_{1(0)}^{\alpha\dot\alpha}\,,
\label{phizero} \\
&& \delta  A^{\gamma\dot{\gamma}} = \frac12 \partial_{\alpha\dot\alpha}\Big[\Sigma_-^{(\alpha\gamma)(\dot\alpha\dot\gamma)}
- \frac{4i}{9}\,\Box \,\rho_{(0)}^{(\alpha\gamma)(\dot\alpha\dot\gamma)}\Big], \label{Azero} \\
&& \Sigma_-^{(\alpha\gamma)(\dot\alpha\dot\gamma)} :=
\chi_{2(0)}^{(\alpha\gamma)(\dot\alpha\dot\gamma)} - 8i\psi_{2(0)}^{(\alpha\gamma)(\dot\alpha\dot\gamma)}\,,\nn
&&\delta D^{-2} = -4i\partial_{\alpha\dot\alpha}\Big[ \ell^{-2\alpha\dot\alpha}_{(0)} + \frac{15i}{16}\Box \chi^{-2\alpha\dot\alpha}_{(0)}
- \frac{3}{2}\,\Box \psi^{-2\alpha\dot\alpha}_{(0)}\Big], \label{Dzero}\\
&& \delta\xi^-_\rho = \partial^{--}\ell^+_{\rho (0)} +  \ldots\,,
\label{xizero}
\eea
where ellipses stand for some terms with $x$-derivatives.

Thus we see that all bosonic fields in \eqref{WZappr} are shifted by divergences of the appropriate vector parameters. Since the parameters
in \eqref{phizero}, \eqref{Azero} and \eqref{Dzero}
are unconstrained and independent (they are new compared to those which were used earlier in fixing various WZ gauges),
these parameters are capable to gauge away all bosonic fields  in \eqref{WZappr}. The fermionic field $\xi^i_\rho$ is shifted by an unconstrained parameter $\ell^i_\rho$ (defined in \eqref{FermParam}),
so one can also choose $\xi^i_\alpha = 0$. As a result, one can fix the gauge $h^{++}=0$.

Note that the gauge transformations of the form $\delta h_{\alpha(s)\dot{\alpha}(s)} = \partial^{\beta\dot{\beta}}\lambda_{\beta\alpha(s)\dot{\beta}\dot{\alpha}(s)}$
are frequently  encountered  in the free theory of massless higher spins \cite{Fronsdal:1978rb, Fang:1978wz} (see also \cite{BK, Hutomo:2017nce} for review).

\medskip

\section{On residual parameters and reparametrization freedom of free hypermultiplet}
\label{app: residual}

Note that among the parameters of the spin $\mathbf{3}$ transformations
 \eqref{eq: s=3 gauge transformations} there are special parameters which do not appear in transformations of the gauge potentials $h^{++MN}$ and $h^{++}$.

For instance, the transformation with parameter
\begin{equation}
    \lambda^{(+4)} = (\theta^+)^4 e(x)
\end{equation}
acts only on the hypermultiplet\footnote{One can define even more general off-shell symmetry transformation of the hypermultiplet:
    $$
    \delta_{c_{(ab)}} q^{+a}= (\theta^+)^{4} c^a_{\;b}(x) (\partial^{--})^2q^{+b}
    $$
    with an arbitrary symmetric matrix $c_{(ab)}(x)$.}:
\begin{equation}
    \delta_{e(x)} q^{+a} = (\theta^+)^4 e(x) \partial^{--} \partial^{--} J q^{+a}.
\end{equation}
This is the exact off-shell symmetry of the free part of the hypermultiplet action, $\delta_{e(x)} S_{free} = 0$.
This transformation and other symmetries of similar kind mix the auxiliary fields of the hypermultiplet,
\begin{equation}
    q^{+a} = \dots + K^{(ijk)a} u^+_i u^+_j u^-_k
    +
    \dots
    +
    (\theta^+)^4 F^{(ijk)a} u^-_i u^-_j u^-_k
    +
    \dots\,,
\end{equation}
\begin{equation}
    \delta _{e(x)} K^{(ijk)a}(x) = e(x) J F^{(ijk)a}(x),
\end{equation}
and seemingly have no impact on the structure of superconformal vertices.

\medskip

\section{Superconformal transformations of ${\cal N}=2$ superspace derivatives and gauge potentials}
\label{app: N2SCDeriv}

For checking the transformation properties of various analytic vielbeins under the rigid ${\cal N}=2$ superconformal group, it is useful
to be aware of the superconformal transformation laws of the partial derivatives with respect to the co-ordinates of the analytic harmonic
superspace. It suffices to know such laws for rigid ${\cal N}=2$ supersymmetry and special conformal transformations, since the whole superconformal group is the closure
of these two.

Using the infinitesimal superconformal coordinate shifts \eqref{eq:superconformal symmetry}, we obtain\\

\noindent{\bf Supersymmetry:}
\begin{equation}
\delta_{\epsilon} \partial^-_{\alpha} = 4i \bar\epsilon^{-\dot\beta} \partial_{\alpha\dot\beta}\,, \; \delta_{\epsilon} \bar\partial^-_{\dot\alpha}
= - 4i \epsilon^{- \beta} \partial_{\beta\dot\alpha}\,, \; \delta_{\epsilon}\partial^{--}
= -\epsilon^{-\alpha}\partial^-_{\alpha} - \bar{\epsilon}^{-\dot\alpha}\bar\partial{-}_{\dot\alpha}\,, \;\delta_{\epsilon}\partial_{\alpha\dot\alpha} = 0\,; \label{SUSYDer}
\end{equation}

\noindent{\bf Special conformal transformations:}
\begin{eqnarray}
&& \delta_k \partial_{\alpha\dot\alpha} =- (k_{\alpha\dot\beta} x^{\gamma\dot\beta}\partial_{\gamma\dot\alpha} + k_{\gamma\dot\alpha} x^{\gamma\dot\beta}\partial_{\alpha\dot\beta} )
- k_{\beta\dot\alpha}\theta^{+\beta} \partial^-_\alpha - k_{\alpha\dot\beta}\bar\theta^{+\dot\beta} \bar\partial^-_{\dot\alpha}\,, \quad \delta_k \partial^{--} = 0\,,  \nn
&& \delta_k \partial_\alpha^- = - k_{\alpha\dot\beta} x^{\beta\dot\beta}\,\partial^-_\beta - 4i k_{\alpha\dot\beta}\bar\theta^{+\dot\beta} \partial^{--}\,, \quad
\delta_k \bar\partial_{\dot\alpha}^- = - k_{\gamma\dot\alpha} x^{\gamma\dot\beta}\,\bar\partial^-_{\dot\beta} + 4i k_{\gamma\dot\alpha}\theta^{+\gamma} \partial^{--}\,.
\label{ConfDer}
\end{eqnarray}

It is straightforward to calculate the corresponding transformation properties of various products of these derivatives, e.g. of the bilinear products $\partial_N\partial_M$ appearing in
\eqref{eq: s=3 most general}, \eqref{eq: s=3 most general-1}. As an example we first present the passive form (without ``transport'' term) of the transformation rules of the analytic
gauge potentials of the spin ${\bf 2}$ case:
\bea
&& \delta^*_\epsilon h^{++\alpha\dot\alpha} = -4i \big(\epsilon^{-\alpha}\bar{h}^{++\dot\alpha +} - \bar{\epsilon}^{-\dot\alpha} h^{++\alpha+}
 \big)\,, \quad  \delta^*_\epsilon h^{+4} = 0\,, \nn
 && \delta^*_\epsilon h^{++\alpha+} = \epsilon^{-\alpha} h^{+4}\,, \quad \delta^*_\epsilon h^{++\dot\alpha+} = \bar\epsilon^{-\dot\alpha} h^{+4}\,, \label{Su2} \\
&&\delta^*_k h^{++\alpha\dot\alpha} = k_{\gamma\dot\beta} \big(h^{++ \gamma\dot\alpha}\,x^{\alpha\dot\beta} + h^{++ \alpha\dot\beta}\, x^{\gamma \dot\alpha}\big)\,,
\quad  \delta^*_k h^{+4} = 4ik_{\gamma\dot\beta}\big(h^{++\gamma +} \bar\theta^{+\dot\beta} -  h^{++\dot\beta +}\theta^{+\gamma} \big), \nn
&& \delta^*_k h^{++\alpha+} = k_{\gamma\dot\beta}\big( h^{++\gamma+} \,x^{\alpha\dot\beta} - h^{++\alpha\dot\beta} \theta^{+\gamma}\big), \nn
&& \delta^*_k h^{++\dot\alpha+} = k_{\gamma\dot\beta}\big( h^{++\dot\beta+} \,x^{\gamma\dot\alpha} - h^{++\gamma\dot\alpha} \bar\theta^{+\dot\beta}\big). \label{Kon2}
\eea

It is also useful to explicitly give how the ${\bf s}=3$ gauge potentials defined in \eqref{s3Split} are transformed by ${\cal N}=2$
superconformal group (before any gauge-fixing). We skip the passive transformation rules of the products of various partial derivatives and
quote at once the transformation laws of the analytic potentials\\

\noindent{\bf Supersymmetry:}
\bea
&& \delta^*_\epsilon h^{++\alpha\dot\alpha\, \beta\dot\beta} = 4i \big[\bar\epsilon^{-\dot\beta} h^{++\beta+\alpha\dot\alpha} - \epsilon^{-\beta}h^{++\dot\beta+\alpha\dot\alpha}\big] +
(\alpha, \dot\alpha \Leftrightarrow \beta, \dot\beta)\,, \nn
&& \delta^*_\epsilon h^{++[\beta+\gamma]+} = 2 \epsilon^{-[\beta}\,h^{++++\,\gamma]+}\,, \quad \delta^*_\epsilon h^{++[\dot\beta+\dot\gamma]+}
= 2 \bar\epsilon^{- [\dot\beta}\, h^{++++\,\dot\gamma]+}\,,
\nn
&& \delta^*_\epsilon h^{++\beta+\alpha\dot\alpha} = \epsilon^{-\beta}\,h^{++\alpha\dot\alpha ++} +4i \big(\epsilon^{-\alpha}\,h^{++\beta+\dot\alpha+}
-\bar\epsilon^{-\dot\alpha} h^{++[\beta+\alpha]+}\big) \nn
&& \delta^*_\epsilon h^{++\dot\beta+\alpha\dot\alpha} = \widetilde{\delta^*_\epsilon h^{++\beta+\alpha\dot\alpha}}\,, \quad
\delta^*_\epsilon h^{(+6)} = 0\,, \nn
&& \delta^*_\epsilon h^{++++\beta+} = \epsilon^{-\beta} h^{(+6)}\,, \quad \delta^*_\epsilon h^{++++\dot\beta+} =\widetilde{\delta^*_\epsilon h^{++++\beta+}}\,,   \nn
&&  \delta^*_\epsilon h^{++\alpha\dot\alpha ++} = -4i\big(\epsilon^{-\alpha} h^{++++\dot\alpha +} - \bar\epsilon^{-\dot\alpha}h^{++++\alpha +}\big)\,, \nn
&& \delta^*_\epsilon h^{++\alpha +\dot\alpha +} = \epsilon^{-\alpha} h^{++++\dot\alpha +} - \bar\epsilon^{-\dot\alpha}h^{++++\alpha +}\,.
\label{su33}
\eea

\noindent{\bf Special conformal transformations:}
\bea
&& \delta^*_k h^{++\alpha\dot\alpha\, \beta\dot\beta} = k_{\lambda\dot\rho}x^{\alpha\dot\rho}h^{++\lambda\dot\alpha\, \beta\dot\beta} +
k_{\rho\dot\lambda}x^{\rho\dot\alpha}h^{++\alpha\dot\lambda\, \beta\dot\beta} + (\alpha, \dot\alpha \Leftrightarrow \beta, \dot\beta)\,, \nn
&& \delta^*_k h^{++[\beta+\gamma]+} = (k\cdot x) h^{++[\beta+\gamma]+} + 2 k^\rho_{\dot\alpha}\theta^{+}_\rho h^{++[\beta+\gamma] \dot\alpha}\,, \quad
\delta^*_k h^{++[\dot\beta+\dot\gamma]+} = -\widetilde{(\delta^*_k h^{++[\dot\beta+\dot\gamma]+})} \nn
&& \delta^*_k h^{(+6)} = 8i\big(k_{\gamma\dot\beta}\theta^{+\gamma} \,h^{++++\dot\beta+} - k_{\beta\dot\beta}\bar\theta^{+\dot\beta} \,h^{++++\beta+}\big),\nn
&& \delta^*_k h^{++\beta+\alpha\dot\alpha} = k_{\lambda\dot\rho}x^{\beta\dot\rho}h^{++\lambda + \alpha\dot\alpha} + k_{\lambda\dot\rho}x^{\alpha\dot\rho}
h^{++\beta + \lambda\dot\alpha} + k_{\rho\dot\lambda} x^{\rho\dot\alpha}h^{++\beta + \alpha\dot\lambda} + k_{\rho\dot\beta}\theta^{+\rho}\,h^{++\alpha\dot\alpha\, \beta\dot\beta}\,,  \nn
&& \delta^*_k h^{++\dot\beta+\alpha\dot\alpha} = \widetilde{\delta^*_k h^{++\beta+\alpha\dot\alpha}}\,, \nn
&& \delta^*_k h^{++++\beta+} = k_{\lambda\dot\rho}x^{\beta\dot\rho}\,h^{++++\lambda+}
+ 4i k_{\gamma\dot\rho} \bar\theta^{+\dot\rho}\, h^{++[\beta+\gamma]+}\,, \; \delta^*_k h^{++++\dot\beta+} =\widetilde{(\delta^*_k h^{++++\beta+} )}\,,    \nn
&&  \delta^*_k h^{++\alpha\dot\alpha ++} = k_{\lambda\dot\rho}x^{\beta\dot\rho}\,h^{++\lambda\dot\alpha ++} + k_{\rho\dot\lambda}x^{\rho\dot\alpha}\,
h^{++\alpha\dot\lambda ++} +4i\big(k_{\lambda\dot\rho}\theta^{+\lambda}\, h^{++\dot\rho + \alpha\dot\alpha}\epsilon^{-\alpha} - k_{\lambda\dot\rho}\bar\theta^{+\dot\rho}\,
 h^{++\lambda + \alpha\dot\alpha}\big), \nn
&& \delta^*_k h^{++\alpha +\dot\alpha +} = k_{\lambda\dot\rho}x^{\beta\dot\rho}\,h^{++\lambda+\dot\alpha +} + k_{\rho\dot\lambda}x^{\rho\dot\alpha}\,
h^{++\alpha+\dot\lambda +} + k_{\lambda\dot\rho}\theta^{+\lambda}\, h^{++\dot\rho + \alpha\dot\alpha}\epsilon^{-\alpha} - k_{\lambda\dot\rho}\bar\theta^{+\dot\rho}\,
 h^{++\lambda + \alpha\dot\alpha}.\nn
&& \label{conf33}
\eea

A curious feature of the realization \eqref{su33} and \eqref{conf33} is that, with respect to it, the set of analytic gauge superfields is divided into
an invariant subset and a quotient  over this subset. The invariant subspace is spanned by the potentials
\bea
&& \hat{h}^{++}:= \epsilon_{\alpha\beta} \epsilon_{\dot\alpha\dot\beta} h^{++\alpha\dot\alpha\, \beta\dot\beta}\,, \; \hat{h}^{+3\dot\alpha} := \epsilon_{\alpha\beta} h^{++\alpha+\beta\dot\alpha},
\; \bar{h}^{+3 \dot\alpha}, \; \hat{h}^{+4} :=\frac12\epsilon_{\alpha\beta}h^{++[\beta + \alpha]+}\,, \, \bar{\hat{h}}^{+4}\,, \nn
&& h^{(+6)}\,, \; h^{++++\alpha+}, \;h^{++++\dot\alpha+}
= \widetilde{h^{++++\alpha+}} , \; \hat{g}^{+ 4 \alpha\dot\alpha} := h^{++\alpha\dot\alpha++} - 4i h^{++\alpha +\dot\alpha +}, \label{InvSub}
\eea
while the quotient by
\bea
h^{++(\alpha\beta)\,(\dot\alpha\dot\beta)},\, h^{+++(\alpha\beta)\dot\alpha},\, \widetilde{h^{+++(\alpha\beta)\dot\alpha}}, \,
{g}^{+ 4 \alpha\dot\alpha} := h^{++\alpha\dot\alpha++} + 4i h^{++\alpha +\dot\alpha +}. \label{Quot}
\eea
The closedness of \eqref{InvSub} under both \eqref{su33} and \eqref{conf33} can be readily checked. The remaining set \eqref{Quot} transforms
through \eqref{InvSub} and itself.

Inspecting the linearized gauge transformations \eqref{eq: s=3 gauge transformations vector} and \eqref{eq: spin 3 b}, we observe that the gauge potentials
from the set \eqref{su33} are transformed through the restricted set of gauge parameters
\bea
&&\epsilon_{\alpha\beta} \epsilon_{\dot\alpha\dot\beta} \lambda^{\alpha\beta\dot\alpha\dot\beta}, \quad \epsilon_{\alpha\beta}\lambda^{\alpha\dot\beta\beta}\, ({\rm and} \;c.c.), \quad
\lambda^{(+4)}\,, \quad \epsilon_{\alpha\beta}\lambda^{[\alpha +\beta]+}\, ({\rm and} \;c.c.), \nn
&& \lambda^{++\alpha +}\, ({\rm and} \;c.c.),\quad  \lambda^{++\alpha\dot\alpha} - 8i \lambda^{\alpha+\dot\alpha+}\,. \label{Restrlambda1}
\eea
Based on this observation, we can choose the gauge in which all potentials from the set \eqref{InvSub} are equal to zero and
end up with \eqref{Quot} as encoding the irreducible gauge ${\bf s}=3$
supermultiplet. Such a gauge does not break rigid superconformal symmetry at all. Note that, instead of choosing the gauge $h^{++\alpha\dot\alpha++} - 4i h^{++\alpha +\dot\alpha +} =0$,
in section 6.3 (and Appendix A) we imposed the equivalent gauge $h^{++\alpha +\dot\alpha +} =0$, which is technically more convenient.  Looking at the $\epsilon$ and $k$-transformations of
$h^{++\alpha+\dot\alpha}$ in \eqref{su33} and \eqref{conf33} we observe that in the latter case the r.h.s. of the $k$-transformation contains the ``physical'' non-zero gauge potentials
$h^{+3 (\dot\rho \dot\alpha)\alpha}$ and $h^{+3 (\lambda\alpha) \dot\alpha}$. So this gauge seemingly breaks superconformal covariance. However, it is easy to check that in the WZ gauge
\eqref{eq: spin 3 GF} for these gauge potentials the sum of the problematic terms in $\delta^*_k h^{++\alpha+\dot\alpha}$ vanishes. So the breaking just mentioned is in fact fictitious.

As the last topic of this Appendix, we discuss  the modification (before imposing any gauge) of the superconformal properties of $h^{++}$ compared to the standard
superconfomal law \eqref{eq: superconformal vector} of the spin ${\bf 1}$ analytic gauge potential. As before, we will deal with the passive form of the conformal transformations.
The modification appears only in the realization of special conformal
transformations due to the property that such transformations of the bilinear products of partial derivatives in $\hat{\cal H}_{s=3}$ contain terms with one derivative.
After integrating by parts, with taking into account that $\Omega_{sc}$ defined in \eqref{eq:conformal weight} is reduced to $2 (x\cdot k)$ for $k$-transformations,
we obtain  the following addition to the conformal  transformation of $h^{++}$:
\bea
\delta_{k}^{*} h^{++} = 2k_{\alpha\dot\alpha}\big(\partial^-_\beta h^{+++(\beta\alpha)\dot\alpha} + \partial^-_{\dot\beta} h^{+++(\dot\beta\dot\alpha)\alpha}
-\partial_{\beta\dot\beta} h^{++(\alpha\beta)(\dot\alpha\dot\beta)}
- \frac12\partial^{--}h^{++\alpha\dot\alpha++} \big). \nonumber
\eea
It is easy to find the compensating gauge transformation of $h^{++}$ of the type \eqref{eq: spin 3 c}, which ensures the $k$-invariance of the gauge $h^{++}=0$
(with WZ gauge \eqref{eq: spin 3 GF} for all other potentials).



\begin{thebibliography}{99}

    \bibitem{FP}
    E.~S.~Fradkin, M.~Ya.~Palchik, {\it Conformal Quantum Field Theory in D-dimensions}, Springer,
    1996, 466 p.

    \bibitem{FMS}
    P.~Di Francesco, P.~Mathieu, D.~Senechal, {\it Conformal field theory}, Springer, 1997, 911 p.

    \bibitem{G}
    M.~R.~Gaberdiel, {\it An introduction to conformal field theory}, Rept. Prog. Phys. {\bf 63} (2000) 607-667,
    [arXiv: hep-th/9910156].

    \bibitem{Rychkov:2016iqz}
    S.~Rychkov,
    {\it EPFL Lectures on Conformal Field Theory in D $\geq$ 3 Dimensions},
    [arXiv:1601.05000 [hep-th]].

    \bibitem{BK}
    I.~L.~Buchbinder, S.~M.~Kuzenko, {\it Ideas and Methods of
    Supersymmetry and Supergravity or a Walk Through Superspace}, IOP Publishing, 1996,
    656 p.

    \bibitem{Freedman:2012zz}
    D.~Z.~Freedman and A.~Van Proeyen,
    {\it Supergravity},
    Cambridge Univ. Press, 2012, 626 p.

    \bibitem{Vasiliev:2003cph}
    M.~A.~Vasiliev,
    {\it Higher spin gauge theories in various dimensions},
    PoS \textbf{JHW2003} (2003) 003,
    [arXiv:hep-th/0401177 [hep-th]].

    \bibitem{Bekaert:2004qos}
    X.~Bekaert, S.~Cnockaert, C.~Iazeolla and M.~A.~Vasiliev,
    {\it Nonlinear higher spin theories in various dimensions},
    [arXiv:hep-th/0503128 [hep-th]].

    \bibitem{Bekaert:2022poo}
    X.~Bekaert, N.~Boulanger, A.~Campoleoni, M.~Chiodaroli, D.~Francia, M.~Grigoriev, E.~Sezgin and E.~Skvortsov,
    {\it Snowmass White Paper: Higher Spin Gravity and Higher Spin Symmetry},
    [arXiv:2205.01567 [hep-th]].

    \bibitem{Tseytlin:2002gz}
    A.~A.~Tseytlin,
    {\it On limits of superstring in $AdS(5) \times S^5$},
    Theor. Math. Phys. \textbf{133} (2002) 1376-1389,
    [arXiv:hep-th/0201112 [hep-th]].

    \bibitem{Gaberdiel:2015wpo}
    M.~R.~Gaberdiel and R.~Gopakumar,
    {\it String Theory as a Higher Spin Theory},
    JHEP \textbf{09} (2016) 085,
    [arXiv:1512.07237 [hep-th]].

    \bibitem{Ponomarev:2022vjb}
    D.~Ponomarev,
    {\it Basic introduction to higher-spin theories},
    [arXiv:2206.15385 [hep-th]].

    \bibitem{Fronsdal:1978rb}
    C.~Fronsdal,
    \textit{Massless Fields with Integer Spin},
    Phys. Rev. D \textbf{18} (1978) 3624.

    \bibitem{Fang:1978wz}
    J.~Fang and C.~Fronsdal,
    \textit{Massless Fields with Half Integral Spin},
    Phys. Rev. D \textbf{18} (1978) 3630.


    \bibitem{Fradkin:1985am}
    E.~S.~Fradkin and A.~A.~Tseytlin,
    {\it Conformal supergravity},
    Phys. Rept. \textbf{119} (1985) 233-362.

    \bibitem{Metsaev:2007rw}
    R.~R.~Metsaev,
    \textit{Ordinary-derivative formulation of conformal totally symmetric arbitrary spin bosonic fields},
    JHEP \textbf{06} (2012) 062,
    [arXiv:0709.4392 [hep-th]].

    \bibitem{Vasiliev:2009ck}
    M.~A.~Vasiliev,
    {\it Bosonic conformal higher-spin fields of any symmetry},
    Nucl. Phys. B \textbf{829} (2010) 176-224,
    [arXiv:0909.5226 [hep-th]].



    \bibitem{Joung:2012qy}
    E.~Joung and K.~Mkrtchyan,
    \textit{A note on higher-derivative actions for free higher-spin fields},
    JHEP \textbf{11} (2012) 153,
    [arXiv:1209.4864 [hep-th]].

    \bibitem{Nutma:2014pua}
    T.~Nutma and M.~Taronna,
    \textit{On conformal higher spin wave operators},
    JHEP \textbf{06} (2014) 066,
    [arXiv:1404.7452 [hep-th]].

    \bibitem{Basile:2018eac}
    T.~Basile, X.~Bekaert and E.~Joung,
    \textit{Conformal Higher-Spin Gravity: Linearized Spectrum = Symmetry Algebra},
    JHEP \textbf{11} (2018) 167,
    [arXiv:1808.07728 [hep-th]].


    \bibitem{Grigoriev:2016bzl}
    M.~Grigoriev and A.~A.~Tseytlin,
    {\it On conformal higher spins in curved background},
    J. Phys. A \textbf{50} (2017) 125401,
    [arXiv:1609.09381 [hep-th]].

    \bibitem{Beccaria:2017nco}
    M.~Beccaria and A.~A.~Tseytlin,
    \textit{On induced action for conformal higher spins in curved background},
    Nucl. Phys. B \textbf{919} (2017) 359-383,
    [arXiv:1702.00222 [hep-th]].

    \bibitem{Manvelyan:2018qxd}
    R.~Manvelyan and G.~Poghosyan,
    \textit{Geometrical structure of Weyl invariants for spin three gauge field in general gravitational background in $d=4$},
    Nucl. Phys. B \textbf{937} (2018) 1-27,
    [arXiv:1804.10779 [hep-th]].


    \bibitem{Kuzenko:2019ill}
    S.~M.~Kuzenko and M.~Ponds,
    {\it Conformal geometry and (super)conformal higher-spin gauge theories},
    JHEP \textbf{05} (2019) 113,
    [arXiv:1902.08010 [hep-th]].

    \bibitem{Kuzenko:2019eni}
    S.~M.~Kuzenko and M.~Ponds,
    {\it Generalised conformal higher-spin fields in curved backgrounds},
    JHEP \textbf{04} (2020) 021,
    [arXiv:1912.00652 [hep-th]].

    \bibitem{Kuzenko:2020jie}
    S.~M.~Kuzenko, M.~Ponds and E.~S.~N.~Raptakis,
    {\it New locally (super)conformal gauge models in Bach-flat backgrounds},
    JHEP \textbf{08} (2020) 068,
    [arXiv:2005.08657 [hep-th]].

    \bibitem{Basile:2022nou}
    T.~Basile, M.~Grigoriev and E.~Skvortsov,
    {\it Covariant action for conformal higher spin gravity},
    [arXiv:2212.10336 [hep-th]].



    \bibitem{Fradkin:1989md}
    E.~S.~Fradkin and V.~Y.~Linetsky,
    {\it Cubic Interaction in Conformal Theory of Integer Higher Spin Fields in Four-dimensional Space-time},
    Phys. Lett. B \textbf{231} (1989) 97-106.

    \bibitem{Fradkin:1990ps}
    E.~S.~Fradkin and V.~Y.~Linetsky,
    \textit{Superconformal Higher Spin Theory in the Cubic Approximation},
    Nucl. Phys. B \textbf{350} (1991) 274-324.

   \bibitem{Segal:2002gd}
   A.~Y.~Segal,
   {\it Conformal higher spin theory},
   Nucl. Phys. B \textbf{664} (2003) 59-130,
   [arXiv:hep-th/0207212 [hep-th]].

    \bibitem{Bekaert:2010ky}
    X.~Bekaert, E.~Joung and J.~Mourad,
    {\it Effective action in a higher-spin background},
    JHEP \textbf{02} (2011) 048,
    [arXiv:1012.2103 [hep-th]].

    \bibitem{Bonezzi:2017mwr}
    R.~Bonezzi,
    {\it Induced Action for Conformal Higher Spins from Worldline Path Integrals},
    Universe \textbf{3} (2017) 64,
    [arXiv:1709.00850 [hep-th]].

    \bibitem{Kuzenko:2022hdv}
    S.~M.~Kuzenko, M.~Ponds and E.~S.~N.~Raptakis,
    \textit{Conformal Interactions Between Matter and Higher-Spin (Super)Fields},
    Fortsch. Phys. \textbf{71} (2023) 1,
    [arXiv:2208.07783 [hep-th]].






    \bibitem{Kuzenko:2017ujh}
    S.~M.~Kuzenko, R.~Manvelyan and S.~Theisen,
    \textit{Off-shell superconformal higher spin multiplets in four dimensions},
    JHEP \textbf{07} (2017) 034,
    [arXiv:1701.00682 [hep-th]].

    \bibitem{Buchbinder:2017nuc}
    I.~L.~Buchbinder, S.~J.~Gates and K.~Koutrolikos,
    \textit{Higher Spin Superfield interactions with the Chiral Supermultiplet: Conserved Supercurrents and Cubic Vertices},
    Universe \textbf{4} (2018) no.1, 6
    [arXiv:1708.06262 [hep-th]].

    \bibitem{Koutrolikos:2017qkx}
    K.~Koutrolikos, P.~Ko\v{c}\'\i{} and R.~von Unge,
    \textit{Higher Spin Superfield interactions with Complex linear Supermultiplet: Conserved Supercurrents and Cubic Vertices},
    JHEP \textbf{03} (2018), 119
    [arXiv:1712.05150 [hep-th]].

    \bibitem{Buchbinder:2018gle}
    I.~L.~Buchbinder, S.~J.~Gates and K.~Koutrolikos,
    \textit{Integer superspin supercurrents of matter supermultiplets},
    JHEP \textbf{05} (2019), 031
    [arXiv:1811.12858 [hep-th]].

    \bibitem{Kuzenko:1993jp}
    S.~M.~Kuzenko, A.~G.~Sibiryakov and V.~V.~Postnikov,
    \textit{Massless gauge superfields of higher half integer superspins},
    JETP Lett. \textbf{57} (1993) 534-538.

    \bibitem{Kuzenko:1993jq}
    S.~M.~Kuzenko and A.~G.~Sibiryakov,
    \textit{Massless gauge superfields of higher integer superspins},
    JETP Lett. \textbf{57} (1993) 539-542.

    \bibitem{Kuzenko:1994dm}
    S.~M.~Kuzenko and A.~G.~Sibiryakov,
    \textit{Free massless higher superspin superfields on the anti-de Sitter superspace},
    Phys. Atom. Nucl. \textbf{57} (1994) 1257-1267,
    [arXiv:1112.4612 [hep-th]].

    \bibitem{Gates:2013rka}
    S.~J.~Gates, Jr. and K.~Koutrolikos,
    \textit{On 4D, $\mathcal{N} = 1$ massless gauge superfields of arbitrary superhelicity},
    JHEP \textbf{06} (2014), 098
    [arXiv:1310.7385 [hep-th]].

    \bibitem{Koutrolikos:2020tel}
    K.~Koutrolikos,
    {\it Superspace formulation of massive half-integer superspin},
    JHEP \textbf{03} (2021), 254
    [arXiv:2012.12225 [hep-th]].


    \bibitem{Buchbinder:2020yip}
    I.~L.~Buchbinder, S.~J.~Gates and K.~Koutrolikos,
    {\it Hierarchy of Supersymmetric Higher Spin Connections},
    Phys. Rev. D \textbf{102} (2020), 125018
    [arXiv:2010.02061 [hep-th]].

    \bibitem{Koutrolikos:2022chj}
    K.~Koutrolikos,
    \textit{Superspace first-order formalism for massless arbitrary superspin supermultiplets},
    Phys. Rev. D \textbf{105} (2022) no.12, 125008
    [arXiv:2204.04181 [hep-th]].

    \bibitem{Howe:1981qj}
    P.~S.~Howe, K.~S.~Stelle and P.~K.~Townsend,
    {\it Suprecurrents},
    Nucl. Phys. B \textbf{192} (1981) 332-352.


    \bibitem{Kuzenko:2020opc}
    S.~M.~Kuzenko, M.~Ponds and E.~S.~N.~Raptakis,
    \textit{Generalised superconformal higher-spin multiplets},
    JHEP \textbf{03} (2021) 183,
    [arXiv:2011.11300 [hep-th]].

    \bibitem{Hutomo:2017nce}
    J.~Hutomo and S.~M.~Kuzenko,
    \textit{The massless integer superspin multiplets revisited},
    JHEP \textbf{02} (2018) 137,
    [arXiv:1711.11364 [hep-th]].

    \bibitem{Kuzenko:2021pqm}
    S.~M.~Kuzenko and E.~S.~N.~Raptakis,
    \textit{Extended superconformal higher-spin gauge theories in four dimensions},
    JHEP \textbf{12} (2021) 210,
    [arXiv:2104.10416 [hep-th]].

    \bibitem{Hutchings:2023iza}
    D.~Hutchings, S.~M.~Kuzenko and E.~S.~N.~Raptakis,
    {\it The $\mathcal{N}=2$ superconformal gravitino multiplet},
    Phys. Lett. B \textbf{845} (2023), 138132
    [arXiv:2305.16029 [hep-th]].

    \bibitem{Butter:2011sr}
    D.~Butter,
    \textit{N=2 Conformal Superspace in Four Dimensions},
    JHEP \textbf{10} (2011) 030,
    [arXiv:1103.5914 [hep-th]].

    \bibitem{Buchbinder:2018wzq}
    I.~L.~Buchbinder, S.~J.~Gates and K.~Koutrolikos,
    \textit{Conserved higher spin supercurrents for arbitrary spin massless supermultiplets and higher spin superfield cubic interactions},
    JHEP \textbf{08} (2018), 055
    [arXiv:1805.04413 [hep-th]].

    \bibitem{Gates:2019cnl}
    S.~J.~Gates and K.~Koutrolikos,
    \textit{Progress on cubic interactions of arbitrary superspin supermultiplets via gauge invariant supercurrents},
    Phys. Lett. B \textbf{797} (2019), 134868
    [arXiv:1904.13336 [hep-th]].

    \bibitem{Buchbinder:2021ite}
    I.~Buchbinder, E.~Ivanov and N.~Zaigraev,
    \textit{Unconstrained off-shell superfield formulation of 4D, $ \mathcal{N} $ = 2 supersymmetric higher spins},
    JHEP \textbf{12} (2021) 016,
    [arXiv:2109.07639 [hep-th]].

    \bibitem{Buchbinder:2022vra}
    I.~Buchbinder, E.~Ivanov and N.~Zaigraev,
    {\it $\mathcal{N} = 2$ higher spins: superfield equations of motion, the hypermultiplet supercurrents, and the component structure}, JHEP {\bf 03} (2023) 036,
    [arXiv:2212.14114 [hep-th]].


    \bibitem{Buchbinder:2022kzl}
    I.~Buchbinder, E.~Ivanov and N.~Zaigraev,
    \textit{Off-shell cubic hypermultiplet couplings to $\mathcal{N}=2$ higher spin gauge superfields},
    JHEP \textbf{05} (2022) 104,
    [arXiv:2202.08196 [hep-th]].



 \bibitem{Buchbinder:2022svx}
 I.~Buchbinder, E.~Ivanov and N.~Zaigraev,
 {\it Unconstrained $\mathcal{N} = 2$ Higher-Spin Gauge Superfields and Their Hypermultiplet Couplings},
 Phys. Part. Nucl. Lett. \textbf{20} (2023) no.3, 300-305
 [arXiv:2211.09501 [hep-th]].

     \bibitem{Ivanov23} E.~Ivanov, {\it Higher Spins in Harmonic Superspace}, Teor. Mat. Fiz. {\bf 217} (2023) 3 515-532
     [Theor. Math. Phys. \textbf{217} (2023) 1855-1869], [arXiv:2306.10401 [hep-th]].

    \bibitem{HSS} A.~Galperin, E.~Ivanov, V.~Ogievetsky, E.~Sokatchev, {\it Harmonic superspace: key to $N=2$ supersymmetric theories},
    Pis'ma ZhETF {\bf 40} (1984) 155 [JETP Lett. {\bf 40} (1984) 912].

    \bibitem{HSS1}
    A.~S.~Galperin, E.~A.~Ivanov, S.~Kalitzin, V.~I.~Ogievetsky, E.~S.~Sokatchev,{\it Unconstrained ${\cal N}=2$ Matter, Yang-Mills
        and Supergravity Theories in Harmonic Superspace}, Class. Quant. Grav. {\bf 1} (1984) 469-498 [Erratum: Class. Quant. Grav. {\bf 2}
    (1985) 127].


    \bibitem{18} A.~S.~Galperin, E.~A.~Ivanov, V.~I.~Ogievetsky, E.~S.~Sokatchev,
    {\it Harmonic superspace}, Cambridge Monographs on Mathematical
    Physics, Cambridge University Press, 2001, 306 p.

    \bibitem{Galperin:1985de}
    A.~Galperin, E.~Ivanov, V.~Ogievetsky and E.~Sokatchev,
    \textit{Hyperkahler Metrics and Harmonic Superspace},
    Commun. Math. Phys. \textbf{103} (1986) 515.

    \bibitem{Ivanov:2022vwc}
    E.~Ivanov,
    {\it ${\cal N}=2\,$ Supergravities in Harmonic Superspace},
    [arXiv:2212.07925 [hep-th]].

    \bibitem{Galperin:1985zv}
    A.~Galperin, E.~Ivanov, V.~Ogievetsky and E.~Sokatchev,
    {\it Conformal invariance in harmonic superspace},
    JINR-E2-85-363.

    \bibitem{Fotopoulos:2007yq}
    A.~Fotopoulos, N.~Irges, A.~C.~Petkou and M.~Tsulaia,
    \textit{Higher-Spin Gauge Fields Interacting with Scalars: The Lagrangian Cubic Vertex},
    JHEP \textbf{10} (2007) 021,
    [arXiv:0708.1399 [hep-th]].

    \bibitem{Galperin:1987em}
    A.~S.~Galperin, N.~A.~Ky and E.~Sokatchev,
    \textit{$\mathcal{N}$=2 Supergravity in Superspace: Solution to the Constraints},
    Class. Quant. Grav. \textbf{4} (1987) 1235.

    \bibitem{Galperin:1987ek}
    A.~S.~Galperin, E.~A.~Ivanov, V.~I.~Ogievetsky and E.~Sokatchev,
    \textit{$\mathcal{N}$=2 Supergravity in Superspace: Different Versions and Matter Couplings},
    Class. Quant. Grav. \textbf{4} (1987) 1255.

    \bibitem{Bergshoeff:1980is}
    E.~Bergshoeff, M.~de Roo and B.~de Wit,
    {\it Extended Conformal Supergravity},
    Nucl. Phys. B \textbf{182} (1981) 173-204.

    \bibitem{Muller:1986ku}
    M.~Muller,
    {\it Minimal $\mathcal{N}=2$ Supergravity in Superspace},
    Nucl. Phys. B \textbf{282} (1987) 329-348.


    \bibitem{Zupnik:1998td}
    B.~M.~Zupnik,
    \textit{Background harmonic superfields in $\mathcal{N}$=2 supergravity},
    Theor. Math. Phys. \textbf{116} (1998) 964-977,
    [arXiv:hep-th/9803202 [hep-th]].


    \bibitem{Kuzenko:1999pi}
    S.~M.~Kuzenko and S.~Theisen,
    \textit{Correlation functions of conserved currents in $\mathcal{N}$=2 superconformal theory},
    Class. Quant. Grav. \textbf{17} (2000) 665-696,
    [arXiv:hep-th/9907107 [hep-th]].



    \bibitem{Rivelles:1981qz}
    V.~O.~Rivelles and J.~G.~Taylor,
    {\it Linearized $\mathcal{N}$=2 superfield supergravity},
    J. Phys. A \textbf{15} (1982) 163.

    \bibitem{Gates:1981qq}
    S.~J.~Gates, Jr. and W.~Siegel,
    {\it Linearized $\mathcal{N}$=2 superfield supergravity},
    Nucl. Phys. B \textbf{195} (1982) 39-60.

    \bibitem{Butter:2010sc}
    D.~Butter and S.~M.~Kuzenko,
    \textit{$\mathcal{N}$=2 supergravity and supercurrents},
    JHEP \textbf{12} (2010) 080,
    [arXiv:1011.0339 [hep-th]].






    \bibitem{Kuzenko:2023vgf}
    S.~M.~Kuzenko and E.~S.~N.~Raptakis,
    \textit{On higher-spin $ \mathcal{N} $ = 2 supercurrent multiplets},
    JHEP \textbf{05} (2023) 056,
    [arXiv:2301.09386 [hep-th]].





    \bibitem{Curtright:1980yj}
    T.~L.~Curtright and P.~G.~O.~Freund,
    {\it Masssive dual fields},
    Nucl. Phys. B \textbf{172} (1980) 413-424.

    \bibitem{Curtright:1980yk}
    T.~Curtright,
    {\it Generalized gauge fields},
    Phys. Lett. B \textbf{165} (1985) 304-308.


    \bibitem{Ogievetsky:1966eiu}
    V.~I.~Ogievetsky and I.~V.~Polubarinov,
    {\it The notoph and its possible interactions},
    Yad. Fiz. \textbf{4} (1966) 216-223.

     \bibitem{CaRa} M.~Kalb and P.~Ramond, {\it Classical direct interstring action}, Phys. Rev. D {\bf 9} (1974) 2273-2294.


    \bibitem{Ivanov:2016lha}
    E.~A.~Ivanov,
    \textit{Gauge Fields, Nonlinear Realizations, Supersymmetry},
    Phys. Part. Nucl. \textbf{47} (2016) no.4 508-539
    [arXiv:1604.01379 [hep-th]].



    \bibitem{BS}
    I.~L.~Buchbinder, I.~L.~Shapiro, {\it Introduction to Quantum Field
    Theory with Applicaions to Quantum Gravity}, Oxford Graduate Texts, Oxford Univ. Press,
    2021, 525 p.

    \bibitem{Bonora1}
    L.~Bonora, M.~Cvitan, P.~Dominis Prester, S.~Giaccari, B.~Lima de
    Souza, T.~Stemberga, {\it One-loop effective actions and higher spins},
    JHEP \textbf{12} (2016) 084, [arXiv:1609.02088 [hep-th]].


    \bibitem{Bonora2}
    L.~Bonora, M.~Cvitan, P.~Dominis Prester, S.~Giaccari,
    T.~Stemberga, {\it One-loop effective actions and higher spins. Part
    II}, JHEP  \textbf{01} (2018) 080, [arXiv:1709.01738 [hep-th]].

    \bibitem{KuLaPo}
    S.M.~Kuzenko, J.~La Fontaine and M.~Ponds,
    \textit{Induced action for superconformal higher-spin multiplets using SCFT techniques},
    Phys. Lett. B \textbf{839} (2023) 137818, [arXiv:2212.00468 [hep-th]].



\bibitem{Raptakis:2023gyu}
E.~S.~N.~Raptakis,
{\it Aspects of superconformal symmetry},
doi:10.26182/9b7c-7c08
[arXiv:2403.02700 [hep-th]].


\end{thebibliography}
\end{document}